\documentclass[11pt,a4paper]{article}

\usepackage{bold-extra}

\usepackage{graphicx}

\graphicspath{ {./Figures/} }

\usepackage{authblk}
\usepackage{comment}

\pdfoutput=1
\usepackage{color}
\usepackage{todonotes}
\usepackage{booktabs}
\usepackage{array}
\usepackage{multirow,rotating}
\usepackage{amsmath,amssymb,bm,cite,graphicx,wasysym,mathrsfs,dsfont,amsfonts}
\usepackage{slashed,relsize}
\usepackage[colorlinks=true,linkcolor=blue,anchorcolor=black,citecolor=blue,filecolor=cyan,menucolor=red,runcolor=filecolor,urlcolor=blue,bookmarks=true,bookmarksnumbered=true]{hyperref}

\usepackage[font=small,labelfont=bf]{caption}
\usepackage{subcaption}

\def\beq{\begin{equation}\displaystyle\displaystyle}
\def\eeq{\end{equation}}
\def\bea{\begin{eqnarray}\displaystyle} 
\def\eea{\end{eqnarray}}

\def\({\left(}
\def\){\right)}
\def\bry{\begin{array}}
\def\ery{\end{array}}

\newcommand\Tau{\mathcal{T}}

\def\data{{\cal{D}}}
\def\N{{\cal{N}}}

\title{\bf Boosting likelihood learning with event reweighting \vspace{0.4cm}} 
\date{}

\author[1]{Siyu Chen}
\author[2]{Alfredo Glioti}
\author[3,4]{Giuliano Panico}
\author[5,6]{Andrea Wulzer}

\affil[1]{\emph{Institut de Th\'eorie des Ph\'enomenes Physiques, EPFL, Lausanne, Switzerland}}
\affil[2]{\emph{Universit\'e Paris-Saclay, CNRS, CEA, Institut de Physique Th\'eorique, 91191, Gif-sur-Yvette, France}}
\affil[3]{\emph{Dipartimento di Fisica e Astronomia, Universit\`a di Firenze, \protect\\ Via G. Sansone 1, 50019 Sesto Fiorentino, Italy}} 
\affil[4]{\emph{INFN, Sezione di Firenze, Via G. Sansone 1, 50019 Sesto Fiorentino, Italy}}
\affil[5]{Institut de F\'{\i}sica d'Altes Energies (IFAE), The Barcelona Institute of Science and Technology (BIST),
Campus UAB, 08193 Bellaterra, Barcelona, Spain}
\affil[6]{ICREA, Instituci\'o Catalana de Recerca i Estudis Avan\c{c}ats, 
Passeig de Llu\'{\i}s Companys 23, 
08010 Barcelona, Spain}

\begin{document}
\baselineskip=14pt

\maketitle

\begin{abstract}
Extracting maximal information from experimental data requires access to the likelihood function, which however is never directly available for complex experiments like those performed at high energy colliders. Theoretical predictions are obtained in this context by Monte Carlo events, which do furnish an accurate but abstract and implicit representation of the likelihood. Strategies based on statistical learning are currently being developed to infer the likelihood function explicitly by training a continuous-output classifier on Monte Carlo events. In this paper, we investigate the usage of Monte Carlo events that incorporate the dependence on the parameters of interest by reweighting. This enables more accurate likelihood learning with less training data and a more robust learning scheme that is more suited for automation and extensive deployment. We illustrate these advantages in the context of LHC precision probes of new Effective Field Theory interactions.
\end{abstract}

\thispagestyle{empty}

\newpage

\begingroup
\tableofcontents
\endgroup 

\setcounter{equation}{0}
\setcounter{footnote}{0}
\setcounter{page}{1}

\newpage

\section{Introduction}\label{sec:intro}

The problem of extracting maximal information from precise experimental measurements compared with precise theoretical predictions is of prime relevance in several domains of science. In the context of high-energy collider physics, the problem should be addressed for the exploitation of the data collected by the Large Hadron Collider (LHC) experiments and of those of its forthcoming High Luminosity (HL-LHC) upgrade. Precision physics is also a major component of proposed future collider projects such as an $e^+e^-$ Higgs factory or a high-energy muon collider. The corresponding sensitivity projection studies can thus benefit from advances in precision physics analysis methodologies.

High energy physics data sets, $\data=\{x_i\}_{i=1}^{\N}$, consist of repeated measurements of a multi-component statistical variable $x$ of observables. The number of data points $\N$ is also a statistical variable, which follows a Poisson distribution with expected $N$. Both the probability distribution of $x$ and the expected number of events are controlled by the microscopic laws of fundamental interactions, which in turn can depend on a number of parameters of interest, $c$. Specifically, the differential cross section $d\sigma(x;c)$ depends on the parameters of interest and so it does the total cross section $\sigma(c)$, which is the integral of $d\sigma$ over the support of the variable $x$. The probability density function of $x$ is the ratio $d\sigma(x;c)/\sigma(c)$. The total number of expected events is $N(c)=L\,\sigma(c)$, where the proportionality factor $L$---the integrated luminosity---is determined by the properties of the collider including its run time. The task of the analysis is to extract information on the parameters of interest by comparing the observed data with their parameter-dependent expected distribution.

The parameters of interest $c$ could be free parameters of the currently established theoretical description of fundamental interactions, the Standard Model (SM) theory. Or, they could parametrize deformations of the SM due to additional or different interactions. In the latter case, which is by far the most common one for LHC and HL-LHC applications, the first goal of the analysis is to establish whether the data favor the SM point $c=0$ in the parameter space, or if instead $c\neq0$ is preferred hinting to non-SM fundamental physics laws. In the former case, one would like to set an exclusion limit, namely to quantify the maximal value that the $c$ parameters can conceivably assume given that the data have not revealed their presence. In the latter case, one would first aim at quantifying the degree of confidence for the discovery of non-SM physical laws, i.e. the confidence level for SM exclusion. The measurement of the value of the $c$ parameters will become relevant at a later stage.

Regardless of the specific goal of the analysis, any classical or Bayesian statistical inference methodology aimed at optimality is based on the knowledge of the likelihood function associated with the experimental data. This is given in our case by the extended likelihood
\beq
\mathfrak{L}(c;\data)={\textrm{Pois}}[\N|N(c)]\prod\limits_{x\in\data}p(x|c)=
\frac{L^\N}{\N !} e^{-N(c)}\prod\limits_{x\in\data}d\sigma(x;c)\,.
\eeq
More precisely, what is truly needed is the dependence of the likelihood on the $c$ parameters, up to a $c$-independent normalization factor. It is natural in our context to employ the likelihood at the SM point $c=0$ for normalization. Optimal statistical inference can thus be attained by the knowledge of the likelihood log-ratio
\beq\label{eq:log-lik}
\lambda(c;{\cal D}) \equiv \log \frac{\mathfrak{L}(c;\data)}{\mathfrak{L}(0;\data)}
= N(0) - N(c) + \sum_{x \in {\cal D}} \log r(x; c)\,,
\eeq
having defined the ratio of differential cross sections
\beq\label{eq:r}
r(x; c) = \frac{d \sigma(x; c)}{d \sigma(x; 0)}\,.
\eeq

In high-energy physics, the theoretical comprehension of fundamental interactions is translated into predictions for the outcome of experiments by a sophisticated chain of analytical and numerical tools that eventually produces Monte Carlo events. In the simplest setup, the events are \emph{unweighted}. In this case, the events are statistical samples that follow the $x$ variable distribution $d\sigma(x;c)/\sigma(c)$. An estimate of the total cross section $\sigma(c)$---and in turn of the expected number of events $N(c)$---is also available at the end of the event generation process as the result of the Monte Carlo integration. Equal weights are conventionally assigned to unweighted events, given by $\sigma(c)$ divided by the number of generated events. Notice that since the differential cross section depends on $c$, independent sets of Monte Carlo events need to be generated at different points in the $c$ parameter space. This limitation can be avoided by employing instead \emph{weighted} Monte Carlo events, to be described later.

Importantly enough, the Monte Carlo event generators in high energy physics do not sample from the distribution of $x$ directly. The differential cross section $d\sigma(x;c)$ can never be computed theoretically and is unknown. What is instead computed and available is the differential cross section $d\sigma(\xi;c)$ in a space of latent variables $\xi$. The $\xi$ variables are in general completely unrelated with the measured variables $x$. The Monte Carlo code generates samples in the $\xi$ space and obtains events in the $x$ space by propagating $\xi$ samples through a series of steps that eventually entail dimensionality reduction. In fact, several of these steps---such as the QCD and QED radiation showering and the simulation of the detector response---are performed by randomized algorithms. The random variables they draw are effectively additional components of the large latent variable vector that is ultimately projected onto the space of observable variables $x$. It is normally impossible to model this complex process analytically such as to obtain a closed form for the distribution in the $x$ space starting from the known distribution $d\sigma(\xi;c)$ in the latent space. One should perform multiple convolutions, and integrate over the unobserved components of the latent vector. These integrals can not be solved analytically and numerical approaches are not viable because the integration should be performed point-by-point in the $x$ space. In this paper, we employ a simple Monte Carlo generator for the validation of our method in a fully controlled ``ideal'' setup, described in Appendix~\ref{app}. This generator exemplifies in practice the role of latent variables. By design, our ideal setup enables a simple integration---by a finite sum, since the variables are discrete---over the latent space. In the realistic setups to be employed for real data analyses, it is on the contrary never possible to perform the integration. The $x$ distribution can not be determined and consequently, we do not have access to the likelihood ratio~(\ref{eq:log-lik}). It is normally possible to determine $N(c)=L\,\sigma(c)$ by the Monte Carlo integration. What is missing is the distribution ratio $r(x;c)$ defined in eq.~(\ref{eq:r}).

Several groups~\cite{Brehmer:2018kdj,Brehmer:2018eca,Brehmer:2018hga,Brehmer:2019xox,Chen:2020mev,Chatterjee:2022oco,Chatterjee:2021nms,Arganda:2022qzy,Arganda:2022zbs,GomezAmbrosio:2022mpm,Kong:2022rnd,Hollingsworth:2020kjg} recently investigated the possibility of extracting $r(x;c)$ from Monte Carlo simulations by employing statistical learning techniques. The goal is to address the limitations of traditional methods such as the matrix element method~\cite{Kondo:1988yd,Artoisenet:2010cn,Fiedler:2010sg,Martini:2015fsa,Martini:2017ydu,Campbell:2012cz,Prestel:2019neg} and similar techniques~\cite{Atwood:1991ka,Diehl:1993br,Dunietz:1990cj,Dighe:1998vk,Durieux:2015zwa,Durieux:2018tev,Pretz:2018bze,Bortolato:2020zcg,Faroughy:2019ird,Gao:2010qx}. In particular, the novel methodologies do not rely on approximate phenomenological modeling of the distribution. Hence, they promise to be universally applicable, simpler and faster to set up and to run, as well as systematically improvable by employing more accurate Monte Carlo generators and larger training data sets. They could be automated to a large extent, enabling their extensive deployment. Attempts are  MadMiner~\cite{Brehmer:2019xox} and  ML4EFT~\cite{GomezAmbrosio:2022mpm}. 

A striking use case for the deployment of these methodologies at the LHC and beyond is the problem of testing new physics effects described by interaction operators of dimension larger than four in an Effective Field Theory (EFT) framework. Most of the work performed so far, and the one presented in the present paper, is specifically tailored to EFT applications. However, many of the results obtained in this context, including ours, are arguably portable to other problems. 

In this paper, we demonstrate the advantages of learning $r(x;c)$ by training the statistical model with weighted Monte Carlo events, as obtained from generators that incorporate the dependence on the $c$ parameters by the technique of \emph{event reweighting}. Weighted Monte Carlo events, unlike unweighted ones, are not samples of the $x$ variable, though they cover the same support with a similar distribution. They come with their own weights, to be employed for weighted sums in the calculation of expectation values. The sum of the weights in the data set is equal to the total cross section, and the sum of the weights of those events that fall in a certain bin (i.e., a region of the $x$ variable space) equals the cross section in that bin. In general, the weighted sum of any observable ${\mathcal{O}}(x)$ evaluated over the events provides an estimate of the expectation value of the observable over $x$, multiplied by the total cross section. It is often merely a matter of convenience whether to employ weighted or unweighted events.\footnote{Notice however that weighted events are a necessity for precise Monte Carlo generators that include radiative loop corrections because the function to be sampled is not manifestly positive by an artifact of the loop expansion. Therefore, the weights can assume negative values and the events cannot be unweighted.} Weighted events obtained by reweighting are more useful in our context.

Event reweighting exploits the knowledge of the differential cross section $d\sigma(\xi;c)$ in the latent space. The generator code has access to this function, as well as to the value of the $\xi$ variables for each event that is sampled. After the sampling in the $\xi$ space, all the steps performed by the generator code---including QCD and QED radiation showering and detector simulation, and the projection from the latent to the $x$ space---are independent of the value assumed by the $c$ parameters. One can thus proceed as follows. First, generate a set of Monte Carlo events at the point $c=0$ in the parameter space. We denote as $w_{\textrm{e}}(0)$ the weight of each event ``e'' in this set. The $c=0$ data set could be an unweighted one, in which case the $w_{\textrm{e}}(0)$ weights are all equal, or be a weighted set with non-trivial $w_{\textrm{e}}(0)$ weights. Next, run through the list of generated events and assign them a weight $w_{\textrm{e}}(c)=w_{\textrm{e}}(0)\,d\sigma(\xi;c)/d\sigma(\xi;0)$. These weights account for the dependence on $c$ of the $\xi$ sampling, which in turn captures the entire dependence on $c$ of the observables because all the subsequent steps of the generation are independent of $c$. The single event data set ${\textrm{S}}=\{{\textrm{e}}_i\}_{i=1}^{n({\textrm{S}})}$---where each event is a pair ${\textrm{e}}=(x_{\textrm{e}},\,w_{\textrm{e}}(c))$---generated by reweighting a single run of the Monte Carlo code with $c$ equal to zero, as previously described, is thus a valid weighted set that enables the prediction of expectation values for arbitrary $c$.

Event reweighting entails enormous practical advantages, which are well recognized in the literature~\cite{Gainer:2014bta,Frixione:2002ik,Alioli:2010xd,Frederix:2014hta,REPOLO,Mattelaer:2016gcx}. The advantages are particularly striking for EFT applications. Therefore, event reweighting is well-developed in this context and is now fully automated. In particular, the {\sc{MadGraph}} framework enables to generate EFT reweighted Monte Carlo samples also at the Next to Leading Order (NLO) accuracy in the QCD loop expansion~\cite{Mattelaer:2016gcx,Alwall:2014hca,Degrande:2020evl}. 

The advantages of event reweighting are straightforwardly portable to the likelihood learning problem. The generation of the Monte Carlo data sets employed for training---rather than training itself---is a major or dominant component of the computational cost, especially at NLO. Event reweighting reduces this computational cost strongly because only one Monte Carlo data set is needed instead of the several data sets that are required to populate the $c$ parameter space in the unweighted approach. In fact, unweighted event generation could become unfeasible in certain EFT applications where a large number of $c$ parameters---denoted as Wilson coefficients in the EFT context---has to be considered because the number of required simulations grows rapidly with the dimensionality of $c$. 

If the effect of the $c$ parameters on the distribution is small, event reweighting is also beneficial for the accuracy of the predictions. More specifically, it improves the determination of the dependence on $c$ of the predictions. In the unweighted approach, this dependence is extracted by comparing the SM prediction, for $c=0$, with the one for $c\neq0$ estimated from two independent sets of Monte Carlo events. The uncertainties on these predictions are ultimately determined by the resources that are invested in the generation, and in particular by the number of events in the Monte Carlo data set, which controls the Monte Carlo statistical uncertainties. The effect of the parameters is typically small in EFT analyses for realistic values of the Wilson coefficients. Very small uncertainties are thus needed in order to be sensitive to the dependence on $c$ of the observables, which in turn requires very large Monte Carlo data sets, often beyond what is feasible in practice. In Ref.~\cite{Chen:2020mev} we solved this problem using unphysically large value of the Wilson coefficients for event generation and extrapolating down to realistic values exploiting the analytic knowledge of the (quadratic) dependence of the $r$ ratio on $c$. This is a viable approach, which however requires a careful choice of the Wilson coefficient values. The choice is problem-specific and arguably difficult to automate. 

Reweighted data enable a precise determination of the Wilson coefficients effects on observables, even if these effects are small, because the predictions are not affected by independent statistical uncertainties in the $c=0$ and in the $c\neq0$ data sets. Consider the simplest prediction of the cross section in a bin. We can determine the $c=0$ cross section by summing the $w_{\textrm{e}}(0)$ weights of the events that fall in the bin, the cross section for $c\neq0$ by summing $w_{\textrm{e}}(c)$, or even directly the difference between the two by summing up $w_{\textrm{e}}(c)-w_{\textrm{e}}(0)$ on the same events. In all cases, the finite Monte Carlo statistics will entail---if the weights are not vastly different---a relative uncertainty that is of the order of one over the square root of the number of events that fall in the bin. No matter how small the cross section difference is, a good relative accuracy on its prediction can be attained with manageable Monte Carlo statistics. We expect a similar advantage using reweighted data sets for the determination of the $r$ ratio.

In this paper, we investigate the advantages of event reweighting through a direct comparison with Ref.~\cite{Chen:2020mev}, by proceeding as follows. First, in Section~\ref{sec:meth}, we describe a straightforward adaptation to reweighted training sets of the methodology we developed in~\cite{Chen:2020mev} for likelihood learning. Next, in Section~\ref{sec:PS}, we describe the application of the novel strategy to the case studies considered in Ref.~\cite{Chen:2020mev} and compare the performances. In Section~\ref{sec:meth} we also introduce novel techniques---expanding ideas from Ref.~\cite{Chen:2020mev}---for the assessment of the quality of the $r$ ratio reconstruction, which is crucial for hyper-parameters selection. Somewhat outside the main line of development of the paper, in Section~\ref{sec:PLS} we describe a frequentist proposal based on asymptotic formulas and the Asimov trick~\cite{Cowan:2010js} for setting limits on the Wilson coefficients using the learned likelihood ratio, and outline its Bayesian interpretation. Our conclusions are in Section~\ref{sec:Conclusions}.

\section{Methodology}\label{sec:meth}

\subsection{Learning from weights}\label{subsec:lfw}

A standard result in statistical learning theory---known as the likelihood-ratio trick---is that a continuous-output classifier trained to tell apart two data sets approximates the ratio between the probability distribution of the two training sets up to a given monotonic transformation. More precisely, the statement is that the classification function $f(x)\in (0,1)$ that minimizes the expectation value of the loss function---i.e., the loss of an infinite training set---is in one-to-one correspondence with the distribution ratio. The classification function that is actually obtained by training does not correspond to the exact distribution ratio because training sets are finite (estimation error), because the class of functions does not contain the exact ratio (approximation error), and because the optimization algorithm might not converge to the actual minimum of the loss. Monitoring and reducing these sources of error down to a satisfactory level is the universal goal of all practical applications of statistical learning including those of the present paper.

\subsubsection{The simple classifier}\label{subsec:sc}

All the methods~\cite{Brehmer:2018kdj,Brehmer:2018eca,Brehmer:2018hga,Brehmer:2019xox,Chen:2020mev,Chatterjee:2022oco,Chatterjee:2021nms,Arganda:2022qzy,Arganda:2022zbs,GomezAmbrosio:2022mpm,Kong:2022rnd,Hollingsworth:2020kjg} to extract the distribution ratio $r(x;c)$ in eq.~(\ref{eq:r}) from Monte Carlo events are implementations of the likelihood-ratio trick. If the Monte Carlo data consists of a single set ${\textrm{S}}=\{{\textrm{e}}_i\}_{i=1}^{n({\textrm{S}})}$, where the dependence of the parameters of interest $c$ is included by reweighting in the events ${\textrm{e}}=(x_{\textrm{e}},\,w_{\textrm{e}}(c))$, as described in the Introduction, a straightforward adaptation of these ideas works as follows. Consider first the simpler task of learning the ratio, $r(x;{\bar{c}})$, at a fixed point $c=\bar{c}$ in the parameter space. This can be achieved with the loss function
\beq\label{eq:lossw}
\ell[f(\cdot)]=
\sum\limits_{{\textrm{e}}\in {\textrm{S}}}w_{\textrm{e}}({\bar{c}})
\left[
f(x_{\textrm{e}})
\right]^2
+\sum\limits_{{\textrm{e}}\in {\textrm{S}}}w_{\textrm{e}}(0)
\left[
f(x_{\textrm{e}})-1
\right]^2\,.
\eeq
Notice that the two summations are performed on the same data set. The classification function $f$ is thus evaluated on the same points in the two terms. 

The events in ${\textrm{S}}$ are by construction such that a weighted sum over them approaches, if the data set is large, the expectation value over the $x$ variable multiplied by the total cross section. Given that the probability distribution function of $x$ is equal to $d\sigma(x;c)/\sigma(c)$, the loss function for infinitely large ${\textrm{S}}$ approaches
\beq\label{eq:loss-ls}
\ell[f(\cdot)] \to \int d\sigma(x;{\bar{c}})
\left[
f(x)
\right]^2
+
\int d\sigma(x;0)
\left[
f(x)-1
\right]^2\,.
\eeq
This functional attains its absolute minimum for
\beq\label{eq:opc}
f(x)=\frac1{1+d\sigma(x;{\bar{c}})/d\sigma(x;0)}=\frac1{1+r(x;{\bar{c}})}\,,
\eeq
which is in one-to-one correspondence with the ratio $r(x;{\bar{c}})$. By inverting the above equation we can thus turn the trained model, which minimizes the loss~(\ref{eq:lossw}), into an estimate of $r(x;{\bar{c}})$. 

It is interesting to compare eq.~(\ref{eq:lossw}) with the loss function that one would employ instead in order to learn $r$ from a Monte Carlo generator---either weighted or unweighted---that does not implement event reweighting. The expression would be very similar (see for instance eq.~(6) of Ref.~\cite{Chen:2020mev}), but the two sums would be evaluated on two different event data sets, ${\textrm{S}}_0$ and ${\textrm{S}}_1$. The two sets are generated by independent runs, with the parameters $c$ set to zero and to ${\bar{c}}$, respectively. They do not contain the same $x_{\textrm{e}}$ points. 

As described in the Introduction, event reweighting is in general beneficial for the accuracy of the predictions thanks to a reduced sensitivity to the Monte Carlo statistical fluctuations. Similar advantages are expected for the determination of the distribution ratio, most strikingly when ${\bar{c}}$ is small, enabling a more accurate determination of the small departures of $r(x;{\bar{c}})$ from one. A concrete verification and quantification of these advantages is postponed to Section~\ref{sec:PS}. In the rest of the present section, we describe our theoretical understanding of this behavior.

When ${\bar{c}}$ is small such that $r$ is close to one, the optimal classification function~(\ref{eq:opc}) is close to $1/2$. In order to study this regime it is thus convenient to express $f(x)=1/2+\delta\hspace{-1pt}f(x)$, where $\delta\hspace{-1pt}f(x)$ is small in the optimal configuration~(\ref{eq:opc}), accounting for the small departures of $r$ from one. The trained model configuration, which minimizes the loss function, is also characterized by a small $\delta\hspace{-1pt}f$. The question is whether this small learned $\delta\hspace{-1pt}f$ is a good approximation of the optimal $\delta\hspace{-1pt}f$, producing an accurate determination of the departure of the $r$ ratio from one. The loss function~(\ref{eq:lossw}) reads
\begin{eqnarray}\label{eq:loss12}
\displaystyle
&&\sum\limits_{{\textrm{e}}\in {\textrm{S}}}w_{\textrm{e}}({\bar{c}})
\,\delta\hspace{-1pt}f(x_{\textrm{e}})
-\sum\limits_{{\textrm{e}}\in {\textrm{S}}}w_{\textrm{e}}(0)
\,\delta\hspace{-1pt}f(x_{\textrm{e}})
+
\sum\limits_{{\textrm{e}}\in {\textrm{S}}}w_{\textrm{e}}({\bar{c}})\,
\delta\hspace{-1pt}f(x_{\textrm{e}})^2
+\sum\limits_{{\textrm{e}}\in {\textrm{S}}}w_{\textrm{e}}(0)
\,\delta\hspace{-1pt}f(x_{\textrm{e}})^2
\nonumber
\\
&&
\displaystyle
=
\sum\limits_{{\textrm{e}}\in {\textrm{S}}}
\left[w_{\textrm{e}}({\bar{c}})-
w_{\textrm{e}}(0)
\right]
\delta\hspace{-1pt}f(x_{\textrm{e}})
+
\sum\limits_{{\textrm{e}}\in {\textrm{S}}}
\left[w_{\textrm{e}}({\bar{c}})+
w_{\textrm{e}}(0)
\right]
\delta\hspace{-1pt}f(x_{\textrm{e}})^2
\,,
\end{eqnarray}
up to an additive constant that is irrelevant for the minimization. If  ${\bar{c}}$ is small, $w_{\textrm{e}}({\bar{c}})$ and $w_{\textrm{e}}(0)$ are almost identical and the term in the loss function which is linear in $\delta\hspace{-1pt}f$ is strongly suppressed. The suppression of the linear term in comparison with the quadratic one eventually makes $\delta\hspace{-1pt}f$ small at the minimum of the loss function. Importantly enough, the linear term is small at each individual training point $x_{\textrm{e}}$ at which $\delta\hspace{-1pt}f$ is evaluated. This is outlined by the second line of eq.~(\ref{eq:loss12}), where we collected under a single summation the linear and the quadratic terms. If the training sample is large, the summations provide good approximations of the $\delta\hspace{-1pt}f$ and $\delta\hspace{-1pt}f^2$ terms of the loss function expectation~(\ref{eq:loss-ls}). The relative accuracy of these approximations scales like one over the square root of the number of training points regardless of whether these terms are small or large. We will attain a similarly small relative accuracy in the determination of the optimal $\delta\hspace{-1pt}f$ by the minimization of the loss function both if ${\bar{c}}$ is large or if it is small and the linear term is suppressed.

We now compare eq.~(\ref{eq:loss12}) with the analogous expression that we would obtain instead when the two independent data sets ${\textrm{S}}_0$ and ${\textrm{S}}_1$ are used for training. We would get the terms on the first line of the equation, but they will be evaluated on the different data sets. In particular, the linear term will emerge from a cancellation between a summation over ${\textrm{S}}_0$ and one over ${\textrm{S}}_1$, with opposite sign. The relative statistical uncertainties of order one over the square root of the number of points will affect the two summations independently entailing, if ${\bar{c}}$ is small, a degradation of the accuracy in the reconstruction of the linear term. A good determination of $\delta\hspace{-1pt}f$ would thus require large training samples and eventually become unfeasible for extremely small ${\bar{c}}$, preventing a determination of the departure of $r$ from one. Using reweighted training data avoids this problem.

While presented in the case of quadratic loss, the considerations above hold for other choices of the loss function including the binary cross-entropy that is most often employed for classification. We tested the usage of the binary cross-entropy in our experiments, finding essentially identical results as for the quadratic loss. Since we did not encounter a case where it makes a difference, the impact of the choice of the loss function is not discussed further, and the quadratic loss is employed 
throughout this paper.

An interesting peculiarity of the training scheme based on reweighted data concerns the origin of overfitting. In regular training based on independent and unweighted data sets ${\textrm{S}}_0$ and ${\textrm{S}}_1$, the classifier function $f(x)$ receives, from the loss function minimization, a push to approach zero at the $x_{\textrm{e}}$ points that belong to the ${\textrm{S}}_0$ set, and a push to approach one at the ${\textrm{S}}_1$ points. Since the $x_{\textrm{e}}$ points in ${\textrm{S}}_0$ and in ${\textrm{S}}_1$ are distinct, this encourages the development of overfitted configurations where the model $f$ wildly oscillates from zero to one in correspondence of individual training points. In the case of reweighted training data instead, there is only one set of $x_{\textrm{e}}$ points where the loss function~(\ref{eq:lossw}) is evaluated. Overfitting thus emerges from a different mechanism. We explained in the Introduction that the weights $w_{\textrm{e}}(c)=w_{\textrm{e}}(0)\,d\sigma(\xi;c)/d\sigma(\xi;0)$ are functions of latent variables, $\xi$, that are not in one-to-one correspondence with the observables $x$. Two events in the training set that are very near in the $x$ space are typically far apart in the $\xi$ space and hence their weights can be vastly different. The loss~(\ref{eq:lossw}) pushes the classifier towards zero at the $x_{\textrm{e}}$ point where $w_{\textrm{e}}({\bar{c}})$ is large, and towards one at the nearby point where it is small. This can produce overfitted configurations, by a mechanism that is slightly different than the one at work in regular classifier training. Avoiding overfitting in regular training requires regularization, which prevents the model to develop overly sharp features, and large training sets, which populate the $x$ space densely. These same overfitting mitigation strategies turn out to be effective also for training with reweighted events.

\subsubsection{The quadratic classifier for EFT}\label{qc-1}

Learning the ratio between two specific distributions can be of practical interest. However, in most cases one needs instead the ratio $r(x;c)$ as a function of the parameters of interest $c$. The simple classifier described above can only learn $r(x;c)$ point-by-point in the $c$ space, making extremely demanding or impossible to reconstruct the dependence on $c$, especially if there are several parameters of interest, $c=(c_1,\ldots,c_d)$. It is possible to overcome this limitation if the parametric dependence of $r(x;c)$ on $c$ is known, by employing a \emph{parametrized classifier}~\cite{Chen:2020mev}.

Parametrized classifiers are useful in cases where the distribution ratio can be parametrized in terms of a known function of $c$ with $x$-dependent coefficient functions $\Gamma(x)$, namely if $r$ reads
\beq\label{eq:parr}
r(x;c)={\mathcal{P}}\left(
\Gamma(x);c
\right)\,.
\eeq
In this case, we know from eq.~(\ref{eq:opc}) the dependence on $c$ of the optimal classification function $f=1/(1+r)$. We can thus make an Ansatz for the functional form of the classification function
\beq\label{eq:clafu}
f(\gamma(x);c)=\frac1{1+
{\mathcal{P}}\left(
\gamma(x);c
\right)
}\,,
\eeq
by employing a flexible class of functions---such as neural networks---to model the coefficient functions $\gamma(x)$. The absolute minimum of the loss function will be attained for $f=1/(1+r)$, i.e. for $\gamma(x)=\Gamma(x)$. By training, we can thus produce $\gamma$ functions that approximate the true $\Gamma$ coefficient functions. By using the trained $\gamma$ functions in eq.~(\ref{eq:parr}) in place of $\Gamma$, we eventually obtain an estimate of the ratio $r$.

The configuration $\gamma(x)=\Gamma(x)$ is a global minimum of the expectation value of any conceivable loss function, including eq.~(\ref{eq:lossw}) with arbitrarily chosen ${\bar{c}}$. However, the minimum needs to be unique in order for the loss function minimization to determine all the different coefficient functions. The loss in eq.~(\ref{eq:lossw}) possesses instead a family of degenerate global minima~\footnote{This can be readily seen by rewriting the large-S limit of the loss, given by eq.~(\ref{eq:loss-ls}), like in eq.~(45) of Ref.~\cite{Chen:2020mev}.}, because it only contains information on the likelihood ratio at a single point $c={\bar{c}}$. It is thus minimized by any of the many configurations that reproduce the likelihood ratio at that point. More points in the $c$ space are needed for a unique determination of all the coefficient functions. We thus consider a set ${\cal{C}}=\{{\bar{c}}^{(1)},\ldots,{\bar{c}}^{(\kappa)}\}$ of $\kappa$ distinct points and we define another loss function 
\beq\label{eq:losswpc}
\ell[\gamma(\cdot)]=
\sum\limits_{{\textrm{e}}\in {\textrm{S}}}
\sum\limits_{{\bar{c}}\in {\cal{C}}}
\left\{
w_{\textrm{e}}({\bar{c}})
\left[
f(\gamma(x_{\textrm{e}});{\bar{c}})
\right]^2
+w_{\textrm{e}}(0)
\left[
f(\gamma(x_{\textrm{e}});{\bar{c}})
-1
\right]^2\right\}\,.
\eeq
This loss in the large-S limit has a unique global minimum for $\gamma(x)=\Gamma(x)$, provided the number $\kappa$ of points in $\cal{C}$ is greater or equal than the number of coefficient functions to be determined.

The systematic exploration of the effect of heavy new physics described through the SM EFT is a prime target of the LHC, the HL-LHC, and future collider projects. In this context, the parameters of interest $c$ are the Wilson coefficients of the new EFT interactions. In general, their effect on the differential cross sections $d\sigma(x;c)$ can be captured by a second-order polynomial. We could thus consider a parametrization
\beq\label{eq:qc}
{\mathcal{P}}\left(
\gamma(x);c
\right)=1+\sum\limits_{i=1}^dc_i \gamma_i(x)+\sum\limits_{j\geq i}^d c_i c_j \gamma_{ij}(x)\,,
\eeq
with a total of $d(d+3)/2$ $\gamma$ functions, when $d$ Wilson coefficients are present. The constant term equals one because $r(x;0)=1$ by definition. 

It is worth mentioning that the quadratic parametrization in eq.~(\ref{eq:qc}) does not capture all possible EFT effects. The EFT might affect parameters, like the mass or the width of SM particles, whose effect on the differential cross section is not polynomial. These contributions are often too small to be seen at the LHC. Still, they could be incorporated where needed by learning the dependence of the $\gamma$ functions on masses or widths with a point-by-point approach. Including EFT contributions beyond the quadratic order, as they emerge from loop diagrams involving more than one EFT operator, would require generalizing eq.~(\ref{eq:qc}). However, these contributions are negligible. The quadratic parametrization accounts instead for relevant loops involving one EFT vertex and SM vertices, such as those associated with NLO QCD corrections.

A clear advantage~\cite{Brehmer:2018eca} of employing the quadratic parametrization is that the number of distinct points in the $c$ space to be considered for learning the complete $c$-dependent ratio $r(x;c)$ scales quadratically with the dimensionality of $c$. It scales instead exponentially in the point-by-point approach where the dependence on $c$ is not imposed and needs to be reconstructed. 

Another advantage~\cite{Chen:2020mev} is that, by exploiting the parametrization, strategies can be devised for improving the quality of the likelihood reconstruction. We previously discussed that the ratio is difficult to learn accurately if the Wilson coefficients $c$ are small, at least when unweighted training data are used. But we do not need to employ small values of $c$ for the training of the parametrized classifier, in spite of the fact that the values that are relevant in the actual analysis of the data will eventually be small. The coefficient functions in the parametrization~(\ref{eq:qc}) can be learned by training with much larger $c$. If these functions are reconstructed with good relative accuracy, the departure from one of the $r$ ratio is accurately reconstructed also for small $c$. The exact analytical knowledge of the dependence of $r(x;c)$ on $c$ enables the extrapolation from large to small values. On the other hand, the $c$ values used for training can not be arbitrarily large, otherwise the contribution of the quadratic polynomial terms in eq.~(\ref{eq:qc}) would dominate, the effect of the linear terms would be hidden and could not be learned. A proper selection of the $c$ points in the set ${\cal{C}}=\{{\bar{c}}^{(1)},\ldots,{\bar{c}}^{(\kappa)}\}$, used for training, is the main factor that controls the quality of the $r$ ratio reconstruction. The selected $c$ values should take into account the need of learning both the linear and the quadratic terms in all the different regions of the phase space. Since the absolute and relative magnitude of the two terms can vary radically in different regions of the phase space, several different ${\bar{c}}$ values are needed and must be included in the ${\cal{C}}$ set. 

The strong sensitivity of the quality of the likelihood reconstruction to the choice of the training points requires case-by-case optimization, which is an obstruction to the automation of the methodology of Ref.~\cite{Chen:2020mev}. We will verify in Section~\ref{sec:PS} that employing reweighted events reduces this sensitivity because it enables accurate learning of small effects as previously discussed. 

An alternative~\cite{Chatterjee:2022oco} to our strategy, which also employs reweighted events, is to learn the linear and quadratic terms in separate training processes using suitable loss functions that contain individual terms of the polynomial expansion of the weight $w_{\textrm{e}}(c)$. We made some attempts to adapt the strategy of~\cite{Chatterjee:2022oco} to our problem, without attaining satisfactory performances. An extensive comparison with other methods, including also the approach of Ref.~\cite{GomezAmbrosio:2022mpm} possibly adapted to reweighted training data, is left to future work.

\subsubsection{The EFT learning strategy}\label{qc-2}

Several options could be considered for the practical implementation of the quadratic classifier approach. The simplest one would be to adopt the basic quadratic parametrization in eq.~(\ref{eq:qc}) using feed-forward neural networks, with output in $\mathbb{R}$, to model the $\gamma_i(x)$ and $\gamma_{ij}(x)$ coefficient functions. All the functions---there are $d(d+3)/2$ of them, for $d$ Wilson coefficients---could be learned simultaneously in a single training using the loss in eq.~(\ref{eq:losswpc}) with $\kappa\geq d(d+3)/2$ training points in $\mathcal{C}$. A potential limitation of this scheme is that the basic parametrization~(\ref{eq:qc}) does not take into account that the $r$ ratio, being the ratio of positive-defined physical cross sections, is itself positive. If the parametrization turns negative at some point in $x$ and $c$ at some stage of the training process, the classification function $f(x;c)$ in eq.~(\ref{eq:clafu}) exits the $(0,1)$ interval and potentially diverges, for ${\mathcal{P}}\left(
\gamma(x);c
\right)=-1$. This risks producing training instabilities, especially if the cross-entropy loss was used in place of the quadratic loss. Furthermore, enforcing the physical constraint $r(x;c)>0$ can be beneficial for the accuracy of the ratio reconstruction.

Following~\cite{Chen:2020mev}, we can enforce cross section positivity using the parametrization
\beq\label{eq:qcpos}
{\mathcal{P}}\left(
\lambda(x);c
\right)=\sum_{I=1}^{d+1} \left[\sum_{J=1}^{d+1} \lambda_{IJ}(x) c_{J-1}\right]^2\,,
\eeq
where we defined $c_0=1$, and $\lambda(x)$ is an upper triangular real $(d+1)$-dimensional squared matrix with $\lambda_{11}(x)=1$. This expression provides, for general $\lambda(x)$, the most general positive quadratic polynomial of $c$. Neural networks with output in $\mathbb{R}$ can be employed to model the non-trivial entries of the $\lambda$ matrix.

The conceptually straightforward approach of learning all the coefficients functions in a single training, using the parametrization in eq.~(\ref{eq:qcpos}), suffers from practical limitations when the number $d$ of Wilson coefficients is large. Typically available GPUs have limited memory. They can hardly accommodate all the gradients that have to be stored for the training of more than around 10 reasonably complex neural network models using the large number of training points required for an accurate learning of the coefficient functions. This prevents using GPUs for more than $2$ or $3$ Wilson coefficients, with a dramatic impact on the training execution time.\footnote{Using mini-batches can circumvent GPU memory limitations, but slows down training. Notice that the mini-batch  gradients need to be accumulated and the weight update step taking only after the whole training data set is processes. This is because an accurate determination of the gradients pointing towards the true minimum of the loss function is needed for an accurate learning.} A different scheme is needed, in which the different polynomial terms of the basic parametrization of eq.~(\ref{eq:qc}) are learned in separate trainings. These individual training stages could be run in parallel, if several GPUs are available, or sequentially on a single GPU still entailing a strong improvement of the execution time in comparison with CPU training. 

A scheme that is suitable for parallelization works as follows. Since the polynomial is quadratic, all its coefficients can be extracted by considering configurations where only two Wilson coefficients---in all possible pairings---are turned on, while the others are set to zero. Let us then start from the case of a 2-dimensional coefficient vector $c=(c_1,c_2)$, for which we can employ the manifestly positive parametrization in eq.~(\ref{eq:qcpos}), with $d=2$. For reasons that will become momentarily clear, we do not employ neural networks to model the entries of the $\lambda$ matrix directly. We instead express the upper triangular matrix $\lambda$ as
\beq\label{eq:lam}
\lambda(x) =
\left(
\begin{array}{c@{\hspace{1.5em}}c@{\hspace{1.5em}}c}
    1 & \rho_1(x) \sin \theta_{11}(x) & \rho_2(x) \sin \theta_{22}(x) \\
    0 & \rho_1(x) \cos \theta_{11}(x) & \rho_2(x) \cos \theta_{22}(x) \sin \theta_{12}(x)\\
    0 & 0 & \rho_2(x) \cos \theta_{22}(x) \cos \theta_{12}(x)
\end{array}
\right)\,,
\eeq
using polar and spherical coordinates for the second and third columns, respectively. With this expression, the parametrization~(\ref{eq:qcpos}) of the distribution ratio becomes
\begin{align}\label{eq:qcpc}
\displaystyle
{\mathcal{P}}(\rho_1,\rho_2,\theta_{11},\theta_{22},\theta_{12};c)=
1+2\,c_1\rho_1\,\sin\theta_{11}+c_1^2\rho_1^2+2\,c_2\rho_2\,\sin\theta_{22}+c_2^2\rho_2^2\; \hspace{15pt} \
\nonumber\\
\displaystyle
+\,2\,c_1c_2\rho_1\rho_2\left(
\cos\theta_{11}\cos\theta_{22}\sin\theta_{12}+\sin\theta_{11}\sin\theta_{22}
\right)\,.
\end{align}
We employ neural networks to parametrize the two radial functions $\rho_{1,2}$ and the three angular functions $\theta_{11}$, $\theta_{22}$ and $\theta_{12}$. We consider networks with unconstrained outputs spanning the whole real axis, and we ignore the periodicity of the angular functions and the positivity of the radial functions. This choice does not invalidate the generality and the positivity of our parametrization. It merely makes it redundant, which is not a limitation: during training, the networks will pick up one of the equivalent configurations that correspond to the correct distribution ratio. No training instability will emerge because each equivalent configuration is a separate global minimum of the loss function.

If only $d=2$ Wilson coefficients are present, the GPU memory is probably sufficient to store all the gradients and the 5 neural networks $\rho_{1,2}$ and $\theta_{11,22,12}$ could be learned in a single training. Consider however the following alternative, which is suited for generalization to the case $d>2$. In eq.~(\ref{eq:qcpc}), the $c_1$ and $c_1^2$ terms are fully determined by the $\rho_1$ and the $\theta_{11}$ networks. These networks can thus be learned separately from the others by training data involving only $c_1$, with vanishing $c_2$. Similarly, we can learn $\rho_2$ and $\theta_{22}$ by training with $c_2$ only. Training data where both $c_1$ and $c_2$ are non-vanishing are only needed in order to learn $\theta_{12}$, which in turn gives access to the mixed polynomial term $c_1c_2$. Learning $\theta_{12}$ requires the knowledge of $\rho_{1,2}$ and of $\theta_{11,22}$, thus the determination of $\theta_{12}$ can not proceed in parallel with the determination of the other networks. However, while training $\theta_{12}$, the $\rho_{1,2}$ and $\theta_{11,22}$ networks can be kept frozen to their previously-learned configurations and they are not optimized. The only gradients to be stored in the GPU memory are those of the $\theta_{12}$ network parameters.

If $d>2$ Wilson coefficients are present, one can first run $d$ separate trainings with a single non-vanishing Wilson coefficient, $c_i$, using a parametrization
\beq\label{eq:qcd1d}
{\mathcal{P}}(\rho_i,\theta_{ii};c_i)=
1+2\,c_i\rho_i\,\sin\theta_{ii}+c_i^2\rho_i^2\,,
\eeq
which is the one-dimensional version of eq.~(\ref{eq:qcpc}). These training stages can operate in parallel and they give access to the $c_i$ and $c_i^2$ terms of the polynomial, for all $i=1,\ldots,d$. Next, we turn on a pair $(c_i,c_j)$ of Wilson coefficients and consider a 2-dimensional parametrization like the one of eq.~(\ref{eq:qcpc}), with networks $\rho_i$, $\rho_j$, $\theta_{ii}$, $\theta_{jj}$ and $\theta_{ij}$. The $\rho_i$, $\rho_j$, $\theta_{ii}$ and $\theta_{jj}$ networks are those of eq.~(\ref{eq:qcd1d}), and they were previously learned in the one-dimensional trainings. The remaining network, $\theta_{ij}$, can be trained keeping the others fixed, as previously explained. This gives access to the $c_ic_j$ mixed term. Extracting all the mixed terms requires $d(d-1)/2$ separate trainings---corresponding to the distinct pairings of Wilson coefficient---that can run in parallel. Finally, once all the polynomial coefficients are known, they can be pulled together in eq.~(\ref{eq:qc}) providing the full knowledge of the distribution ratio everywhere in the Wilson coefficients space.

For the case studies of the present paper, with $d=2$, the distribution ratio parametrized by eq.~(\ref{eq:qcpc}) can be learned in a single training and the protocol described above is not needed. However, it will be essential in order to deal with the large number of Wilson coefficients that are required, as emphasized in~\cite{GomezAmbrosio:2022mpm}, for global EFT fits. Our methodology offers an ideal solution to this problem. It enables parallelization or serial execution on GPUs with limited memory while, at the same time, rigorously enforcing the physical constraint of distribution ratio positivity at all training stages. The only potential limitation, in comparison with directly learning all the coefficient functions of the general manifestly positive parametrization in eq.~(\ref{eq:qcpos}), is that our strategy based on learning the individual terms of the basic polynomial parametrization~(\ref{eq:qc}) does not produce a distribution ratio that is necessarily positive when more than 2 coefficients are non-vanishing. It is unclear whether negative ratios will be ever encountered. Their occurrence for the small values of the Wilson coefficients we will be interested in probing would signal overfitting because the true effect of the EFT operators should typically produce a small correction to the SM cross section that can hardly cause a strong departure of the ratio from unity.

\subsection{Performance metrics}\label{subsec:val}

Learning the distribution ratio with the strategy described in the previous section is not very different from training a classifier. It requires choosing a model---feed-forward neural networks, in our case---and selecting its parameters as well as training hyper-parameters like the training sample size, the number of epochs and the learning rate plus, if needed, explicit regularization parameters. The loss function~(\ref{eq:losswpc}) also features the number and the values of the Wilson coefficients in the $\mathcal{C}$ set, ${\bar{c}}^{(1)},\ldots,{\bar{c}}^{(\kappa)}$, as additional hyper-parameters. However, when training a classifier for regular classification purposes one operates hyper-parameter selection based on figures of merit that quantify the performances of the trained classification function. Standard performance metrics are the accuracy or the AUC. Our scope is not classification, we thus need to define different performance metrics. This is the goal of the present section.\footnote{Most of what follows is a refinement of ideas we first presented in Ref.~\cite{Chen:2020mev}.}

Our goal is to extract a good approximation of the distribution ratio $r(x;c)$. Denoting as ${\hat{r}}(x;c)$ the reconstructed ratio, the most straightforward performance metrics to assess the quality of our results should thus measure some sort of distance between the true ratio $r(x;c)$ and ${\hat{r}}(x;c)$. Even if the true ratio is of course unknown, valid notions of distance can be constructed as follows. The true distribution ratio relates the SM differential cross section, $d\sigma(x;0)$, to the $c$-dependent cross section, $d\sigma(x;c)$, by 
\beq\label{eq:sigc}
d\sigma(x;c)=r(x;c)\,d\sigma(x;0)\,.
\eeq
We can thus define an approximate cross section
\beq\label{eq:sigchat}
d\hat\sigma(x;c)={\hat{r}}(x;c)\,d\sigma(x;0)\,,
\eeq
by employing $\hat{r}$ in place of $r$. Comparing $d\sigma(x;c)$ with $d\hat\sigma(x;c)$ measures the distance between $r$ and $\hat{r}$. We do not have access to the differential cross section, but we can compute the cross section $\Delta\sigma(c)$ integrated in some bin using Monte Carlo events. If the events are reweighted, the cross section is the sum of the weights of the events that fall into the bin
\beq\label{eq:ds}
\Delta\sigma(c)=\sum\limits_{x_{\textrm{e}}\in {\textrm{bin}}}w_{\textrm{e}}(c)\,.
\eeq
This quantity can be compared with the integral of $d\hat\sigma(x;c)$ in the bin
\beq\label{eq:recsig}
\Delta\hat\sigma(c)=\sum\limits_{x_{\textrm{e}}\in {\textrm{bin}}}w_{\textrm{e}}(0)\,{\hat{r}}(x_{\textrm{e}};c)\,.
\eeq
By performing this comparison for relevant binned univariate marginals, we can visualize the quality of the distribution ratio reconstruction. Results are presented in Section~\ref{sec:PS}. See for instance Figure~\ref{fig:idealrec}.

An alternative measure of the distance between $\hat{r}$ and $r$, which does not rely on the arbitrary selection of marginals, can be defined by exploiting a certain characteristic property of the true distribution ratio $r(x;c)$, seen as a one-dimensional variable depending on $x$, for fixed $c$. We actually work with the logarithm of this variable, $\tau_c$, and with its reconstruction, $\hat\tau_c$, namely
\beq\label{eq:tth}
\tau_c(x)=\log r(x;c)\,,\;\;\;\;\;
\hat\tau_c(x)=\log {\hat{r}}(x;c)\,.
\eeq
The differential cross-section for the $\tau_c$ variable is the integral of $d\hat\sigma(x;c)$ on slices of fixed $r(x;c)$. Hence, using eq.~(\ref{eq:sigc}), we have
\beq\label{eq:ratdist}
{d\sigma}(\tau_c;c)=e^{\tau_c} {d\sigma}(\tau_c;0)\;\;\Rightarrow\;\;
\log\left[
{d\sigma}(\tau_c;c)/{d\sigma}(\tau_c;0)
\right]=\tau_c
\,.
\eeq

The reconstructed log-ratio variable, $\hat\tau_c$, does not obey eq.~(\ref{eq:ratdist}) because it is not the logarithm of the true distribution ratio, but of the reconstructed one. We can thus quantify the quality of the approximation by studying the validity of eq.~(\ref{eq:ratdist}) for the differential distributions of the  $\hat\tau_c$ variable: ${d\sigma}(\hat\tau_c;c)$ and  ${d\sigma}(\hat\tau_c;0)$. We proceed by binning $\hat\tau_c$ and computing, for each bin with boundaries $(\tau_{-},\tau_{+})$, the integrated cross section
\beq\label{eq:sb}
\Delta\sigma(c)=\hspace{-15pt}\sum\limits_{
\tau_c(x_{\textrm{e}})\in(\tau_{-},\tau_{+})} \hspace{-15pt}w_\textrm{e}(c)\,.
\eeq
For each bin we then compute
\beq\label{eq:rcq}
{{\Tau_c}} = \log \bigg[
\Delta\sigma(c)/\Delta\sigma(0)
\bigg]\,.
\eeq
If eq.~(\ref{eq:ratdist}) is approximately verified for the ${d\sigma}(\hat\tau_c;c)$ and ${d\sigma}(\hat\tau_c;0)$ cross sections, and if the bin is sufficiently narrow, $\Tau_c$ approximately follows a straight line, ${{\Tau_c}}\simeq\tau_{\textrm{avg}}$ with $\tau_{\textrm{avg}}=(\tau_++\tau_-)/2$.  By plotting ${{\Tau_c}}$ in the different bins we can thus visualize the validity of eq.~(\ref{eq:ratdist}). One example is shown, for instance, on the left panel of Figure~\ref{fig:scaletta}.

A further refinement exploits that if $\hat\tau_c$ was equal to the true $\tau_c$, the cross section $\Delta\sigma(c)$ in the bin $\hat\tau_c(x)\in (\tau_{-},\tau_{+})$ would be bounded by
\beq\label{eq:bder}
e^{\tau_-}\Delta\sigma(0)<\Delta\sigma(c)=\int_{\tau_{-}}^{\tau_{+}}\hspace{-4pt} d\sigma(\hat\tau;c)
<e^{\tau_+}\Delta\sigma(0)\,,
\eeq
using $\hat\tau(x)=\tau(x)$, and eq.~(\ref{eq:ratdist}). Therefore, if $\hat\tau_c$ was equal to $\tau_c$, $\Tau_c$ would be bounded as
\beq\label{eq:taubounds}
\tau_{-}<\Tau_c<\tau_{+}\;\;\rightarrow\;\;
|\Tau_c-\tau_{\textrm{avg}}|<\frac{\tau_{+}-\tau_{-}}2
\,.
\eeq
Verifying if $\Tau_c$ sits in this window---see for instance the right panel of Figure~\ref{fig:scaletta}---gives an indication of the validity of eq.~(\ref{eq:ratdist}) for the reconstructed ratio that is more quantitative than checking qualitatively the relation ${{\Tau_c}}\simeq\tau_{\textrm{avg}}$. 
\subsubsection[{The Neyman--Pearson \emph{p}-value: definition}]{The Neyman--Pearson $\boldsymbol{p}$-value: definition}\label{sec:nppdef}

The previously-described strategies provide an important assessment of the quality of the $r$ ratio reconstruction. We have found them effective in practice to discriminate between different trained models and therefore helpful for hyper-parameters selection. However, they do not constitute our prime performance metric, because of two reasons. First, because it is difficult to condense the information they provide into a single quality indicator. Second, because they do not offer an objective criterion to quantify the improvement obtained by a certain configuration in comparison with another one. One can improve indefinitely the quality of the reconstruction by employing more computational resources i.e., typically, by using more training data and bigger neural networks. In order to balance performances against computational resources, we must be able to judge the significance of the improvement attained by a given configuration, relative for instance to a configuration with smaller networks and less data. 

Our prime performance indicator is still sensitive to the distance between $r(x;c)$ and ${\hat{r}}(x;c)$, but less directly than the other ones. Ultimately, we seek access to $r$ in order to model the likelihood log-ratio $\lambda$, in eq.~(\ref{eq:log-lik}), of a given collider experiment. In turn, its knowledge would enable us to extract maximal---i.e., optimal---statistical information on the parameters of interest from the experimental data. It is thus natural to measure the quality of the reconstructed ratio ${\hat{r}}$ in terms of its statistical performances, if used in place of $r$ in eq.~(\ref{eq:log-lik}), on the collider experiment under examination. We thus define the reconstructed likelihood log-ratio
\beq\label{eq:log-lik-hat}
\hat\lambda(c;{\cal D}) 
= N(0) - N(c) + \sum_{x \in {\cal D}} \log {\hat{r}}(x; c)=
 N(0) - N(c) + \sum_{x \in {\cal D}} {\hat{\tau}}_c(x)\,,
\eeq
where, as explained in the Introduction, $N(c)=L\,\sigma(c)$, with $L$ the integrated luminosity of the experiment. The total cross section, $\sigma(c)$, is taken from the Monte Carlo with negligible error.

Several different statistical analyses could be potentially performed by employing the reconstructed likelihood log-ratio $\hat\lambda$~(\ref{eq:log-lik-hat}), ranging from setting a limit on the allowed size of the $c$ parameters, excluding the SM point $c=0$, or measuring the parameters. Classical or Bayesian methodologies could be employed. Without committing to any of these options for the analysis of the actual data, here we pick up one statistical analysis that could be performed---in line of principle---using $\hat\lambda$ and that is endowed with a sharp guarantee of statistical optimality if the learned likelihood log-ratio $\hat\lambda$ was exactly equal to the true log-ratio $\lambda$. The statistical performances of this analysis improve as $\hat\lambda$ approaches $\lambda$, providing another way to quantify the agreement between $\hat{r}$ and $r$. The performances saturate when the agreement between $\hat{r}$ and $r$ is sufficient, and do not improve indefinitely. When saturation occurs and the performance gain stops being significant, it means that the reconstructed $\hat\lambda$ contains the same information as $\lambda$ on the parameters of interest and no further improvement of the ratio reconstruction is needed.

The Neyman--Pearson Lemma~\cite{Neyman:1933wgr} guarantees the optimality of employing the true likelihood ratio in order to discriminate between two different values $c_0$ and $c_1$ of the parameters of interest. The one considered in the Lemma is a test of hypothesis that could in principle be used to exclude the existence of EFT interaction operators with a specific value $c$ of the Wilson coefficients. The performances of such exclusion analysis would be quantified---prior to the experiment---in terms of its expected sensitivity to non-vanishing $c$ under the SM hypothesis that no EFT interaction exists and thus $c$ is equal to $0$. In the standard Neyman--Pearson notation, we should thus identify $c_0=c$ and $c_1=0$ as the null and the alternative hypothesis, respectively.

A generic test of hypothesis works by defining a ``test statistic'' variable, $t(\data)$, which depends collectively on the whole set of data $\data=\{x_i\}_{i=1}^{\N}$ that are collected in the experiment. Any quantity could be employed as a test statistic, in line of principle. However, a meaningful test statistic is one that is typically small when the null hypothesis $H_0$ ($c\neq0$, in our case) is true, and large if instead the alternative hypothesis $H_1$ (i.e., $c=0$) is true. Observing on the data a value of $t$ that is way larger than the typical values of $t$ attained in the presence of the EFT interactions disfavors the presence of the new interactions. A statistical notion of typicality is provided by the $p$-value
\beq\label{eq:pval0}
p_c(t)=\int_t^\infty\hspace{-7pt} dt^\prime {\textrm{pdf}}(t^\prime|c)\,.
\eeq
The $p$-value relates the observed value of $t$ to the probability that an even larger value is observed when the EFT interactions are present in the data distribution. Small $p_c$ signals that EFT interactions with Wilson coefficient $c$ are unlikely to be present.

An \emph{efficient}~\cite{Neyman:1933wgr} hypothesis test for $c$ exclusion is one that excludes with high confidence, i.e. with low $p$-value, if the SM hypothesis $c=0$ is true. The metric that quantifies the test efficiency thus considers the typical value of $t(\data)$ in the SM hypothesis, and the corresponding $p$-value~(\ref{eq:pval0}). We use  the median of $t$ and, since $p_c(t)$ is monotonic in $t$, define the median $p$-value
\beq\label{eq:medp}
p(c)={\textrm{Median}}\big[p_c(t)|c=0\big]\,.
\eeq
The median $p$-value is the figure of merit that quantifies the efficiency of hypothesis tests. Lower $p(c)$ indicates better performances.

Any choice of the test statistic variable $t(\data)$ defines a valid test, whose efficiency is evaluated by the median $p$-value~(\ref{eq:medp}). However, the Lemma~\cite{Neyman:1933wgr} identifies the most efficient hypothesis test as the one that employs as test statistic minus the logarithm of the ratio between the likelihood in the null and in the alternative hypotheses, times a conventional factor of two
\begin{equation}\label{eq:tc}
t_c({\mathcal{D}})=-2\,\lambda(c;\data)=2\bigg[N(c) - N(0) - \sum_{x \in {\cal D}} {{\tau}}_c(x)\bigg]\,.
\end{equation}
The Lemma guarantees that the test based on $t_c(\data)$ (or a monotonic function of it) has the lowest possible median $p$-value. Any other variable has inferior performances, namely a larger median $p$-value. In particular, this means that the performances of the reconstructed likelihood log-ratio
\begin{equation}\label{eq:tchat}
{\hat{t}}_c({\mathcal{D}})=-2\,\hat\lambda(c;\data)= 2\bigg[N(c) - N(0) - \sum_{x \in {\cal D}} {\hat{\tau}}_c(x)\bigg]\,,
\end{equation}
are not optimal, but they approach the optimum as the reconstructed $\hat\lambda$ approaches the true $\lambda$. We can thus employ the median $p$-value of the ${\hat{t}}_c({\mathcal{D}})$ test statistic as a metric to evaluate the performances of the ratio reconstruction. We denote this quantity as $\hat{p}(c)$.

Extensive use of $\hat{p}(c)$ is made in Section~\ref{sec:PS} to compare different models and eventually select the hyper-parameters. Various types of performance studies are conducted. For instance, the evolution of $\hat{p}(c)$ during the training of a certain model is displayed on the right panel of Figure~\ref{fig:evolution}. The value of $c$ considered in the figure has been chosen such that $\hat{p}(c)\approx{\textrm{few}}\%$, close to the threshold of $5\%$ that is often conventionally considered to set an exclusion limit. This is because we want to probe the quality of the ratio reconstruction when $c$ is close to the values that will be eventually relevant for the statistical analysis of the data. Alternatively, we can identify the region where $\hat{p}(c)$ is exactly equal to $5\%$ and draw exclusion contours, as in Figure~\ref{fig:idealContour}. Figure~\ref{fig:hyperparameters} exemplifies instead the plots we made for hyper-parameter selection. It shows, among other things, the saturation of $\hat{p}(c)$ for increasingly complex networks and more training data. Section~\ref{sec:PS} describes these results extensively.

It should be noted that observing the saturation of $\hat{p}(c)$ does not guarantee that optimal performances are attained: the $p$-value evolution might decrease very slowly towards its (unknown) global minimum. However, a very slow decrease suggests that a significant performance improvement, if any, would require radically larger training data sets and neural networks, beyond what is feasible in practice. We can thus stop improving the quality of the ratio reconstruction as soon as saturation is observed. The same criterion is routinary adopted for the training of regular classifiers. Also in that context, the model improvement is stopped when the performances saturate, without guarantee that optimal performances have been attained.

\subsubsection[{The Neyman--Pearson \emph{p}-value: calculation}]{The Neyman--Pearson $\boldsymbol{p}$-value: calculation}\label{sec:nppcalc}

The determination of the median $p$-value $\hat{p}(c)$ is conceptually straightforward, but numerically cumbersome. The direct approach is to generate artificial instances of the experimental data set $\data=\{x_i\}_{i=1}^{\N}$ that follow the $H_0$ hypothesis with non-vanishing $c$, and data that follow $H_1$, with $c=0$. These pseudo-data---called toy data---are build by first drawing the total number $\N$ from a Poisson distribution with expected $N(c)$ or $N(0)$, as in the hypothesis under examination, and next extracting $\N$ instances of the $x$ variable, following the appropriate distribution, from Monte Carlo data. Toy data with $c\neq0$ serve to determine empirically the distribution of the $\hat{t}_c$ variable~(\ref{eq:tchat}) and in turn the $p_c(t)$ function. From the $c=0$ toys one computes the median and $\hat{p}(c)$~(\ref{eq:pval0}). This procedure is feasible, but too slow to repeatedly compute $\hat{p}(c)$ for performance evaluation and hyper-parameters scan. Two alternative approaches are described below.

The first approach~\cite{Chen:2020mev} is to model analytically the probability density functions of $\hat{t}_c$ under the two hypotheses $c\neq0$ and $c=0$. The former probability function determines $p_c(t)$~(\ref{eq:tchat}), while the knowledge of the latter one enables to compute the median in eq.~(\ref{eq:pval0}). Such analytical modeling is possible because $\hat{t}_c$~(\ref{eq:tchat}) is trivially related to the sum over the data set of the variable ${\hat{\tau}_c}(x)$. Since the data set is large, the Central Limit theorem ensures that the distribution of $\hat{t}_c$ is approximately Gaussian. Departures from Gaussianity can be taken into account by modeling the $\hat{t}_c$ distribution with a skew-normal distribution~\cite{Chen:2020mev}. Its free parameters, namely the mean, variance and skewness are related to the moments of ${\hat{\tau}_c}(x)$ by 
\beq\label{eq:momt}
\mu(\hat{t}_c)=
2\big[N(c)-N(0)-N\,\langle
{\hat{\tau}_c}
\rangle\big]\,,\;\;\;\;\;
\sigma^2(\hat{t}_c)=4\,{N}\,\langle
{\hat{\tau}_c^2}
\rangle\,,
\;\;\;\;\;
\mu_3(\hat{t}_c)=\frac1{\sqrt{N}}\,
\frac{\langle
{\hat{\tau}_c^3}
\rangle^{\phantom{3/2}}}{\langle
{\hat{\tau}_c^2}
\rangle^{3/2}}
\,.
\eeq
In the equation, $N$ denotes the expected number of events when either $c\neq0$ or $c=0$. The expectation value, denoted as $\langle\ldots\rangle$, is taken either under the $c\neq0$ or the  $c=0$ hypotheses in order to determine the two distributions of $\hat{t}_c(\data)$. 

This semi-analytical approach to the calculation of ${\hat{p}}(c)$ is extremely fast, especially when using reweighted events that enable the determination of the averages in eq.~(\ref{eq:momt}) using the same Monte Carlo sample for any value of $c$. Implemented on a GPU, it can be run during training enabling online monitoring of the performances. It provides results that are normally accurate, as one can verify by comparing with the empirical evaluation of ${\hat{p}}(c)$ based on toy experiments. In some rare cases, however, the semi-analytical estimate of ${\hat{p}}(c)$ fails due to the failure of the quasi-Gaussian approximation for the distribution of $\hat{t}_c$. This typically occurs in networks that slightly overfit producing overly sharp peaks in $\hat{\tau}_c(x)$. The contribution of the peaks to $\hat{t}_c$ emerges from a small region in the $x$ space, where few events are present. This violates the Central Limit theorem even if the total number $\N$ of events is large. When the Central Limit theorem violation occurs, the estimate of ${\hat{p}}(c)$ can be either much larger or much smaller than the true $p$-value. 

A determination of ${\hat{p}}(c)$ that is more robust, but computationally more demanding, is obtained as follows. We consider the discretization of the variable $\hat{\tau}_c(x)$ in a large number $n_{\textrm{bin}}$ of non-overlapping bins. Namely, we approximate $\hat{\tau}_c(x)$ with a piecewise constant function:
\beq\label{eq:appthat}
\hat{\tau}_c(x)\;\simeq\;\left\{
\tau_{c}{[{\textrm{b}}]}\,,\;{\text{for}}\;
x\;\;{\textrm{s.t.}}\;\hat{\tau}_c(x)\in (\tau_{\textrm{b}-1},\,\tau_{\textrm{b}})
\right\}\,,
\eeq
where $\tau_0=-\infty$ and $\tau_{n_{\textrm{bin}}}=+\infty$ in order to cover the whole space. The approximately constant value of $\hat\tau_c$ in each bin, $\tau_{c}{[{\textrm{b}}]}$, equals the central point of the bin apart from the extremes, where we set $\tau_{c}{[1]}=\tau_1$ and $\tau_{c}{[n_{\textrm{bin}}]}=\tau_{n_{\textrm{bin}}-1}$. Using this approximation, eq.~(\ref{eq:tchat}) becomes
\beq\label{eq:tchatbinned}
{\hat{t}}_c(\data) \simeq 
2\,\bigg[N(c)-N(0)-\sum_{\textrm{b}}
\N_{\textrm{b}}\tau_{c}{[{\textrm{b}}]}\bigg]\,,
\eeq
where $\N_{\textrm{b}}$ is the number of points in $\data$ that fall in the bin. $\N_{\textrm{b}}$ follows a Poisson distribution with expected $L\,\Delta\sigma_{\textrm{b}}(c)$, or $L\,\Delta\sigma_{\textrm{b}}(0)$, with $L$ the luminosity of the experiment. Highly optimized computer packages exist to generate Poisson-distributed numbers. We can thus efficiently determine ${\hat{p}}(c)$ empirically by generating toy data in the $c\neq0$ and in the $c=0$ hypotheses.

The binned determination of ${\hat{p}}(c)$ relies on the choice of a binning strategy, and of the number of bins $n_{\textrm{bin}}$. Binning is performed by ensuring that the cross sections of all bins are equal under the SM hypothesis $c=0$. The number of bins must be large enough to ensure that the binned test statistic in eq.~(\ref{eq:tchatbinned}) is a good enough approximation of un-binned ${\hat{t}}_c$ in eq.~(\ref{eq:tchat}). At the same time, using too many bins reduces the accuracy of the determination of $\Delta\sigma_{\textrm{b}}$ due to the finite Monte Carlo statistics. For the studies performed in Section~\ref{sec:PS}, we found that $n_{\textrm{bin}}=1000$ is a good compromise. 
\section{Performance studies}\label{sec:PS}

We illustrate the performances of our methodology on the same case study considered in Ref.~\cite{Chen:2020mev}. This will enable a direct assessment of the advantages of training with reweighted Monte Carlo data, rather than populating the EFT Wilson coefficients parameter space by independent data sets as done in~\cite{Chen:2020mev}. We consider the production of a Z and of a W boson at the 14~TeV LHC, and their decay to leptons. We restrict our analysis to the high energy regime, with a cut of 300~GeV on the transverse momentum of the bosons. We study the sensitivity of this process, with the full integrated luminosity $L=3\,{\textrm{ab}}^{-1}$ of the HL-LHC, to two specific dimension-six EFT interactions
\beq
{\cal O}_{\varphi} = G_{\varphi} \left(\overline Q_L \sigma^a \gamma^\mu Q_L\right)
(i H^ \dagger {\scriptstyle \overleftrightarrow{\rule{0pt}{.75em}}}\hspace{-.95em}{D}_\mu H)\,,\qquad\quad
{\cal O}_{W} = G_{W} \varepsilon_{abc} {W^{a\,\nu}_{\mu}} {W^{b\,\rho}_{\nu}} {W^{c\,\mu}_\rho}\,.
\label{operators0}
\eeq
See Appendix~\ref{app} for an extensive description of the process and of the two Monte Carlo generators, namely the {ideal} and {NLO} generators, that we employ to model the distributions and the effect of the Wilson coefficients $c=(G_{\varphi},\, G_{W})$.

We do not aim at fully realistic sensitivity projections, nor at a comparative assessment of the ZW process sensitivity to the EFT operators at hand and its role in a global fit. Nevertheless, it is worth emphasizing~\cite{Falkowski:2015jaa,Green:2016trm,Butter:2016cvz,Franceschini:2017xkh,Panico:2017frx,Azatov:2017kzw,Azatov:2019xxn,Baglio:2019uty} that the ZW process at high energy is a promising probe of the operators ${\cal O}_{\varphi }$ and ${\cal O}_{W}$, because these operators produce growing-with-energy effects in the ZW scattering amplitudes as displayed in eq.~(\ref{eq:amplitudes}). The unique ability of the LHC to probe the operators in the high energy regime, where their effects are enhanced, can boost the sensitivity to their Wilson coefficients well beyond the current bounds from measurements performed at lower energy. Furthermore, the three leptons from the ZW decay define a sufficiently complex final state to expect a gain in sensitivity from an unbinned multivariate analysis in comparison with a more standard approach based on binned measurements of one- or two-dimensional distributions. A sensitivity gain by a factor more than 2 was demonstrated in~\cite{Chen:2020mev} for the $G_W$ Wilson coefficient, while the gain on $G_{\varphi}$ is more moderate. The different behavior of the two operators is due---see~\cite{Panico:2017frx,Chen:2020mev}---to the different role that is played by the kinematical variables that describe the decay of the vector bosons: their measurement is essential in order to access the growing-with-energy linear term in $G_W$, while the growing-with-energy linear $G_{\varphi}$ term is present already in the differential di-boson cross section integrated over the decay angles. A simple binned analysis that does not exploit the distribution of the decay angles fully is thus nearly optimal for $G_{\varphi}$ and vastly sub-optimal in the case of $G_W$. At the technical level, this difference makes the $G_W$ contribution to the distribution ratio $r(x;c)$ a much more intricate function of the kinematical variables $x$ than the $G_{\varphi}$ contribution. Learning the $G_W$ contribution is thus a harder problem than learning the $G_{\varphi}$ contribution.

We consider the problem of learning the distribution ratio based on two different Monte Carlo generators: the ideal and the NLO generators. The ideal generator offers a simplified description of the process. It is based on approximations that are not sufficiently accurate for the actual analysis of the data, but enable a simple analytical determination of the true distribution ratio $r(x;c)$. Applying our methodology to the ideal setup provides a validation of the performances against ground-truth knowledge, on a learning problem of realistic complexity. The NLO generator provides instead an accurate state-of-the-art description of the ZW process. Studying the problem with the NLO generator validates our methodology in a realistic setup and tests its ability to deal with events with negative weight, whose presence is unavoidable for event generation beyond the tree level. Appendix~\ref{app} provides an extensive technical description of the ideal and NLO Monte Carlo generators and of how they are employed to produce reweighted data sets. 

The implementation of our learning strategy (defined in Section~\ref{subsec:lfw}) and the study of its performances (based on the metrics of Section~\ref{subsec:val}) on ideal and on NLO data is described in the next two sections in turn. The direct comparison between the reconstructed and true distribution ratios will be also employed as a performance indicator in the case of ideal data.

All our models are implemented in \texttt{PyTorch} version 1.11.0~\cite{NEURIPS2019_9015} and CUDA 11.3. All trainings were performed on NVIDIA A30 GPU and employing the Adam optimization algorithm~\cite{kingma2017adam}.

\subsection{Ideal data}\label{sec:idd}

Reweighted ideal Monte Carlo data sets ${\textrm{S}}=\{{\textrm{e}}_i\}_{i=1}^{n({\textrm{S}})}$, with ${\textrm{e}}=(x_{\textrm{e}},\,w_{\textrm{e}}(c))$, are sampled from the ideal Monte Carlo generator---described in Appendix~\ref{app}---implemented in a dedicated code. A total of 20 million events have been generated to produce the results that follow. Ten million are used for testing purposes, namely for the evaluation of the performance metrics introduced in Section~\ref{subsec:val}. Training is performed with $n({\textrm{S}})=3{\textrm{M}}$ if not specified otherwise. An independent sample of the same size is employed for validation during training. The weights $w_{\textrm{e}}(c)$ are computed by reweighting in the latent space as in eq.~(\ref{rw:lat}). 

Each event is characterized by seven independent observable variables $x$, listed in eq.~(\ref{feat}). It facilitates the learning task to pre-process the input by performing change of variables that avoid overly sharp one-dimensional marginal distributions, and by introducing redundancies. We pre-process with the transformation 
\beq\label{eq:truef}
x\;\to\; \bigg\{ \log[s/{\textrm{GeV}}^2],\,\Theta,\, 
\theta_Z,\,
\theta_W,\, 
\log[p_T/{\textrm{GeV}}],\,
Q,\,
\sin\varphi_Z,\,
\sin\varphi_W,\,
\cos\varphi_Z,\,
\cos\varphi_W
\bigg\}\,.
\eeq
The neural networks we employ for our analysis thus receive a total of 10 features as input, ordered as in the above equation. A normalization layer in the network shifts and scales each variable to have zero mean and unit variance on the training sample. Our pre-processing~(\ref{eq:truef}) follows relatively standard practice: the steeply-falling distribution of the center of mass energy squared, $s$, is smoothed out by taking the logarithm. The $2\pi$ periodicity of the azimuthal angles is enforced explicitly by giving the sine and the cosine, rather than the angle itself, as input to the network. The variable $p_T=\sqrt{s}/2\,\sin\Theta$ can be more useful to the network than $s$ and $\Theta$ to model the effect of the EFT operators in some kinematical regimes. 

\subsubsection{The simple classifier}\label{sec:scper}

We start from the problem of learning the distribution ratio $r(x;{\bar{c}})$ at a specific point $c=\bar{c}$ of the Wilson coefficients parameter space. As explained in Section~\ref{subsec:sc}, this is achieved using the loss function in eq.~(\ref{eq:lossw}) to train a classifier $f(x)\in(0,1)$. We consider the classification function
\beq
f(x)=\frac1{1+e^{\rho(x)}}\,,
\eeq
where $\rho(x)$ is a feed-forward neural network with real output. By comparing with eq.~(\ref{eq:opc}) we see that, after training, the neural network provides an approximation, ${\hat{r}}(x,{\bar{c}})$, of the true distribution ratio ${{r}}(x,{\bar{c}})$. More precisely
\beq
\log{\hat{r}}(x,{\bar{c}})=\hat\tau_{\bar{c}}(x)=\rho(x)\,.
\eeq
The value of ${\bar{c}}$ chosen for illustration has $G_{\varphi}=0$ and $G_W=10^{-2}\,{\textrm{TeV}}^{-2}$. Learning the distribution ratio with the simple classifier for such a small value of $G_W$ was found to be possible in Ref.~\cite{Chen:2020mev}, but only modest accuracy could be attained. Better performances are expected using reweighted training data for the reasons explained in Section~\ref{subsec:sc}.

We use 3M training points, a $\rho$ neural network with $(10,64,64,1)$ architecture---namely, two hidden layers with 64 neurons each---and sigmoid activation functions. Pre-processing is performed as in eq.~(\ref{eq:truef}). Training employs an initial learning rate of $3\cdot10^{-3}$ for the first 20k epochs, after which the initial learning rate parameter is lowered to $10^{-3}$ without re-initializing the optimizer. This 2-step training scheme with decreasing initial learning rate is found to be convenient in general: the first stage reduces the loss quickly, but the precise optimization with a lower rate performed at the second stage is required in order to attain a deeper minimum of the validation loss.
Each training step is performed computing the gradients of the network parameters on the whole training set.
Updating the networks with mini-batches
is found, consistently with Ref.~\cite{Chen:2020mev}, to degrade the accuracy of the distribution ratio reconstruction strongly.
A validation loss is computed on 3M independent Monte Carlo data. Training is stopped at the minimum of the validation loss, which is reached in this case after around 45k training epochs. However, the quality of the ratio reconstruction is very stable during training. The behavior is similar to the one displayed in Figure~\ref{fig:evolution}, discussed in the next section. 

\begin{figure}[t]
    \centering
    \includegraphics[width=0.29\columnwidth]{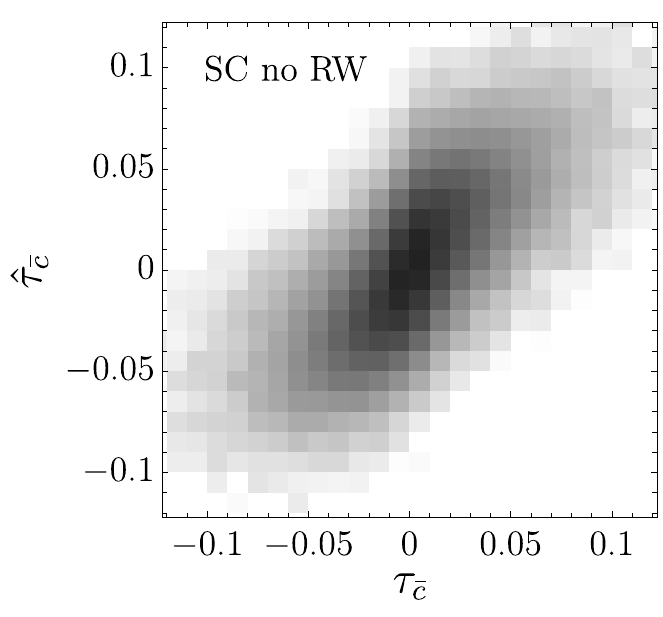}
    \hfill
    \includegraphics[width=0.29\columnwidth]{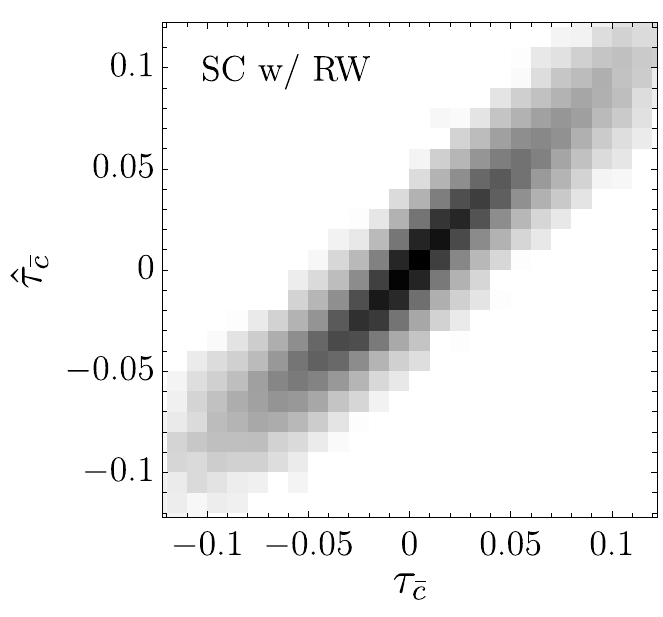}
    \hfill
    \includegraphics[width=0.353\columnwidth]{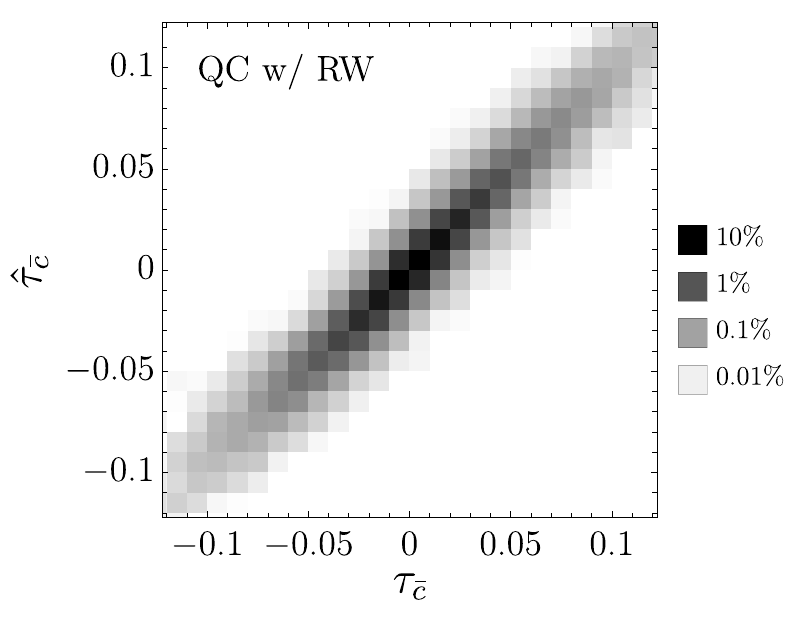}
    \caption{Density histograms displaying the correlation between the true $\tau_{\bar{c}}$ and the reconstructed $\hat\tau_{\bar{c}}$ obtained with three different strategy. From left to right: the simple classifier (SC) trained in Ref.~\cite{Chen:2020mev} on independent Monte Carlo samples for $c={\bar{c}}$ and $c=0$; the simple classifier trained with reweighted (RW) data; the quadratic classifier (QC) trained as explained in Section~\ref{sec:qcper}. The ${\bar{c}}$ point has $G_{\varphi}=0$ and $G_W=10^{-2}\,{\textrm{TeV}}^{-2}$.} \label{fig:scatter}
\end{figure}

The central panel of Figure~\ref{fig:scatter} (labeled ``SC w/ RW'') displays the quality of the log-ratio reconstruction by direct comparison with the true log-ratio $\tau_{\bar{c}}=\log r(x,{\bar{c}})$. The reconstruction is more accurate than the one, shown on the left panel with label ``SC no RW'', of the neural network trained in Ref.~\cite{Chen:2020mev} with the same Monte Carlo statistics but using independent data sets for $c=0$ and for $c={\bar{c}}$. Using reweighted Monte Carlo data for training is evidently beneficial.

The visual comparison between the left and the middle panels of Figure~\ref{fig:scatter} conclusively shows that a more accurate reconstruction of the distribution ratio is possible using reweighted training data. On the other hand, the comparison does not offer a quantitative measure of the advantages of reweighting, ultimately because it is unclear, a priori, which accuracy is needed for a satisfactory reconstruction of the ratio. As explained in Section~\ref{sec:nppdef}, a useful quantitative performance metric can be defined only in relation to the actual experimental conditions in which the reconstructed ratio will be employed for statistical inference. We target the HL-LHC collider, with an integrated luminosity of $3~{\textrm{ab}}^{-1}$. The ideal Monte Carlo predicts a total cross section $\sigma(0)=1.06~{\textrm{fb}}$, and $\sigma({\bar{c}})=\sigma(0)+4.0~{\textrm{ab}}$, corresponding to an expected data statistics $N(0)=3180$, within the SM, and to $N({\bar{c}})-N(0)=12$. Using this information we compute the median $p$-value $\hat{p}({\bar{c}})$ defined in Section~\ref{sec:nppdef}. Ten million ideal Monte Carlo data are employed to determine $\hat{p}({\bar{c}})$ using the binned strategy described in Section~\ref{sec:nppcalc}. An error is assigned to $\hat{p}({\bar{c}})$ by repeating the determination on ten partitions of the 10M data set and computing the standard deviation. The model from Ref.~\cite{Chen:2020mev}, corresponding to the left panel of Figure~\ref{fig:scatter}, has $\hat{p}({\bar{c}})=(3.8\pm0.3)\%$, while using reweighted events (middle panel) we reach $\hat{p}({\bar{c}})=(1.78\pm0.04)\%$. 

The right panel of Figure~\ref{fig:scatter} (labeled ``QC~w/~RW'') displays the reconstruction performances of the quadratic classifier strategy defined in Sections~\ref{qc-1} and~\ref{qc-2}, trained with reweighted data with the benchmark hyper-parameters described in the following section. The reconstruction further improves in comparison with the ``SC~w/~RW'' setup, but the $p$-value only drops by $20\%$: $\hat{p}({\bar{c}})=(1.40\pm0.04)\%$. A $20\%$ improvement of the $p$-value is appreciable but modest, especially in comparison with the reduction of a factor more than two of the ``SC~w/~RW'' $p$-value relative to the $p$-value in the ``SC~no~RW'' setup. We discussed in Section~\ref{sec:nppdef} that the saturation of the $p$-value is expected to occur when the reconstructed distribution ratio is so close to the true ratio that the test of hypothesis performed with the reconstructed likelihood attains nearly-optimal performances. The optimal median $p$-value, obtained using the knowledge of the true distribution ratio that is available for ideal data, is ${p}({\bar{c}})=(1.12\pm0.04)\%$. The ``SC~w/~RW'' $p$-value is quite close to the optimal $p$-value. The improvement in the ratio reconstruction that is achieved in the ``QC~w/~RW'' setup can lower the $p$-value, but obviously not push it below the optimal $p$-value. The performance gain is thus unavoidably moderate.

The slight gap in performances between the ``SC w/ RW'' and the ``QC w/ RW'' models probably emerges from the combination of two factors. First, the quadratic classifier strategy is advantageous because it exploits the quadratic dependence of the ratio on the Wilson coefficients in order to learn the ratio simultaneously from several different points in the $c$ parameter space. Second, no hyper-parameters optimization was performed in the case of the simple classifier, while the quadratic classifier hyper-parameters are optimized as discussed in Section~\ref{sec:hyps}. Regardless of performance, the simple classifier is not a viable strategy for the EFT likelihood learning. We will thus not investigate it further, nor optimize its performances, and devote the rest of this section to the quadratic classifier.

\subsubsection{The quadratic classifier}\label{sec:qcper}

Statistical inference requires the knowledge of the distribution ratio $r(x;c)$ as a function of the Wilson coefficients $c$. This can be learned efficiently by exploiting the known (quadratic) dependence of the ratio on $c$ as explained in Sections~\ref{qc-1} and~\ref{qc-2}. The implementation of this strategy on ideal Monte Carlo data is presented below.

We are interested in the dependence of $r$ on the two Wilson coefficients $c=(G_{\varphi},G_W)$, however, we start from the simpler one-dimensional problem of learning the ratio in the direction of each of the two coefficients, setting the other one to zero. This is a valid starting point in general because the terms in the distribution ratio that are linear in the Wilson coefficients are typically harder to reconstruct, and often more important for the sensitivity because they account for the leading corrections to the SM distributions. It is thus convenient to study and optimize the reconstruction of each of them separately in the different one-dimensional problems, where they contribute individually. Furthermore, separate trainings in the direction of each Wilson coefficient are the first step of the efficient parallelizable learning strategy described in Section~\ref{qc-2}. Learning the dependence of the ratio on $G_{\varphi}$ turns out---as anticipated---to be rather simple in the case at hand. We thus describe the one-dimensional learning problem only in the $G_{\varphi}=0$ direction and consider non-vanishing $G_{\varphi}$ only for the two-dimensional study. 

We employ the one-dimensional parametrization in eq.~(\ref{eq:qcd1d}) using two feed-forward neural networks to model the two coefficient functions $\rho(x)$ and $\theta(x)$. The input is pre-processed by the transformation in eq.~(\ref{eq:truef}). In the benchmark configuration, $(10,24,24,1)$ architecture and sigmoid activation functions are considered for the two networks. The parametrization is inserted in the classification function defined by eq.~(\ref{eq:clafu}), out of which the loss function in eq.~(\ref{eq:losswpc}) is constructed. At the end of training, the reconstructed distribution ratio ${\hat{r}}(x;c)$ is given by
\beq\label{reccr}
{\hat{r}}(x;c)={\mathcal{P}}(\rho(x),\theta(x);c)=
1+2\,c\,\rho(x)\,\sin\theta(x)+c^2\rho^2(x)\,,
\eeq
as a function of the Wilson coefficient $c=G_W$.

The quadratic classifier loss function~(\ref{eq:losswpc}) depends on a list ${\cal{C}}=\{{\bar{c}}^{(1)},\ldots,{\bar{c}}^{(\kappa)}\}$ of points in the Wilson coefficient parameter space. These points can be freely chosen and they are among the hyper-parameters associated with the training of the quadratic classifier. In the benchmark configuration, we use the following values
\beq\label{gwtv}
{\cal{C}}=\{\bar{G}_W^{(1)},\ldots,\,
\bar{G}_W^{(6)}\} = \{\pm 5, \pm 2.5, \pm 1.25\}\times 10^{-2}\,\textrm{TeV}^{-2}\,.
\eeq
The criteria for selecting these hyper-parameters, and the (very mild) effect of departures from the benchmark choice, are discussed in Section~\ref{sec:hyps}.

\begin{figure}[t]
    \centering
    \includegraphics[width=0.45\columnwidth]{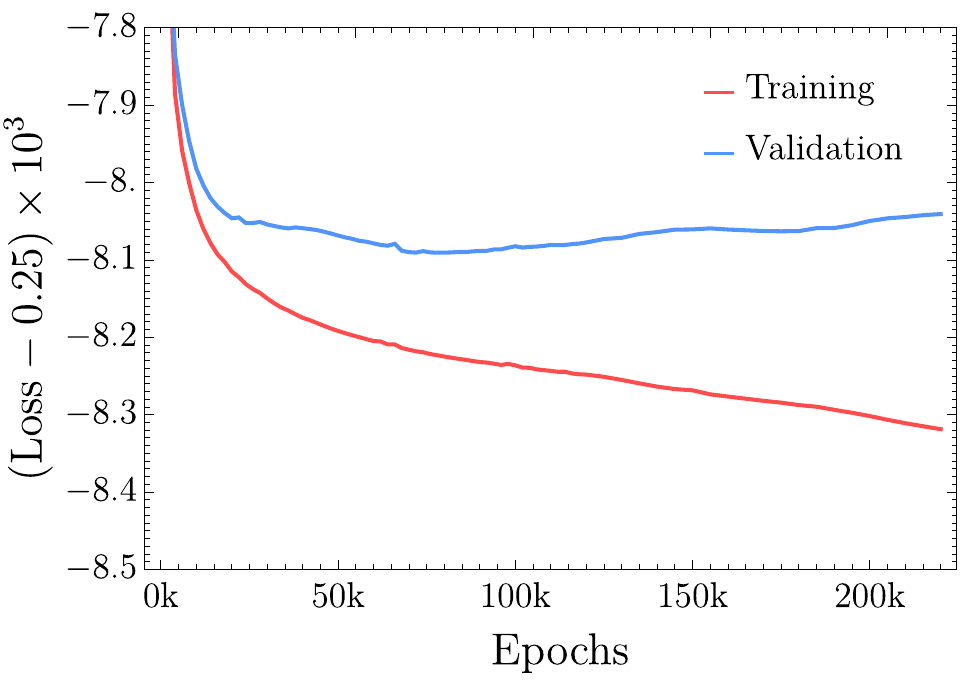}
    \hfill
    \includegraphics[width=0.445\columnwidth]{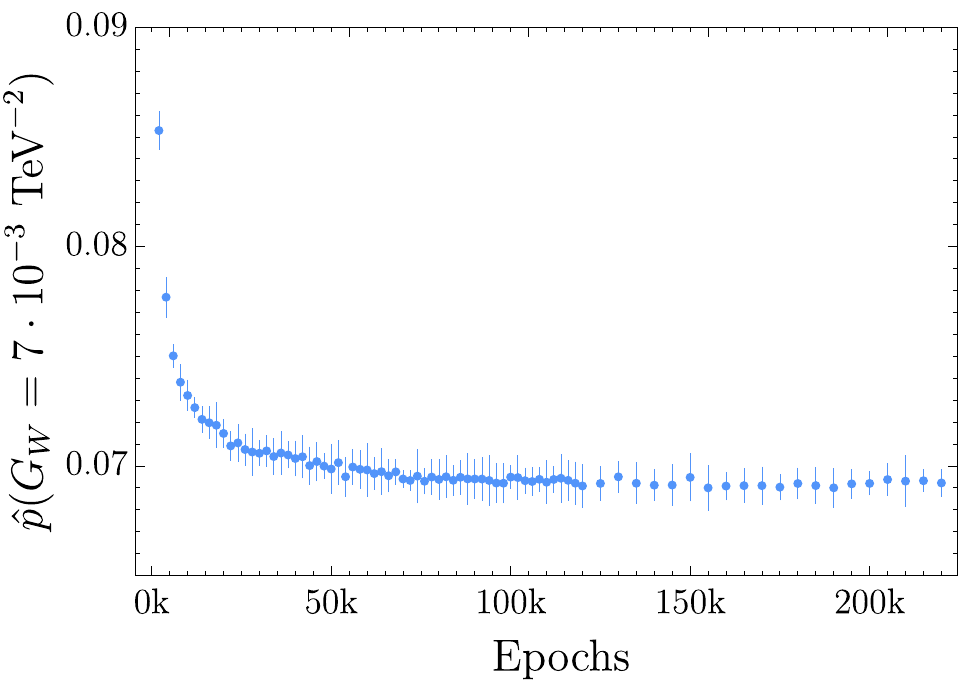}
    \caption{Evolution of the training loss and validation loss (left panel), and of the median $p$-value evaluated at $G_W=7\cdot10^{-3}\,{\textrm{TeV}}^{-2}$ (right panel) for the benchmark network} \label{fig:evolution}
\end{figure}

Training is performed on 3M re-weighted Monte Carlo events, and 3M independent data are used for validation. We adopt the 2-step training strategy with decreasing learning rate described in Section~\ref{sec:scper}. The left panel of Figure~\ref{fig:evolution} shows the evolution of the loss function during training. Actually, the one reported in the figure is a re-scaled loss function, obtained by dividing eq.~(\ref{eq:losswpc}) by the sum of all the weights that appear in the equation, namely by $w_{\textrm{e}}(\bar{c})+w_{\textrm{e}}(0)$ summed over ${\textrm{e}}$ and over $\bar{c}\in{\cal{C}}$. With this normalization, an indecisive classification function $f=1/2$ gives a loss of exactly $1/4$. Our classification function will be close to $1/2$ because the distribution ratio is close to one. Plotting the normalized loss minus $0.25$, times a large factor, thus provides a better representation of the training evolution. Notice that the training and the validation loss are normalized separately. This avoids the emergence of spurious differences due to the different total weight of the training and of the validation data sets. Also notice that the normalized loss and not the original loss~(\ref{eq:losswpc}) is passed to the optimizer in order to prevent possible issues associated with a loss that is numerically very far from unity.

Training is run for many epochs, waiting for an increase in the validation loss that signals overfitting. The trained model configuration is the one that minimizes the validation loss. In Figure~\ref{fig:evolution}, the minimum is attained after around 65k training epochs. We also evaluate, during training, the median $p$-value $\hat{p}(\bar{c})$ for a representative value $c=\bar{c}$ of the Wilson coefficient. This is chosen to be close to the conventional threshold for exclusion of $5\%$. $G_W=7\cdot10^{-3}\,{\textrm{TeV}}^{-2}$ is employed in Figure~\ref{fig:evolution}. The $p$-value computed at run time is obtained with the skew-normal approximation of the test statistic distribution described in Section~\ref{sec:nppcalc}. In Figure~\ref{fig:evolution} we report instead the binned determination of the $p$-value, evaluated off-line on the saved models. The two determinations are actually in good agreement in the specific case under examination.

Figure~\ref{fig:evolution} displays a remarkable stability of the $p$-value during training, which extends deeply in the overfitted region where the validation loss (moderately) increases. We also observe a precise correspondence between the minimum of the validation loss---at 65k epochs---and the onset of the $p$-value stability region. The figure also shows that good performances could have been obtained also with less training epochs: after 10k epochs the $p$-value is only $10\%$ larger than at the end of training. We are not limited by the training time because training for 1000 epochs takes around 1 minute in the benchmark setup, on the NVIDIA A30 GPU we used to produce our results. A reduction of the number of training epochs could however be considered in computationally more demanding problems.

We discussed in Section~\ref{subsec:val} that the median $p$-value is our prime performance metric. It is particularly powerful in the case of ideal data because the optimal $p$-value can be computed owing to the knowledge of the $r$ ratio. For the value $G_W=7\cdot10^{-3}\,{\textrm{TeV}}^{-2}$, considered in Figure~\ref{fig:evolution}, the optimal $p$-value is $(6.5\pm0.1)\%$, while the quadratic classifier in the benchmark configuration gives $(6.95\pm0.04)\%$. We can thus conclude that the quadratic classifier in the benchmark configuration has reached effectively optimal performances and no further improvement is needed in the reconstruction of the distribution ratio. However, the $r$ ratio and in turn the optimal $p$-value is never known in realistic problems. A more direct validation of the quality of the distribution ratio reconstruction must thus accompany the calculation of the median $p$-value.

\begin{figure}[t]
    \centering
    \includegraphics[width=0.45\columnwidth]{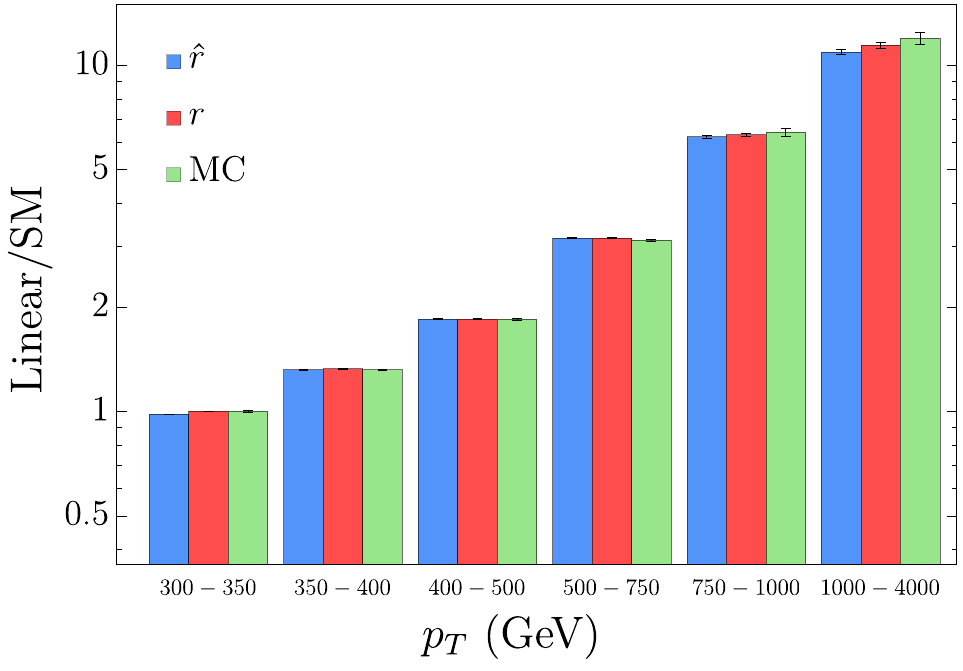}
    \hfill
    \includegraphics[width=0.46\columnwidth]{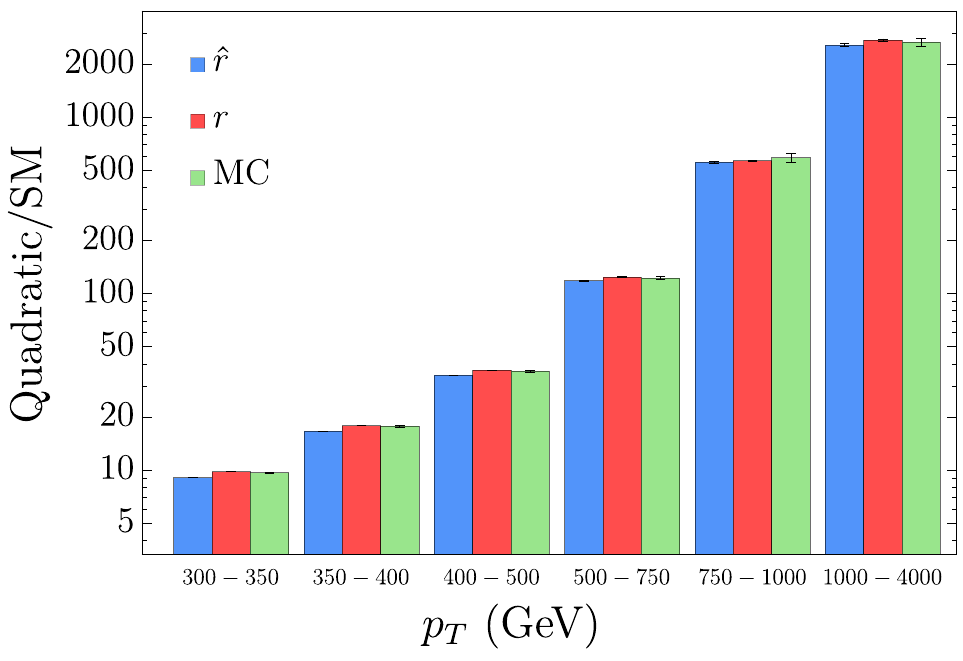}
    \caption{Linear (left) and quadratic (right) contributions of the ${\cal O}_W$ operator to the cross section in $p_{T}$ bins, in the ideal setup. The direct determination from Monte Carlo data, by eq.~(\ref{eq:ds}) is shown in green. The determination from the reconstructed ratio (see~(\ref{eq:recsig})) in the benchmark configuration is shown in blue. The one from the exact ratio is in red. Angular cuts $|\cos \varphi_{W/Z}| > 0.6$ are applied.} \label{fig:idealrec}
\end{figure}

The simplest test of the $r$-ratio reconstruction quality is to compare the predictions for the cross section in bins obtained from the reconstructed ratio $\hat{r}(x;c)$, as in eq.~(\ref{eq:recsig}), with direct Monte Carlo estimates. Particularly significant binned distributions must be selected for the comparison. In Figure~\ref{fig:idealrec} we consider the transverse momentum of the Z boson, $p_{T}$, with cuts $|\cos \varphi_{W/Z}| > 0.6$ on the W and Z  decay angles. The selections are needed to access---in order to validate its reconstruction---the growing-with-energy linear term in $G_W$, which cancels in observable integrated over the angles. The linear and quadratic contributions to the distribution, relative to the SM, are reported separately in the figure. The reconstructed predictions (in blue) are in good agreement with the Monte Carlo predictions (in green). Both predictions are affected by errors due to the finite Monte Carlo statistics. Errors are obtained by splitting the test data in 10 subsets of 1M points each, repeating the determination of the observables on each subset and computing the standard deviation. The figures also reports the predictions obtained using, in eq.~(\ref{eq:recsig}), the true $r$ in place of the reconstructed $\hat{r}$. The perfect agreement with the Monte Carlo predictions provides a cross-check and supports the credibility of our error estimate.

\begin{figure}[t]
    \centering
    \includegraphics[width=0.45\columnwidth]{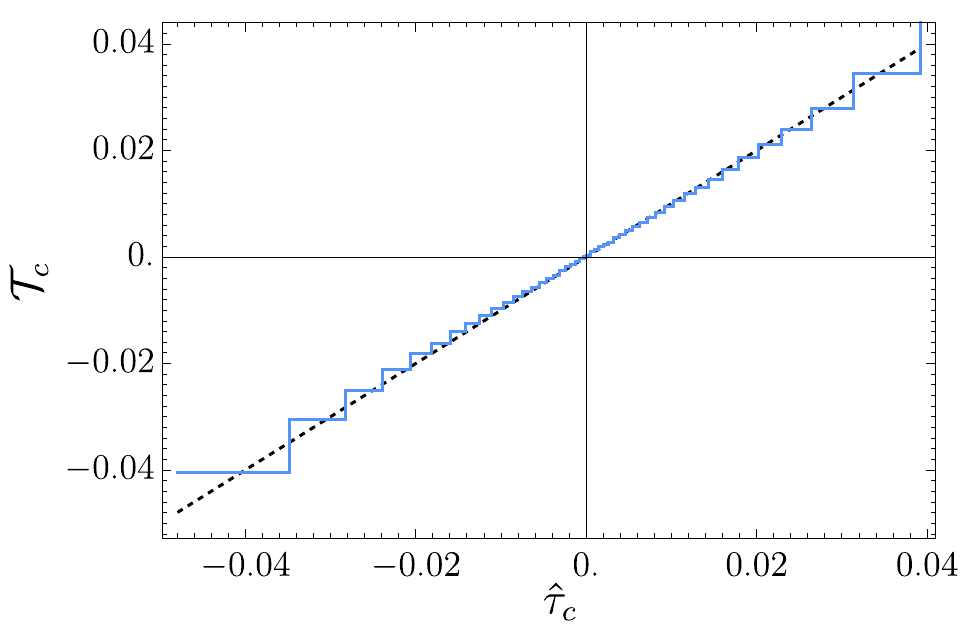}
    \hfill
    \includegraphics[width=0.45\columnwidth]{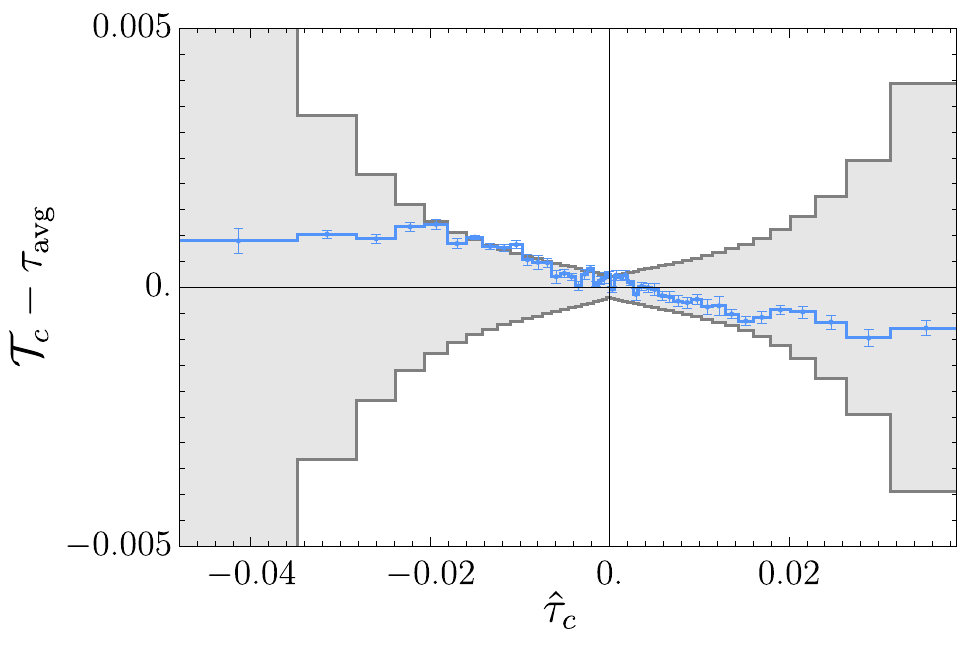}
    \caption{The $\Tau_c$ quantity of eq.~(\ref{eq:rcq}), for the reconstructed $\hat{r}$ ratio obtained with the quadratic classifier in the benchmark configuration, with ideal data.  $G_W = 7 \cdot 10^{-3}\;\textrm{TeV}^{-2}$ is considered. The right panel illustrates the approximate validity of the bounds in eq.~(\ref{eq:taubounds}), as explained in the main text.} \label{fig:scaletta}
\end{figure}

We can also monitor the quality of the reconstruction by exploiting the peculiar property of $\Tau_c$, defined in eq.~(\ref{eq:rcq}) as the logarithm of the cross section ratio in bins of $\hat\tau_c(x)=\log\hat{r}(x;c)$. We consider $c=G_W = 7 \cdot 10^{-3}\;\textrm{TeV}^{-2}$ and we employ $50$ bins, defined in such a way that each bin contains $1/50$ of the SM cross section. A finer binning could provide a more accurate test of the quality of the reconstruction, but it would compromise the accuracy of the $\Tau_c$ prediction due to the finite Monte Carlo statistics. The results, shown on the left panel of Figure~\ref{fig:scaletta}, display the approximate relation $\Tau_c=\tau_{\textrm{avg}}$---with $\tau_{\textrm{avg}}=(\hat\tau_++\hat\tau_-)/2$ the center of the bin---that signals a good agreement of the reconstructed ratio with the true ratio. The right panel of the figure verifies the approximate validity of the bounds in eq.~(\ref{eq:taubounds}) by plotting $\Tau_c-\tau_{\textrm{avg}}$ overlaid with the intervals $[(\hat\tau_--\hat\tau_+)/2,(\hat\tau_+-\hat\tau_-)/2]$, for each bin, represented as a shadowed region. We see that $\Tau_c$ is often far from the center of the interval. It falls preferentially close to the upper or to the lower extreme. This does not signal a poor agreement between the reconstructed and the true ratio: we verified that the same behavior is observed using the true log-ratio $\tau_c(x)=\log{r}(x;c)$ instead of the reconstructed one for the calculation of $\Tau_c$. This is because the $\tau_c$ distribution is sharply peaked at zero. The cross section integral---see for instance eq.~(\ref{eq:bder})---is thus dominated by the lower or upper extreme of the integration region for positive and negative $\tau$, respectively. We thus preferentially saturate the lower/upper extreme of eq.~(\ref{eq:taubounds}) for positive/negative $\tau$, precisely like we see happening for the reconstructed $\hat\tau_c$, in the figure. Because of these considerations, the only indication of a discrepancy between the reconstructed and the true ratio are those bins where the bound is strictly violated. There are very few such bins on the right panel of Figure~\ref{fig:scaletta}.

Since we got satisfactory performances on the one-dimensional problem, we can now address the complete task of learning the distribution ratio in the plane $c=(G_{\varphi},G_W)$. We use the two-dimensional parametrization of eq.~(\ref{eq:lam}), employing 5 feed-forward neural networks to model the two $\rho$ and the three $\theta$ coefficient functions. For the modeling of $\rho$ and $\theta$ in the $G_W$ direction we use the $(10,24,24,1)$ architecture, which was found to perform well on the one-dimensional problem. The architecture for $\rho$ and $\theta$ in the $G_{\varphi}$ direction could in principle have been chosen differently, following a hyper-parameters optimization on the one-dimensional $G_{\varphi}$ problem. However, since learning the dependence of the ratio on $G_{\varphi}$ is a very simple task as previously discussed, no specific optimization is required and the $(10,24,24,1)$ architecture is chosen for simplicity. The same architecture is also used for the third $\theta$ network, which describes the $G_{\varphi}G_W$ mixed term. The parametrization is inserted in the classification function~(\ref{eq:clafu}) and eventually in the loss function~(\ref{eq:losswpc}). After training, the reconstructed $\hat{r}(x;c)$ is obtained from the parametrization similarly to eq.~(\ref{reccr}).

The training points in the $\cal{C}$ set are formed by the $G_W$ values in eq.~(\ref{gwtv}) setting $G_{\varphi} = 0$, plus $G_W=0$ points with $G_{\varphi}$ in the set
\beq
G_{\varphi} \in \{\pm 2.5, \pm 1.25, \pm 0.625\}\times10^{-2}\,\textrm{TeV}^{-2}\,,
\eeq
and six additional points with both $G_W$ and $G_\varphi$ non-vanishing along the diagonal of the grid formed by the two lists of values.

An alternative approach to the determination of the two-dimensional ratio (see Section~\ref{qc-2}) is to learn the diagonal $\rho$ and $\theta$ networks in one dimension, and to perform two-dimensional training only to train the mixed $\theta$ network. This computationally convenient strategy is not needed for our analysis, because our resources are sufficient to train the five neural networks at once directly in the two-dimensional setup. One might have expected a more accurate reconstruction of the ratio with two individual one-dimensional trainings, but in fact no such advantage has been observed in the case under examination. We have verified that one-dimensional trainings give essentially identical results as training directly in two dimension. In particular, the same $p$-values are found in the two cases along the single-coefficient lines $G_\varphi=0$ or $G_W=0$.

 \begin{figure}[t]
    \centering
    \begin{minipage}[b]{.5\textwidth}
    \includegraphics[width=1\columnwidth]{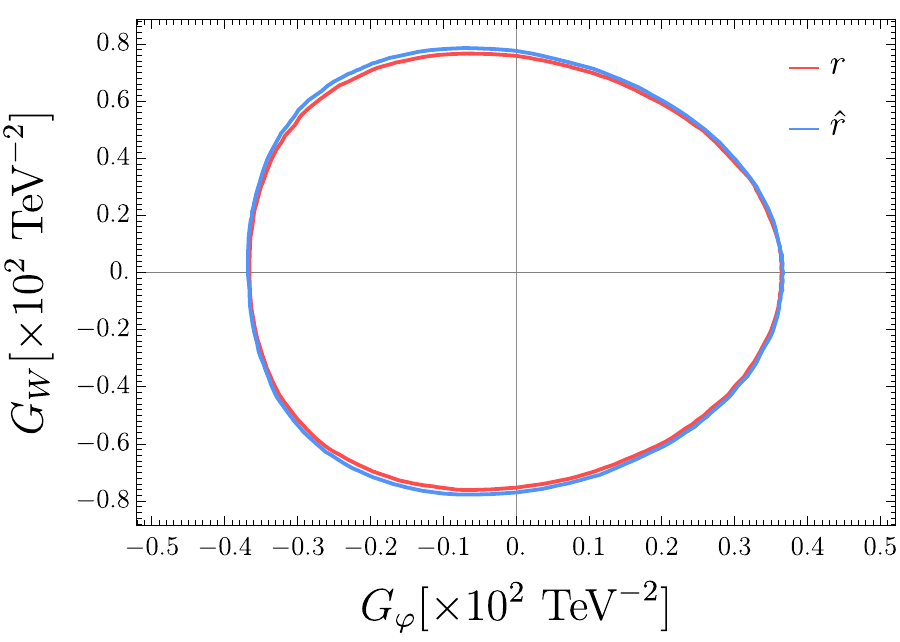}
    \end{minipage}
    \hfill
    \begin{minipage}[b]{.47\textwidth}
    \small
    \begin{tabular}[b]{c|c|c}
    \multirow{2}{*}{\rule{0pt}{1.1em}$G_{W}$} & \rule{0pt}{1.25em}$r$ & $[-0.752(2), 0.756(3)]$ \\
    & \rule{0pt}{1.5em}$\hat r$ & $[-0.770(2), 0.773(2)]$ \\
    \midrule
    \multirow{2}{*}{\rule{0pt}{1.1em}$G_{\varphi}$}  & \rule{0pt}{1.25em}$r$ & $[-0.3657(4), 0.3642(3)]$ \\
    & \rule{0pt}{1.5em}$\hat r$ & $[-0.3671(3), 0.3654(3)]$ \\
    \midrule
    \multirow{2}{*}{\rule{0pt}{1.1em}$G_{W}=G_{\varphi}$} & \rule{0pt}{1.25em}$r$ & $[-0.3409(6), 0.3207(4)]$\\
    & \rule{0pt}{1.5em}$\hat r$ & $[-0.3436(6), 0.3228(4)]$ \\
    \midrule
    \multirow{2}{*}{\rule{0pt}{1.1em}$G_{W}=-G_{\varphi}$} & \rule{0pt}{1.25em}$r$ & $[-0.3232(4), 0.3433(5)]$\\
    & \rule{0pt}{1.5em}$\hat r$ & $[-0.3257(4), 0.3477(5)]$ 
    \end{tabular}
    \vspace{1em}
    \end{minipage}
    \hspace{-2.25em}
    \caption{LEFT: Contour lines $\hat{p}(c)=5\%$ in the two-dimensional parameter space, for ideal data. The blue contours are obtained with the reconstruction of the $r$-ratio obtained by the quadratic classifier trained in two dimensions. The red contours are obtained using the exact ratio. RIGHT: Reach at NLO on $G_{\varphi}$ and $G_{W}$ in 4 directions on the plane. The bounds are given in units of $10^{-2}$ TeV$^{-2}$.} \label{fig:idealContour}
\end{figure}

Several studies were performed to validate the accuracy of the distribution ratio reconstruction along different directions of the 2-dimensional parameter space. The performances are similar---or better, along the $G_{\varphi}$ direction---than the ones previously described for $G_W$ in the one-dimensional case. The corresponding plots are not reported for brevity. In essence, we find that the quadratic classifier attains a basically perfect reconstruction of the distribution ratio, that guarantees nearly-optimal statistical performances. This is shown in Figure~\ref{fig:idealContour} by drawing $95\%$ exclusion contours in the $(G_\varphi,G_W)$ plane. The plot is obtained by computing and interpolating $\hat{p}(c)$ on a grid of points, and drawing the $\hat{p}(c)=5\%$ contours. The red contour is obtained using the exact $r$ ratio in place of the reconstructed ratio $\hat{r}$. By the Neyman--Pearson Lemma, the one that employs the exact ratio is the most powerful hypothesis test, namely the one with the smallest median expected $p$-value that in turn corresponds to the optimal (smallest) exclusion contour. The contour obtained with the reconstructed ratio, in blue, essentially coincides with the optimal contour. The $95$~CL exclusion reach is also reported in the Table in the right panel of Fig.~\ref{fig:idealContour} in different directions of the parameter space.

\subsubsection{Hyper-parameters selection}\label{sec:hyps}

\begin{figure}
    \centering
    \includegraphics[width=\columnwidth]{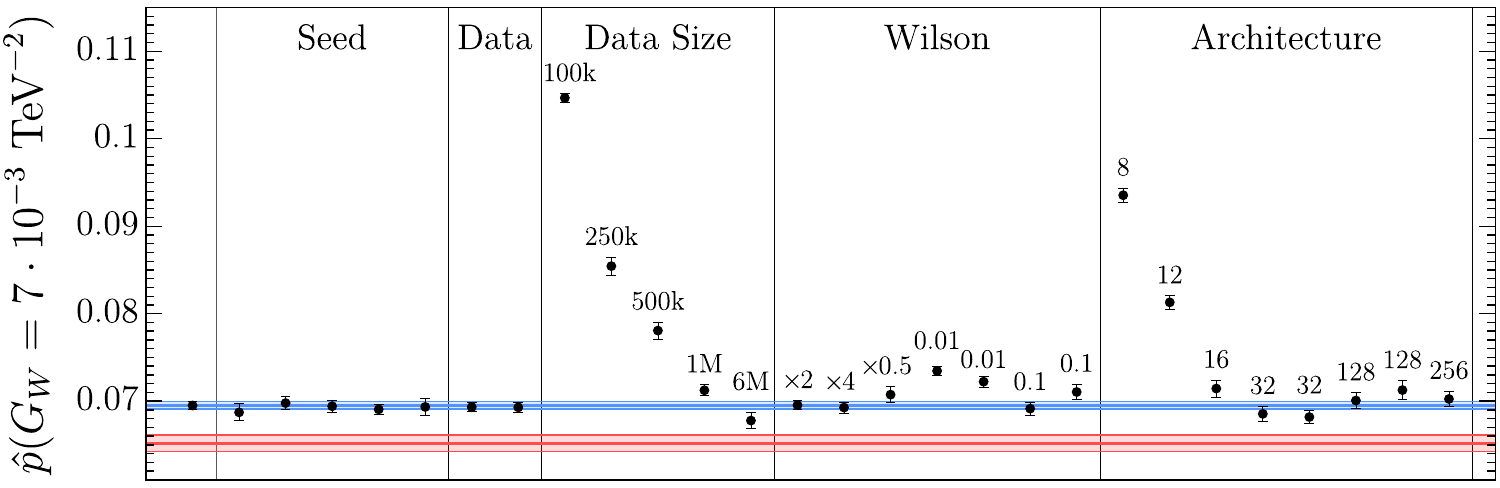}
    \caption{Median $p$-value for $G_W = 7 \cdot 10^{-3}\;\textrm{TeV}^{-2}$ in the configurations described in the main text. The blue band is the result obtained in the benchmark configuration. The red band is the optimal $p$-value. Training is repeated with two different seeds for the 32 and 128 neurons architectures.} \label{fig:hyperparameters}
\end{figure}

The benchmark choice of the hyper-parameters described in the previous section results from an optimization of the performances. The median $p$-value was used as the prime indicator, but other performance metrics such as those in Figures~\ref{fig:idealrec} and~\ref{fig:scaletta} were also employed, finding good correlation with the median $p$-value. The optimization was performed in the one-dimensional setup with $G_\varphi=0$ as previously explained. In this section, we report some of the results of this hyper-parameters scan. In particular, we study the dependence of the performances on each individual hyper-parameter, keeping the other ones fixed to the benchmark values. The performances are measured by the median $p$-value at $G_W = 7 \cdot 10^{-3}\;\textrm{TeV}^{-2}$. The results are reported in Figure~\ref{fig:hyperparameters} and compared with the benchmark $p$-value, reported as a blue band.

We consider variations of the benchmark architecture, employing two hidden layers with a variable number of neurons. We observe, in Figure~\ref{fig:hyperparameters}, the expected saturation of the performances. Small networks are not expressive enough to accommodate an accurate modeling of the true distribution ratio. Large networks can instead model the distribution ratio accurately, fully exploiting the information that is present in the training data. On the other hand, large networks are more exposed to overfitting, which might prevent an accurate learning. Indeed, the training of large networks behaves differently from the one of the benchmark networks, displayed in Figure~\ref{fig:evolution}: overfitting is more pronounced and it is associated with a degradation of the $p$-value. Using the model configuration that minimizes the validation loss eliminates this issue and delivers good performances also with rather large networks, as the figure shows. 

In Figure~\ref{fig:hyperparameters} we also study the dependence on the number of Monte Carlo data points used for training. A very significant degradation of the performances is observed only with a significant reduction of the training data set size relative to the benchmark of 3M points. The $p$-value is only slightly larger than the benchmark $p$-value using 1M training points. Using 6M points the $p$-value is smaller than the benchmark $p$-value, but the gain in performance is too limited to justify doubling the training sample size. The benchmark choice of 3M points is almost equally effective and in fact, very good results would have been obtained already with 1M points. 

We also report in Figure~\ref{fig:hyperparameters} two technical checks of stability. The ``Seed'' column displays the $p$-value obtained by repeating training in the benchmark configuration using 5 different random seeds for the initialization of the neural network parameters. In the ``Data'' column we repeat training on two independent data sets. Since they are independent of the seed and of the specific data set used for training, our results are robust and reproducible.

In the ``Wilson'' column we study the dependence on the values of the Wilson coefficients used for training, namely on the set of values, $\cal{C}$, that defines the loss function of the quadratic classifier in eq.~(\ref{eq:lossw}). The values used in the benchmark configuration---reported in eq.~(\ref{gwtv})---were selected with a very simple criterion. At $95\%$~CL, the analysis is sensitive to values of $G_W$ that are of the order of $10^{-2}\;\textrm{TeV}^{-2}$. We should thus prioritize an accurate reconstruction of the ratio when $G_W$ in this range, because much larger values will be in any case easy to exclude while much lower values are invisible. We thus use for training values of $G_W$ that are of this order of magnitude, with a substantial spacing among them in order to provide more information about the dependence on the ratio on $G_W$. In order to study the sensitivity to the specific benchmark choice, the values~(\ref{gwtv}) used in the benchmark configuration were raised by a factor of 2 or of 4, or divided by 2. The results---labeled as ``$\times2$'', ``$\times4$'' and ``$\times0.5$'' in Figure~\ref{fig:hyperparameters}, are essentially identical to the one in the benchmark configuration. Employing values of $G_W$ that are loosely close to the reach thus ensures good performances and no precise optimization is needed. We also tried a simpler training scheme that employs only 2 distinct equal and opposite values of $G_W$ instead of 6 values as in the benchmark configuration. The results for $G_W = \pm  0.1\;\textrm{TeV}^{-2}$ and $G_W = \pm  0.01\;\textrm{TeV}^{-2}$ are reported in the figure. Two results are reported for each value by employing different random seeds. A very mild degradation of the performances is observed, showing that employing more than 2 values of $G_W$ for training as in the benchmark configuration is beneficial, but only marginally. We also tried to use $G_W = \pm  0.001\;\textrm{TeV}^{-2}$ and $G_W = \pm  1\;\textrm{TeV}^{-2}$, much below and much above the reach, respectively. In the first case, training was found to proceed too slowly, which was expected because the gradients of the loss function become small if the training points are too close to the SM $G_W=0$ point. In the second case, strong overfitting was observed, probably due to the fact that if $G_W$ is overly large the loss function is dominated by the high-energy tail of the training data points, where the EFT contribution from the quadratic terms dominates over the SM. In summary, a good stability of the performances is observed under variations of the values of the Wilson coefficients used for training, provided they are of the order of the expected reach and not vastly above or below.

These results should be contrasted with the findings of Ref.~\cite{Chen:2020mev}, where the quadratic classifier was trained on independent Monte Carlo samples generated at each point in $\cal{C}$ without using reweighting. In that setup, the training points are a major factor controlling the performances. They must be chosen by balancing two competing criteria. First, the values of the Wilson coefficients used for training must be large enough to modify the SM distribution considerably and typically well above the sensitivity reach. Otherwise, their effects would be too small to be seen by comparing independent Monte Carlo data sets. We explained in the Introduction and in Section~\ref{subsec:sc}---and we verified in Section~\ref{sec:scper}---that small departures from the SM are on the contrary effectively reconstructed if reweighted data are used for training. Therefore, we do not need to employ large values of the Wilson coefficients for training. The second criterion for the selection of the Wilson coefficients in the setup of Ref.~\cite{Chen:2020mev} is to avoid overly large values. Otherwise, the contribution to the cross section would be dominated by the quadratic term. The effect of the linear term would be a small correction and could not be learned. Reweighted training is sensitive to small effects, therefore also this second criterion is less relevant to our methodology. The optimal Wilson coefficient that obeys the above-mentioned criteria depends on the phase space region because the size of the EFT contribution to the cross section scales, in particular, with the energy of the process. The optimal Wilson coefficients in three $p_{T}$ regions were identified in Ref.~\cite{Chen:2020mev} and used for training producing good performances. However, the need of optimizing the choice of the training points is an obstruction to the systematic deployment of the methodology of Ref.~\cite{Chen:2020mev}. This optimization is not required because the performances are weakly sensitive to the training points with the methodology of the present paper.

\subsection{Results at NLO}\label{sec:nlod}

We finally describe our results on NLO simulations of the ZW process. Reweighted Monte Carlo samples are generated using the {\sc{MadGraph}} software suite, as detailed in Section~\ref{sec:nlogen}. Like in the ideal case, 3M points are used for training and 3M independent points are used to compute the validation loss. Only 3M points rather than 10M are employed for testing. Ten independent observable variables $x$ characterize the NLO events, listed in eq.~(\ref{featNLO}). These are pre-processed with the transformation
\begin{eqnarray}\label{eq:truefNLO}
x\;\to\; &\bigg\{&\hspace{-8pt} \log[s/{\textrm{GeV}}^2],\,\Theta,\, 
\theta_Z,\,
\theta_W,\, 
\log[p_T/{\textrm{GeV}}],\,
\nonumber\\
&\ &\hspace{-7pt}
\log[1 + p_{T,{\textrm{ZW}}}/{\textrm{GeV}}],\,
Q,\,
\ell_Z,\,
\ell_W\,
\sin\varphi_Z,\,
\sin\varphi_W,\,
\cos\varphi_Z,\,
\cos\varphi_W
\bigg\}\,,
\end{eqnarray}
and normalized to zero mean and unit variance on the training data set by a normalization layer in the neural networks. The discrete labels $\ell_Z$ and $\ell_W$ that describe the flavor of the leptons from the Z and the W boson decays assume the numerical values of 1 or 0 for muons and electrons, respectively.

\begin{figure}[t]
    \centering
    \includegraphics[width=0.45\columnwidth]{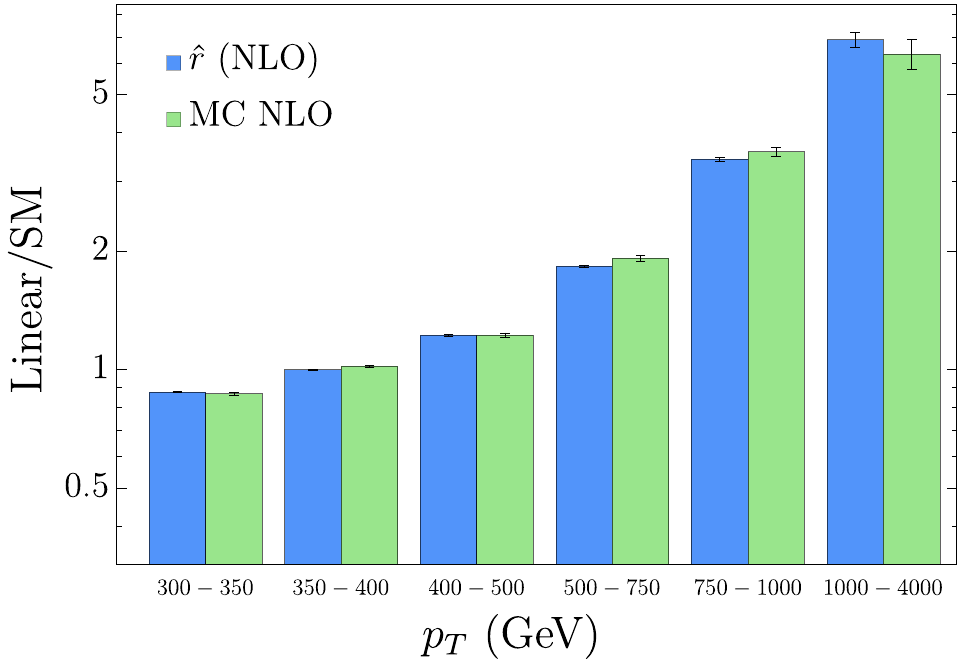}
    \hfill  \includegraphics[width=0.45\columnwidth]{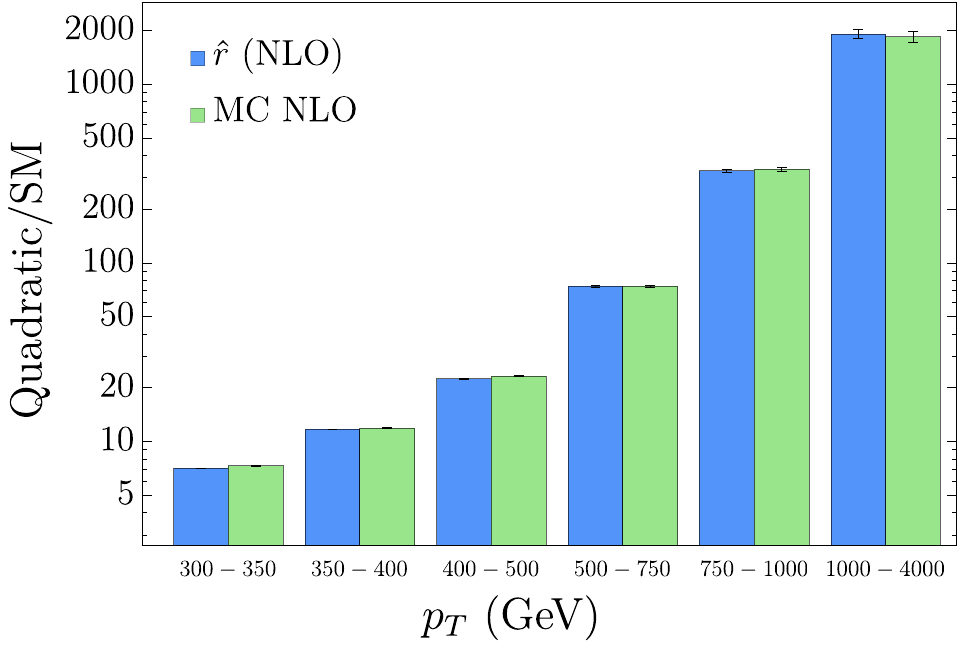} 
    \caption{
    Linear (left) and quadratic (right) contributions of the ${\cal O}_W$ operator to the cross section in $p_{T}$ bins, using NLO simulations. The direct determination from Monte Carlo data, by eq.~(\ref{eq:ds}) is shown in green. The determination from the reconstructed ratio (see~(\ref{eq:recsig})) is shown in blue. Cuts $|\cos \varphi_{W/Z}| > 0.6$ and $p_{T,{\textrm{ZW}}}/p_T < 0.5$ are applied.} \label{fig:nlorec}
\end{figure}

The implementation of the quadratic classifier method proceeds as explained in Section~\ref{sec:qcper} and no significant difference is observed---neither in the behavior of the training process nor in the performances---associated with the usage of NLO rather than ideal Monte Carlo data. This is interesting because NLO Monte Carlo data sets contain events with negative weights. These events give a negative contribution to the loss function that might encourage overfitting, raising a concern about the possibility of applying our methodology to realistic NLO simulation. Our methodology is instead found to behave on NLO data as well as on ideal data, where no negative weights are present. It should be noted that currently available NLO generators like {\sc{MadGraph}} produce a small fraction of negative-weights events: they are $3\%$ of the total in the Monte Carlo data sets under consideration. A larger fraction of negative weights that might be obtained with a non-optimized generator, or in a different process, could in principle invalidate this conclusion.

\begin{figure}[t]
    \centering
    \includegraphics[width=0.45\columnwidth]{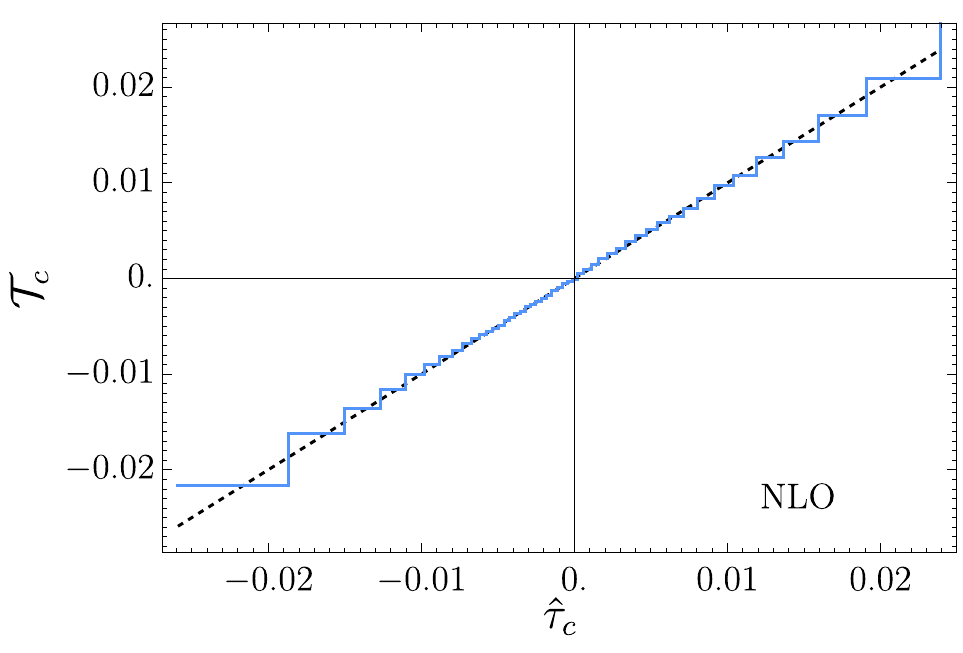}
    \hfill
    \includegraphics[width=0.46\columnwidth]{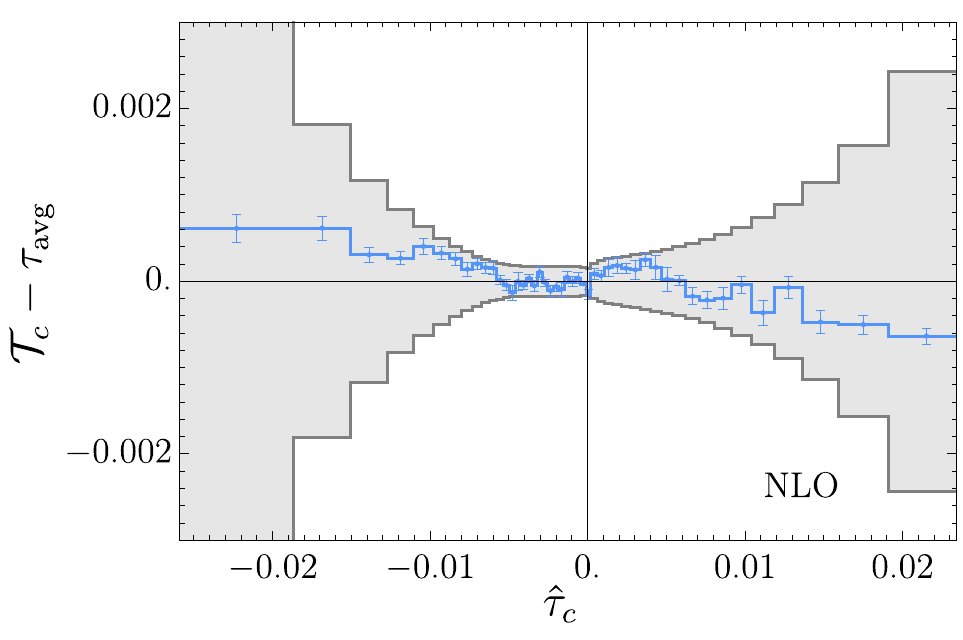}
    \caption{The $\Tau_c$ quantity of eq.~(\ref{eq:rcq}), for the reconstructed $\hat{r}$ ratio obtained with the quadratic classifier, with NLO data. The point $G_W = 7 \cdot 10^{-3}\;\textrm{TeV}^{-2}$ and $G_\varphi=0$ is considered. The right panel illustrates the approximate validity of the bounds in eq.~(\ref{eq:taubounds}), as explained in the main text.} \label{fig:nloscaletta}
\end{figure}

Since no strong difference is found in comparison with the ideal results presented in the previous section, we illustrate our findings directly on the full two-dimensional setup where the distribution ratio---parametrized as in eq.~(\ref{eq:qcpc})---is learned with a single training as a function of the two Wilson coefficients $G_\varphi$ and $G_W$. All the hyper-parameters are set to the benchmark values we used for ideal data in the previous section, apart from the architecture of the neural networks that is now $(13,32,32,1)$ to account for the larger dimensionality of the NLO input features. The performances of the trained model are shown in Figures~\ref{fig:nlorec} and~\ref{fig:nloscaletta}.

Figure~\ref{fig:nlorec} displays the ability of the trained model to reproduce the linear and quadratic contributions of the ${\mathcal{O}}_W$ operator to the cross section in $p_{T}$ bins and with cuts $|\cos \varphi_{W/Z}| > 0.6$ on the azimuthal decay angles and $p_{T,{\textrm{ZW}}}/p_T < 0.5$ on the total transverse momentum. These cuts emphasize the leading growing-with-energy contribution of the ${\mathcal{O}}_W$ operator. The same plot is shown in Figure~\ref{fig:idealrec} for ideal data and the agreement between the reconstructed (blue) and the Monte Carlo (green) predictions is at the same level as in Figure~\ref{fig:idealrec}. Figure~\ref{fig:idealrec} also shows the results obtained using the knowledge of the true distribution ratio~$r$, which is available for ideal simulations and not available for NLO simulations. Figure~\ref{fig:nloscaletta} displays the approximate linear behavior of the $\Tau_c$ variable and the validity of the bounds in eq.~(\ref{eq:taubounds}), at the point $c$ in the parameter space where $G_W = 7 \cdot 10^{-3}\;\textrm{TeV}^{-2}$ and $G_\varphi=0$. The plot is obtained in the same way as Figure~\ref{fig:scaletta} for ideal data. The results confirm that a good reconstruction of the distribution ratio is obtained also at NLO.

\begin{figure}
    \centering
    \begin{minipage}[b]{.47\textwidth}
    \includegraphics[width=1\columnwidth]{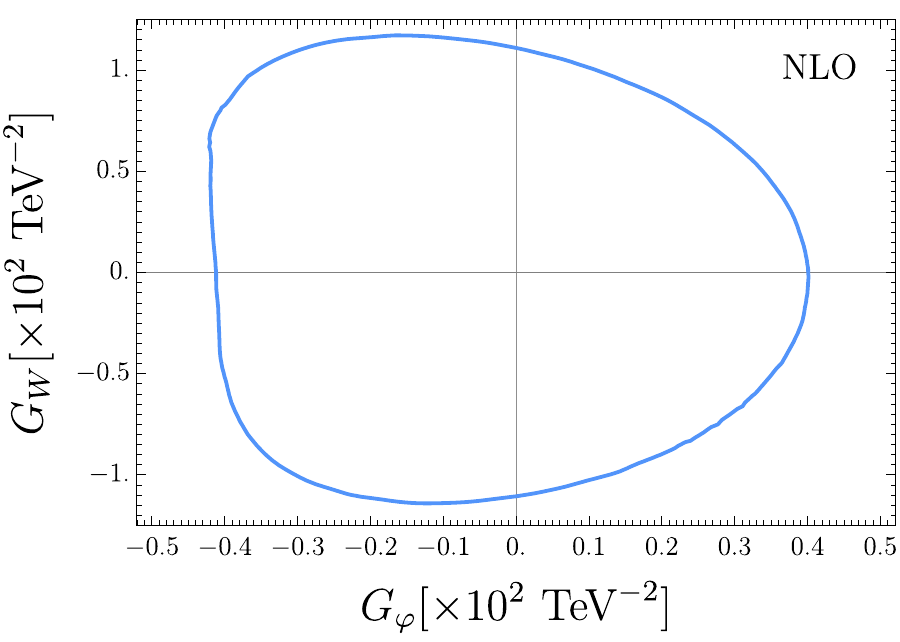}
    \end{minipage}
    \hfill
    \begin{minipage}[b]{.47\textwidth}
    \small
    \begin{tabular}[b]{c|c}
    \rule[-.5em]{0pt}{1.75em}$G_{W}$ & $[-1.106(6), 1.108(7)]$\\
    \midrule
    \rule[-.5em]{0pt}{1.75em}$G_{\varphi}$ & $[-0.412(1), 0.401(1)]$\\
    \midrule
    \rule[-.5em]{0pt}{1.75em}$G_{W}=G_{\varphi}$ & $[-0.405(1), 0.366(1)]$\\
    \midrule
    \rule[-.5em]{0pt}{1.75em}$G_{W}=-G_{\varphi}$ & $[-0.376(1), 0.419(1)]$
    \end{tabular}
    \vspace{4em}
    \end{minipage}
    \hspace{-2.25em}
    \caption{LEFT: Contour lines $\hat{p}(c)=5\%$ in the two-dimensional parameter space, for NLO data. RIGHT: Reach at NLO on $G_{\varphi}$ and $G_{W}$ in the 4 directions of the 2D plane. The bounds are given in units of $10^{-2}$ TeV$^{-2}$} \label{fig:nlocontour}
\end{figure}

Finally, the left panel of Figure~\ref{fig:nlocontour} displays the $5\%$ contours of the median $p$-value $\hat{p}(c)$. For its determination, we proceeded as explained in Section~\ref{sec:nppcalc} and in the previous section. The right panel of the figure summarized the reach along four directions in the parameter space.

\section{A proposal for exclusions}\label{sec:PLS}

We explained in the Introduction and in Section~\ref{sec:nppdef} that our methodology is not tied to the specific task of the statistical analysis one eventually aims at performing on the experimental data, nor to the specific (frequentist of Bayesian) methodology that will be adopted for the analysis. Our ultimate goal is to extract a good approximation of the distribution ratio~(\ref{eq:r}), to be later used for any analysis aimed at approaching optimality exploiting the accurate knowledge of the ratio and in turn of the likelihood. On the other hand, it is worth discussing and addressing the challenges in the design and in the practical implementation of one such nearly-optimal analysis. We focus in particular on the design of a classical test of hypothesis aimed at excluding the existence of EFT operators with a given pre-specified value $c\neq0$ for their Wilson coefficients (the $H_0$ hypothesis) in favor of the SM value $c=0$ (the $H_1$ hypothesis). A Bayesian approach to the same problem is described in Section~\ref{se:bf}.

As explained in Section~\ref{sec:nppdef}, the Neyman--Pearson Lemma guarantees the optimality of the test based on the likelihood ratio test statistic $t_c(\data)$ defined in eq.~(\ref{eq:tc}). Since we are able to reconstruct the likelihood log-ratio, a natural option would be to employ the reconstructed test statistic $\hat{t}_c(\data)$ defined in eq.~(\ref{eq:tchat}) for the actual statistical analysis on the observed data. The test statistic depends on the variable $\hat\tau_c(x)$, which is the logarithm of the reconstructed distribution ratio~(\ref{eq:tth}). If the reconstruction is accurate, we are guaranteed that this test attains optimal performances, as we also verified in Section~\ref{sec:PS}. This theoretical guarantee of optimality is precisely the reason why the median $p$-value of the test based on $\hat{t}_c(\data)$ is used for a robust assessment of the quality of the distribution ratio reconstruction. On the other hand, it would be difficult to apply this test to real data. In this section, we define an alternative test that could be more viable in practice and with expected performances that are not far from optimality.

The main factor that limits the applicability of the $\hat{t}_c(\data)$-based test---or of the test based on the exact ${t}_c(\data)$, if available---is that the distribution of the test statistic variable is in general not known. On the other hand, we need the distribution in the null $c\neq0$ hypothesis in order to obtain the $p$-value. We also need the distribution in the alternative $c=0$ hypothesis, if we want to quantify the expected sensitivity of the test. The determination of the distributions has to be performed numerically, and this requires generating many toy instances of the $\data$ data set, which is numerically expensive. In Section~\ref{sec:nppcalc} we addressed this issue by discretizing the $\hat\tau_c(x)$ variable in a set of bins. This enabled us---in eq.~(\ref{eq:tchatbinned})---to express $\hat{t}_c(\data)$ in terms of the counts $\N_{\textrm{b}}$ in each bin, which are Poisson-distributed and efficient to sample for the generation of toys. Inspired by that result, we consider the possibility of defining an analysis strategy based on binning the $\hat\tau_c(x)$ variable.

\subsection{The optimal binned test}

On general grounds, a \emph{binned} analysis is one that divides the $x$ space in a set of non-overlapping bins $\textrm{b}\in\textrm{B}$ and computes the number of points in $\data$, $\N_{\textrm{b}}$, that fall in each bin. The counts, $\data_{\textrm{B}}=\{\N_{\textrm{b}},\forall\,\textrm{b}\in\textrm{B}\}$, are the aggregate data used for statistical inference, in place of the original data set $\data$. Importantly enough, the $\N_{\textrm{b}}$ variables are independent Poisson distributions with an expected, $N_{\textrm{b}}(c)=L\,\Delta\sigma_{\textrm{b}}$, that can be accurately calculated with the Monte Carlo, at least if the number of bins is not extremely large. Therefore, we have complete analytical access to the distribution of the binned data. On the contrary, an analysis based on variables such as the reconstructed~(\ref{eq:tchat}) or the exact~(\ref{eq:tc}) log-ratio, which depend directly on the observables vector $x\in\data$, is called \emph{unbinned}. The fact that we do not have access to the $x$ data distribution makes unbinned analyses generically more difficult. Binned analyses are easier and way more common in LHC data analysis practice.

The analytical knowledge of the distribution of the binned data $\data_{\textrm{B}}$ enables the deployment of the optimal Neyman--Pearson strategy for hypothesis testing. This test would be based on the negative of the log-ratio between the binned likelihoods in the $c\neq0$ and in the $c=0$ hypotheses. Specifically, the optimal test statistic for binned data would be
\beq\label{eq:blr}
-2\,\log \frac{\mathfrak{L}(c;\data_{\textrm{B}})}{\mathfrak{L}(0;\data_{\textrm{B}})}
=2\sum\limits_{{\textrm{b}}\in{\textrm{B}}}
\left[
N_{\textrm{b}}(c)-N_{\textrm{b}}(0)-
\N_{\textrm{b}}
\log\frac{N_{\textrm{b}}(c)}{N_{\textrm{b}}(0)}
\right]
\,.
\eeq
One should be careful here with the notion of optimality. The guarantee of optimality of the Neyman--Pearson Lemma is obviously tied to the nature of the data upon which the statistical inference is based. The complete data set $\data$ contains all the information that is collected by the experiment, therefore the (unbinned) likelihood ratio associated with the $\data$ data is the truly optimal test statistic. The binned likelihood ratio in eq.~(\ref{eq:blr}) would be the optimal variable if the counts $\data_{\textrm{B}}$ were the only available experimental information, but its performances are in general sub-optimal because binning reduces information. 

Consider however defining the bins in terms of the variable $\tau_c(x)$, which is the logarithm of the true distribution ratio. Namely, the bins are defined by the conditions $\tau_c(x) \in (\tau_{\textrm{b}-1},\tau_{\textrm{b}})$, with $\textrm{b}=1,\ldots,n_{\textrm{bin}}$, $\tau_{0}=-\infty$ and $\tau_{n_{\textrm{bin}}}=+\infty$ in order to cover the whole $x$ space. In this case, the binned likelihood log-ratio~(\ref{eq:blr}) is easily shown to coincide, when the bins are infinitely narrow, with the unbinned likelihood ratio $t_c(\data)$ defined in eq.~(\ref{eq:tc}). In fact, if the bins are very narrow, the $\tau_c(x)$ variable is well-approximated over the entire $x$ space by the following piecewise  constant function
\beq\label{eq:tauapp}
{\tau}_c(x)\;\simeq\;\left\{
\log\frac{\Delta\sigma_{\textrm{b}}(c)}{\Delta\sigma_{\textrm{b}}(0)}\,,\;{\text{for}}\;
x\;\;{\textrm{s.t.}}\;{\tau}_c(x)\in (\tau_{\textrm{b}-1},\,\tau_{\textrm{b}})
\right\}\,,
\eeq
where $\Delta\sigma_{\textrm{b}}(c)$ is the cross section in the bin. This approximation holds because $\tau_c(x)$ is the logarithm of the distribution ratio and thus it can be computed as the logarithm of the ratio of the cross sections integrated in a narrow bin.\footnote{The first and the last bins extend up to infinity and hence are not narrow. Eq.~(\ref{eq:tauapp}) still holds if $\tau_c$ is bounded. If it is not,  extreme tails of the $\tau_c$ distribution can be excluded from the analysis of the data without sensitivity loss because they are not populated by the finite data statistics.} Notice that the value of $\tau_c$ in the bin could have also been approximated by the center of the bin, which in fact coincides with the log cross section ratio up to finite bin effects as we proved in eq.~(\ref{eq:bder}). Inserting eq.~(\ref{eq:tauapp}) into the definition (\ref{eq:tc}) of $t_c(\data)$, we find
\beq
t_c(\data)\simeq 2\,\bigg[
N(c)-N(0)-
\sum\limits_{{\textrm{b}}\in{\textrm{B}}}\N_{\textrm{b}}
\log\frac{N_{\textrm{b}}(c)}{N_{\textrm{b}}(0)}
\bigg]\,.
\eeq
This expression coincides with eq.~(\ref{eq:blr}) since $\sum_{\textrm{b}}N_{\textrm{b}}=N$. Therefore, the binned likelihood ratio~(\ref{eq:blr}) approaches the optimal test statistics, and hence it is nearly-optimal, if the binning is performed on the $\tau_c(x)$ variable and if a sufficiently narrow binning is employed. Clearly, we do not have access to the exact $\tau_c(x)$ variable. The reconstructed $\hat\tau_c(x)$ variable will be used instead to define the bins. Optimality will thus be attained only if the reconstruction is accurate.

The development of methodologies aimed, like the one of the presented paper, at reconstructing the distribution log-ratio $\tau_c(x)$ is often motivated by the need of accessing the unbinned log-likelihood ratio for optimal statistical inference. The previous results however show that binned data contain, up to finite-binning effects, the exact same amount of information as the original data set $\data$. In particular, we have seen that the optimal log-likelihood ratio test on binned data has the same sensitivity as the optimal unbinned test. An unbinned approach is thus not essential to attain optimality. The reconstruction of $\tau_c(x)$ is instead essential because it is crucial that the binning is performed on the $\tau_c(x)$ variable or a good approximation thereof. Binning any other variable would lead to sub-optimal performances even for infinitely narrow binning, barring the possibility of binning all the individual components of the data vector $x$, which is not feasible in more than two or three dimensions. The $\tau_c(x)$ variable needs to be reconstructed because it brings the whole information about the parameters of interest that is available in the data. It can be used for an unbinned analysis, or for a binned analysis like we propose.

\subsection{Maximum likelihood and Asymptotic formulas}

The implementation of the optimal binned hypothesis test defined by eq.~(\ref{eq:blr}) still requires drawing toys to determine the test statistic distributions empirically. Poisson-distributed toys are numerically cheap to generate at one fixed point $c$ in the parameter space, but repeating the toys generation and the determination of the distribution point-by-point in the parameter space can easily become unfeasible if the number of parameters is large. The relevant parameters here are not only the parameters of interest $c$. Imperfections in the Monte Carlo predictions due to the approximate knowledge of the underlying physical model, or of the modeling of the detector response, need to be incorporated in the analysis in the form of nuisance parameters, denoted as $\nu$. Determining the dependence of the likelihood on the nuisance parameters is rather standard practice in LHC data analysis. It requires estimating the dependence of the number of expected events in each bin, $N_{\textrm{b}}(c,\nu)$, not only on the parameters of interest but also on the nuisance parameters. This determination requires dedicated Monte Carlo events, and it can be demanding to achieve but is feasible. It is instead not feasible to scan this large parameter space for the empirical determination of the test statistic distribution, which in general depends on the parameters as the test statistic variable does. Practical LHC data analysis thus needs to rely on analytical approximations of the distributions. These approximations are typically valid in the large-sample---or, Asymptotic---limit. Therefore, they are called Asymptotic formulas.

\begin{figure}
    \centering
    \includegraphics[width=0.25\columnwidth]{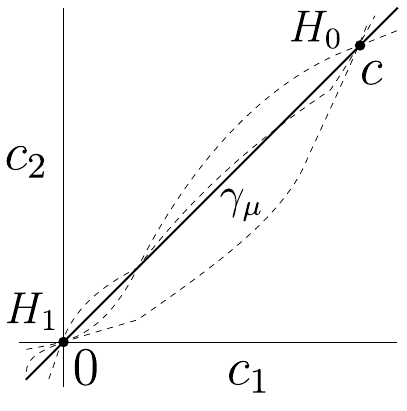}
    \caption{Schematic representation of the interpolation between the $H_0$ hypothesis (the point $c$ in the parameter space) and the $H_1$ hypothesis (the origin, $c=0$). }
    \label{fig:gamma}
\end{figure}

No Asymptotic formula exists to model the distribution of the likelihood log-ratio test statistic~(\ref{eq:blr}). We thus need to employ a different variable, which in turn defines a different (and sub-optimal) hypothesis test. Following the standard LHC data analysis practice~\cite{Cowan:2010js}, we use the \emph{maximum likelihood ratio} test statistics, which is defined as follows. Consider an open curve that interpolates from the SM point $c=0$ to the point $c\neq0$ that we are interested in testing, in the space of the parameters of interest. For definiteness, we choose a straight line but other options would be also viable as depicted in Figure~\ref{fig:gamma}. Be $\gamma_\mu$ an explicit parametrization of the curve, where $\mu\in\mathbb{R}$ would correspond to the ``signal strength'' parameter in the notation of Ref.~\cite{Cowan:2010js}. The curve $\gamma_\mu$ defines a one-parameter family of statistical hypotheses for the distribution of the data, within the larger space span by the Wilson coefficients $c$, which interpolates between $H_0$ and $H_1$. The maximum likelihood test statistics is $-2$ times the logarithm of the ratio between the likelihood at $c$ and the maximum of the likelihood along the $\gamma_\mu$ curve, namely
\beq\label{eq:mlr}
t_c^{\textrm{\sc{ml}}}(\data_{\textrm{B}})=
-2\,\log \frac{\mathfrak{L}(c;\data_{\textrm{B}})}{\mathfrak{L}(\gamma_{\widehat\mu};\data_{\textrm{B}})}
=2\,\sum\limits_{{\textrm{b}}\in{\textrm{B}}}
\left[
N_{\textrm{b}}(c)-N_{\textrm{b}}(\gamma_{\widehat\mu})-\N_{\textrm{b}}\log\frac{N_{\textrm{b}}(c)}{N_{\textrm{b}}(\gamma_{\widehat\mu})}
\right]
\,.
\eeq
The expression is similar to the likelihood log-ratio test statistics~(\ref{eq:blr}). The difference is that the likelihood in the denominator is not computed at the SM nor at any other pre-specified point of the parameter space. It is computed at the point $c=\gamma_{\widehat\mu}$ that maximizes the likelihood along the curve for the specific instance, $\data_{\textrm{B}}$, of the binned data set under examination. The maximal point on the curve, located at $\mu=\widehat\mu$, is thus an implicit function of the data: $\widehat\mu=\widehat\mu(\data_{\textrm{B}})$. 

Unlike the likelihood log-ratio~(\ref{eq:blr}), the maximum likelihood test statistics~(\ref{eq:mlr}) is endowed with Asymptotic formulas. Under the null hypothesis $c\neq0$, it approaches a $\chi^2$ with one degree of freedom in the Asymptotic limit. Under the alternative hypothesis $c=0$ it approaches a non-central $\chi^2$ with one degree of freedom. These formulas allow us to implement the test of hypothesis based on the maximum likelihood ratio test statistic without having to resort to toy data. The $p$-value associated to a given observed value $t$ of the $t_c^{\textrm{\sc{ml}}}(\data_{\textrm{B}})$ test statistic, defined as in eq.~(\ref{eq:pval0}), is simply given by
\beq\label{eq:pvas}
p_c(t)=p[t]=1-{\textrm{CDF}}_{\chi^2_{1}}(t)\,,
\eeq
where CDF denotes the cumulative distribution function. Notice that the association between the value of the test statistic and the $p$-value, namely the $p_c(t)$ function, depends on $c$ in general because the distribution of the test statistics does depend on $c$. The function is instead independent of $c$ in this particular case because the Asymptotic formula for the distribution is a $\chi^2$ independently of $c$. 

We can also use approximations to compute the expected $p$-value under the SM hypothesis $c=0$. Several methods were considered in Ref.~\cite{Cowan:2010js}. The one that is most commonly employed is based on the so-called ``Asimov'' data set, $\data_{\textrm{B}}^{\cal{A}}$. The Asimov data set is one where all the observables are exactly equal to their expected value. In the case at hand
\beq\label{eq:asds}
\data_{\textrm{B}}^{\cal{A}}=\{N_{\textrm{b}}(0),\forall\,\textrm{b}\in\textrm{B}\}\,,
\eeq
where $N_{\textrm{b}}(0)$ is the expected number of events in each bin as predicted by the SM $c=0$ hypothesis. Ref.~\cite{Cowan:2010js} proposed to estimate the median of the $t_c^{\textrm{\sc{ml}}}$ variable as the value it assumes on the Asimov data set. Consequently, the median $p$-value can be estimated as 
\beq\label{eq:pvasy}
p_{\textrm{med}}^{\textrm{\sc{ml}}}(c)=p\left[
t_c^{\textrm{\sc{ml}}}(\data_{\textrm{B}}^{\cal{A}})
\right]\,.
\eeq

A few comments are in order. First, the validity of the Asymptotic formulas relies on the availability of a large data statistics. Giving for granted that the total expected number of events is large, for the validity of the Asymptotic formulas we still have to ensure that the number of expected events in each bin is also large. Experience reveals that $10$ events are sufficient for the Asymptotic formulas to hold accurately.~\footnote{The validity of the Asymptotic formulas can be directly related, in the case at hand, to the accuracy of the Gaussian approximation for the Poisson distribution, which is good for more than 5 or 10 expected events.} This limits the maximum number of bins that can be employed in the analysis, entailing some loss of sensitivity because, as we have seen, the data binned in the $\hat\tau_c$ variable contain the same amount of information as the original data only in the limit of infinitely narrow binning. However, in practice, we do not expect this to be the main source of degradation in comparison with the optimal test sensitivity. In fact, it should be stressed that the one based on maximum likelihood is a different test of hypothesis than the one based on the likelihood ratio test statistic~(\ref{eq:blr}), and hence it is intrinsically sub-optimal. Extensive experience in statistical practice ensures that its performances are typically not far from optimal, but as far as we know this fact is not guaranteed by any rigorous result. Whenever possible, it is useful to compare its sensitivity with the one of the optimal test for a few representative points of the parameter space in order to get an idea of the degradation.

The role played by the curve $\gamma_\mu$ along which the maximum of the likelihood is computed also deserves to be described in detail. First, we notice that the shape of the curve plays no role in sensitivity projections based on the Asymptotic formulas and on the Asimov trick. Namely, the median expected $p$-value in eq.~(\ref{eq:pvasy}) is independent of the shape of the curve. The $p(t)$ function is universally provided by the cumulative of the $\chi^2$, and obviously independent of the $\gamma_\mu$ curve. The value of the test statistic on the Asimov data set is also independent of the curve, because of the following. On the Asimov data set, the logarithm of the likelihood along the $\gamma_\mu$ curve is
\beq
\log\mathfrak{L}(\gamma_{\mu};\data_{\textrm{B}}^{\cal{A}})=
\sum\limits_{{\textrm{b}}\in{\textrm{B}}}
\left[
-N_{\textrm{b}}(\gamma_\mu)+N_{\textrm{b}}(0)\log N_{\textrm{b}}(\gamma_\mu)\right]+\;{\textrm{const.}}\,,
\eeq
and attains its absolute maximum, in each bin, when $N_{\textrm{b}}(\gamma_\mu)=N_{\textrm{b}}(c)$. The point on the curve that maximizes the likelihood is thus the SM point, $\gamma_{\widehat\mu}=0$. Evaluating the maximum likelihood test statistic~(\ref{eq:mlr}) on the Asimov data set thus gives
\beq\label{eq:tasi}
t_c^{\textrm{\sc{ml}}}(\data_{\textrm{B}}^{\cal{A}})
=-2\,\log \frac{\mathfrak{L}(c;\data_{\textrm{B}}^{\cal{A}})}{\mathfrak{L}(0;\data_{\textrm{B}}^{\cal{A}})}
=2\,\sum\limits_{{\textrm{b}}\in{\textrm{B}}}
\left[
N_{\textrm{b}}(c)-N_{\textrm{b}}(0)-N_{\textrm{b}}(0)\log\frac{N_{\textrm{b}}(c)}{N_{\textrm{b}}(0)}
\right]
\,,
\eeq
regardless of the shape of the $\gamma_\mu$ curve. 

The shape of the curve does affect, on the contrary, the determination of $t_c^{\textrm{\sc{ml}}}$ on the real data, and in turn it affects the $p$-value that will be obtained in the analysis of the data collected by the experiment. Each curve formally defines a different test of hypothesis. Since all these tests have the same expected sensitivity as previously discussed, they are all equally valid and each of them could be applied to the data. The most sensible line of action would be to select, prior to the experiment, a simple curve like a straight line and stick to that choice. Otherwise, one could combine the $p$-values obtained with different curves easily, because all curves have the same expected sensitivity.

One might wonder why considering a curve, rather than interpolating between $c$ and the origin with a higher-dimensional surface. A seemingly natural choice would be to employ in fact the whole $c$ space for the interpolation. In the present paper we are dealing with two Wilson coefficients: one might have considered defining the maximum likelihood ratio test statistic~(\ref{eq:mlr}) by minimizing over the whole Wilson coefficient plane rather than on the curve $\gamma_\mu$. Asymptotic formulas are available for the test statistic distribution even if the maximization is performed on a multidimensional family of hypotheses. They involve $\chi^2$ distributions with a number of degrees of freedom that is equal to the dimensionality of the family. In particular, a central $\chi^2$ with two degrees of freedom would have been obtained for the distribution of the test statistic in the null $c\neq0$ hypothesis, if the maximization was performed on the entire plane. In eq.~(\ref{eq:pvas}) we would thus have encountered the cumulative of the $\chi^2$ with two rather than one degree of freedom. On the contrary, the determination of the median expected value of the test statistics, estimated by the Asimov trick, would have remained the same as in eq.~(\ref{eq:tasi}). The $\chi^2$ with two degrees of freedom is broader than the $\chi^2$ with one degree of freedom. Therefore, the median expected $p$-value in eq.~(\ref{eq:pvasy}) would be higher if the maximization was performed on the two variables rather than on the one-dimensional curve. We thus discard this possibility because it would produce a test that is less performant and farther from optimality.

Finally, we notice that the inclusion of nuisance parameters in the maximum likelihood ratio framework is conceptually straightforward, following Ref.~\cite{Cowan:2010js}. It merely amounts to maximizing over the nuisance parameters, independently, the likelihoods in the numerator and in the denominator of eq.~(\ref{eq:mlr}). The Asymptotic $\chi^2$ formulas still hold. On the Asimov data set---defined at the central value of the nuisance parameters $\nu=0$, namely $\N_{\textrm{b}}=N_{\textrm{b}}(0,0)$---the likelihood is still maximal at the SM point $c=\gamma_{\widehat\mu}=0$ and with central-value nuisance. The maximum likelihood-ratio test statistics on the Asimov data is similar to eq.~(\ref{eq:tasi}), though it involves a potentially costly minimization over the nuisance parameters. Namely, in the presence of nuisance parameters $\nu$, eq.~(\ref{eq:tasi}) becomes
\beq\label{eq:tasinu}
t_c^{\textrm{\sc{ml}}}(\data_{\textrm{B}}^{\cal{A}})
=2\min\limits_\nu\left\{\,\sum\limits_{{\textrm{b}}\in{\textrm{B}}}
\left[
N_{\textrm{b}}(c,\nu)-N_{\textrm{b}}(0,0)-N_{\textrm{b}}(0,0)\log\frac{N_{\textrm{b}}(c,\nu)}{N_{\textrm{b}}(0,0)}
-\log\frac{\mathfrak{L}(\nu)}{\mathfrak{L}(0)}
\right]\right\}
\,.
\eeq
In the equation, $\mathfrak{L}(\nu)$ is the likelihood for the nuisance parameters as determined from measurements that are independent of the data under examination. This likelihood is often simply approximated with a multivariate Gaussian centered on the central-value nuisance configuration $\nu=0$.  
The minimization to be performed in eq.~(\ref{eq:tasinu}) and other elements that are peculiar to our proposal could complicate its practical implementation in the presence of many nuisance parameters. This is discussed in the following section.

\subsection{Implementation}

The implementation of our proposal requires, in the first place, to select a suitable set of points ${\tau}_{\textrm{b}}$  on the real axis for binning the ${\hat\tau}_c(x)$ variables. For a given chosen number of bins, $n_{\textrm{bin}}$, these points are selected in such a way that all resulting bins have an equal cross section in the SM hypothesis, namely $\Delta\sigma_{\textrm{b}}(0)=\sigma(0)/n_{\textrm{bin}}$. The points are determined by evaluating ${\hat\tau}_c(x)$ on the Monte Carlo data set, sorting the list and scanning through it cumulating the SM weights $w_{\textrm{e}}(0)$ until when integer multiples of the target cross section $\sigma(0)/n_{\textrm{bin}}$ are reached. We use $n_{\textrm{bin}}=300$ because this produces---both with the ideal and with the NLO predictions---around 10 expected events in each bin, ensuring the validity of the Asymptotic formulas. Once the binning points ${\tau}_{\textrm{b}}$ are determined, the cross sections and the expected number of events for non-vanishing $c$ are determined by summing the $w_{\textrm{e}}(c)$ weights of the events that fall in each bin. Finally, the Asimov estimator, $t_c^{\textrm{\sc{ml}}}(\data_{\textrm{B}}^{\cal{A}})$, is computed by eq.~(\ref{eq:tasi}). The median $p$-value $p_{\textrm{med}}^{\textrm{\sc{ml}}}$ is given by eq.~(\ref{eq:pvasy}).

\begin{figure}[t]
    \centering
    \includegraphics[width=0.45\columnwidth]{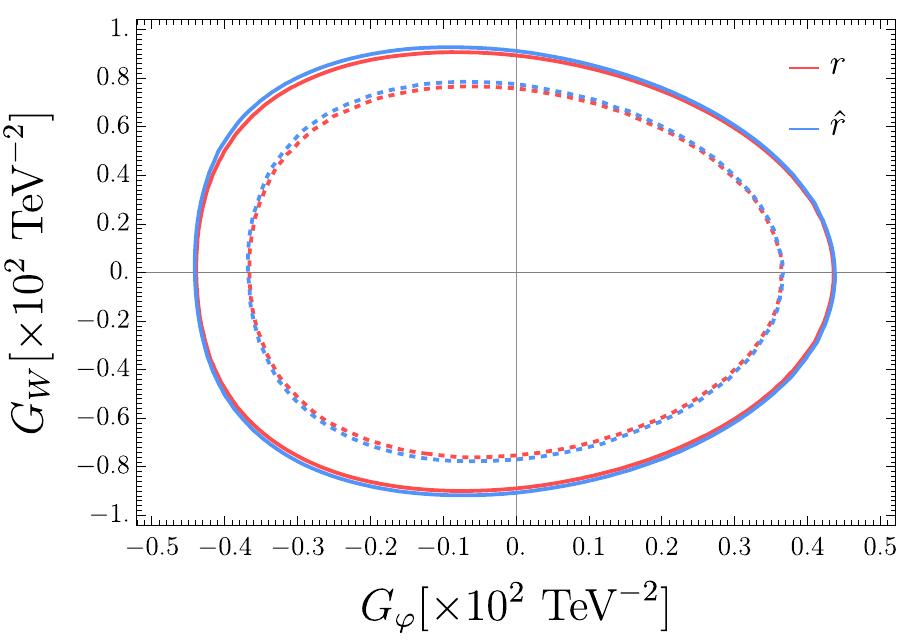}
    \hfill
    \includegraphics[width=0.45\columnwidth]{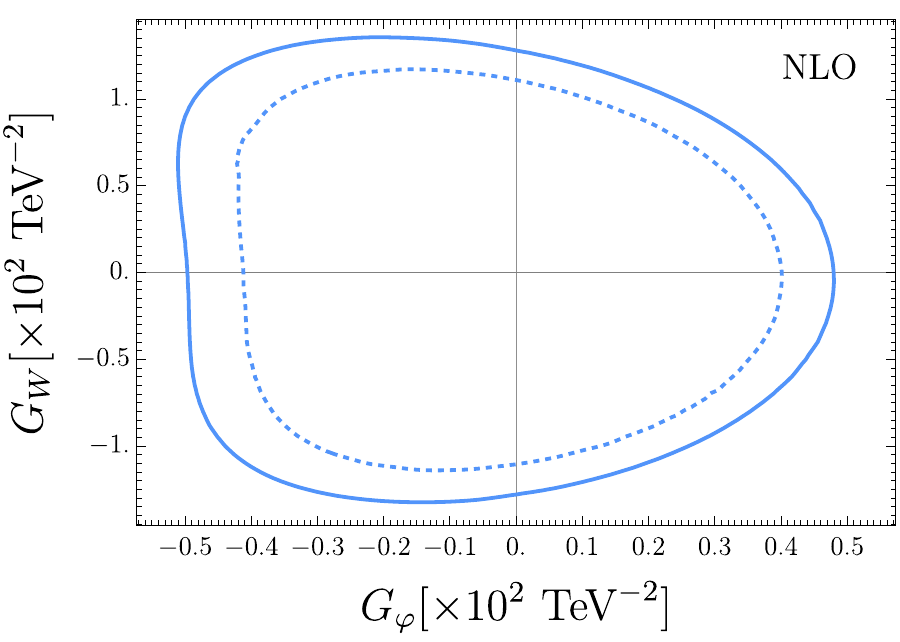}
    \caption{Contours of $p_{\textrm{med}}^{\textrm{\sc{ml}}}(c)=5\%$ (i.e., $t_c^{\textrm{\sc{ml}}}(\data_{\textrm{B}}^{\cal{A}})=3.84$) in the Wilson coefficient plane $c=(G_{\varphi},\, G_{W})$ for ideal (left panel) and NLO (right panel) simulations. The contours obtained using the exact ${\tau}_c$ rather than the reconstructed ${\hat\tau}_c$ are displayed in red color on the left panel. The dashed lines are the contours from Figures~\ref{fig:idealContour} and~\ref{fig:nlocontour}.} \label{fig:asimovContour}
\end{figure}

We applied this algorithm to draw $95\%$~CL expected exclusion contours in the plane of the two Wilson coefficients $c=(G_{\varphi},\, G_{W})$ that we studied in Section~\ref{sec:PS}. The results are displayed as continuous blue lines in Figure~\ref{fig:asimovContour}. The left/right panel of the figure shows the results for ideal/NLO simulations. The reconstructed distribution ratio is obtained from the parameterized classifier trained as explained in Sections~\ref{sec:idd} and~\ref{sec:nlod} for ideal and NLO simulations, respectively, using benchmark hyper-parameters. The plots are obtained computing $t_c^{\textrm{\sc{ml}}}(\data_{\textrm{B}}^{\cal{A}})$ on a grid of points on the plane. The $95\%$~CL contour is the region where $p_{\textrm{med}}^{\textrm{\sc{ml}}}(c)=5\%$, i.e. where $t_c^{\textrm{\sc{ml}}}(\data_{\textrm{B}}^{\cal{A}})=3.84$ because of the $\chi^2$ formula. The contours are drawn from an interpolation of the grid. The exact distribution ratio is available for ideal data. We can thus run the previous algorithm but use the exact ${\tau}_c(x)$ in place of the reconstructed ${\hat\tau}_c(x)$. The result is shown with a red continuous line on the left panel of the figure. The excellent agreement with the blue contour provides an additional cross-check of the accuracy of the distribution ratio reconstruction, though arguably a less powerful one than those we performed in Section~\ref{sec:PS} because it is based on an inherently less sensitive hypothesis test methodology than the likelihood-ratio test we employ as a performance metric.

The figure also displays, in dashed, the contours previously reported in Figures~\ref{fig:idealContour} and~\ref{fig:nlocontour}. These are the $5\%$ contours of the optimal Neyman--Pearson test $p$-value. They provide tighter exclusion bounds than the test based on the maximum likelihood ratio, as expected. However, the contours are relatively close, showing that the maximum likelihood ratio test is not vastly sub-optimal, in the case at hand.

\begin{figure}[t]
    \centering
    \includegraphics[width=0.45\columnwidth]{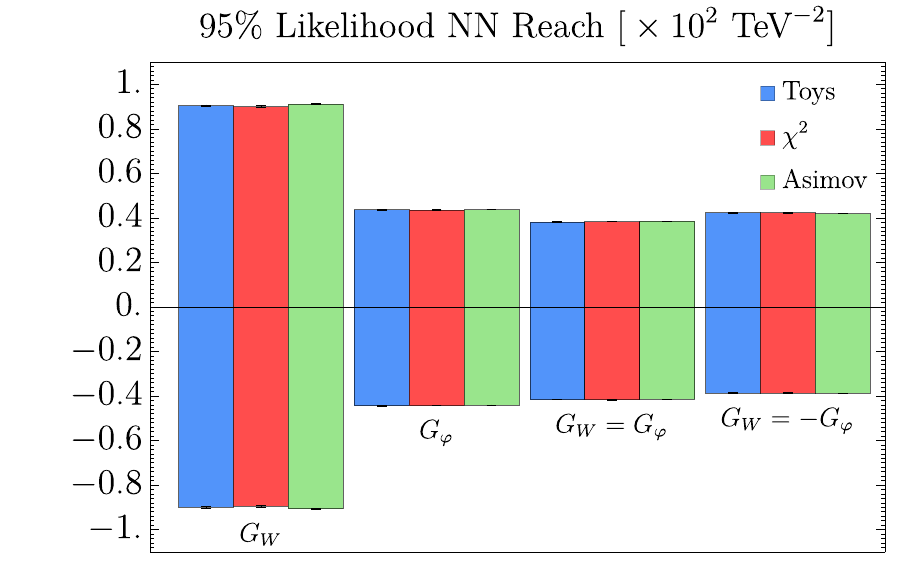}
    \caption{Upper and lower exclusion reach at $95\%$~CL along four directions in the Wilson coefficient plane $c=(G_{\varphi},\, G_{W})$. Blue bars are obtained empirically with toy data sets. Red bands make use of the Asymptotic $\chi^2$ approximation for the test statistics distribution under the null hypothesis, while the median $p$-value is determined empirically on the toys. The green band employs both the Asymptotic and the Asimov approximations as in eq.~(\ref{eq:pvasy}).} \label{fig:idealLikelihood}
\end{figure}

As a cross-check, we verify in Figure~\ref{fig:idealLikelihood} the validity of the Asymptotic and Asimov approximations of the maximum likelihood ratio test sensitivity. For this study, we generated Poisson-distributed toy data sets for the binned data under the null and under the alternative hypotheses, and we computed $t_c^{\textrm{\sc{ml}}}(\data_{\textrm{B}})$ from the definition~(\ref{eq:mlr}), performing the maximization along the $\gamma_\mu$ curve. The curve is taken to be a straight line interpolating from $c=0$ to the test point $c\neq0$. We determine by interpolation the eight points on the $c$ plane where the empirically-determined $p$-value equals $5\%$, along four directions in the plane. Ideal simulations are used for this study. The results are shown as blue bars in Figure~\ref{fig:idealLikelihood}. The green bars are obtained using the Asymptotic $\chi^2$ formula for the distribution under the null hypothesis and the Asimov data set for the determination of the median. They correspond to the intersection between the contour on the left panel of Figure~\ref{fig:asimovContour} and the four selected directions in the plane. The red bars in the figures employ instead the Asymptotic $\chi^2$ formula, but estimate the median by toy experiments performed under the alternative hypothesis. The three sets of results are in excellent agreement, confirming the accuracy of the Asymptotic and Asimov approximations. The result was widely expected because these approximations are well-established and routinely employed for binned data with sufficient statistics in each bin.

Our proposal for setting exclusion limits is based on concepts and approximations that are deeply routed in the LHC statistical practice. However, it should be stressed that an analysis based on our proposal is rather different from a regular LHC binned exclusion analysis, and potentially more demanding computationally. In a regular binned analysis, both the variable or variables used for binning and the bins are pre-specified. Here instead they depend on the point $c$ in the space of the parameters of interest that we are interested in probing. The $c=0$ and $c\neq0$ cross sections in each bin are thus computed once and for all in a regular analysis, while in our case they are computed point-by-point as previously explained, because we have access to the ${\hat\tau}_c(x)$ variable used for binning only after the relevant value of $c$ is specified. This is the price to pay for attaining nearly-optimal sensitivity: the variable to be binned is optimized point-by-point in the $c$ space. 

The additional computational cost, in comparison with the cost of a regular binned analysis on real LHC data, is not easy to quantify. A few considerations are reported below. A realistic LHC analysis is affected by systematic uncertainties incorporated in a typically large number of nuisance parameters. As previously explained, these parameters need to be incorporated in the likelihood by determining their effect on the expected $N_{\textrm{b}}(c,\nu)$ in each bin using dedicated Monte Carlo data sets. This calculation is not a significant addition to the total computational cost in a regular binned analysis, because it is performed once and for all in the pre-defined bins. The most costly part of the analysis is the minimization over the nuisance parameters, for each $c$, that has to be performed for the determination of the Asimov estimate of the median $t_c^{\textrm{\sc{ml}}}$ as in eq.~(\ref{eq:tasinu}). For our strategy, one needs to run the minimization, but also to first determine $N_{\textrm{b}}(c,\nu)$ at each point in the $c$ space. The relative computational cost of these two operations defines the additional cost of our proposal in comparison with a regular binned analysis. The Bayesian counterpart of minimization, i.e.~the integration over the nuisance, would pose a similar challenge in the alternative approach described in the next section. 

\subsection{Bayes factor}\label{se:bf}

The problem of comparing the agreement with data of two hypotheses $H_0$ and $H_1$---in our case, the hypotheses that $c$ assumes one given non-vanishing value, and the one that $c=0$, respectively---could be also addressed in a Bayesian framework without performing a test of hypothesis. In this case, one would need to compute the Bayes factor, which is the ratio of the posterior probabilities for the two hypotheses if the priors $P(H_1)$ and $P(H_0)$ are equal.~\footnote{In general, $P(H_1|\data)/P(H_0|\data)={\cal{B}}_{10}P(H_1)/P(H_0)$.} In the absence of nuisance parameters
\beq\label{eq:bfunb}
{\cal{B}}_{10}(\data)=\frac{P(H_1|\data)}{P(H_0|\data)}
=\frac{\mathfrak{L}(0;\data)}{\mathfrak{L}(c;\data)}=e^{-\lambda(c;{\cal D})}=e^{t_c(\data)/2}\,,
\eeq
with $\lambda$ the unbinned log-likelihood ratio in eq.~(\ref{eq:log-lik}) and $t_c(\data)$ the optimal Neyman–Pearson test statistic in eq.~(\ref{eq:tc}). A large value of ${\cal{B}}_{10}$, which corresponds to large and positive $t_c(\data)$, signals poor agreement of the $H_0$ hypothesis with the data and in turn a low posterior $H_0$ probability.

If the exact distribution log-ratio $\tau_c(x)$ was known, we could simply evaluate $t_c(\data)$, compute the Bayes factor~(\ref{eq:bfunb}) and, based on its value, draw conclusions on the plausibility of the null hypothesis $H_0$. This could be done on a representative data set drawn under the alternative $c=0$ hypotheses---or the median could be computed over $c=0$ toys---in order to determine the expected sensitivity of the analysis. However, we do not have access to $\tau_c(x)$, but only to the reconstructed $\hat\tau_c(x)$ and in turn to the reconstructed likelihood log-ratio $\hat{t}_c(\data)$ in eq.~(\ref{eq:tchat}). We can monitor the agreement between $\hat\tau_c$ and $\tau_c$, but we have no way to control quantitatively the departure of the reconstructed Byes factor ${\hat{\cal{B}}}_{10}=\exp({\hat{t}}_c/2)$ from the true Bayes factor. 

It should be noted that the remark above is an obstruction to employ the reconstructed Bayes factor for the analysis of the actual data, but an even more severe obstruction to employ it as a metric to assess the performances of the distribution log-ratio reconstruction. Errors on the determination of the log ratio can make  ${\hat{\cal{B}}}_{10}$ larger or smaller than the true ${{\cal{B}}}_{10}$. In particular, any overfitting that makes $\hat\tau_c$ large and negative produces a large and positive $\hat{t}_c$ and in turn large ${\hat{\cal{B}}}_{10}$. Therefore, monitoring ${\hat{\cal{B}}}_{10}$ would not give any information on the quality of the reconstruction and a model with larger ${\hat{\cal{B}}}_{10}$ could easily be worse than one with smaller ${\hat{\cal{B}}}_{10}$. The classical Neyman–Pearson median $p$-value is instead sensitive to the quality of the reconstruction owing to the Neyman–Pearson Lemma, which guarantees it to attain its absolute minimum when the true ratio is reconstructed, and obliges it to be larger if the reconstruction is less accurate. This is the reason why a classical hypothesis test was pursued in Section~\ref{sec:nppdef} to define a performance metric and Bayesian inference methods were not considered for that task.

We could, on the other hand, employ Bayesian inference on the analysis of the actual data, by proceeding as follows. The distribution of the binned data is known exactly, hence the Bayes factor can be exactly computed and reads
\beq\label{eq:baybin}
\log{\cal{B}}_{10}^{\textrm{B}}(\data_{\textrm{B}})=\sum\limits_{{\textrm{b}}\in{\textrm{B}}}
\left[
N_{\textrm{b}}(c)-N_{\textrm{b}}(0)-
\N_{\textrm{b}}
\log\frac{N_{\textrm{b}}(c)}{N_{\textrm{b}}(0)}
\right]
\,.
\eeq
We have seen that the binned data contain the same information as the original data set $\data$ if the binning is performed on the $\hat\tau_c(x)$ variable, provided this variable is a good approximation of $\tau_c(x)$. Errors in the $\tau_c(x)$ reconstruction can undermine the optimality of the analysis, entailing some loss of information to occur in the process of binning. However, they do not undermine eq.~(\ref{eq:baybin}) as a consistent calculation of the Bayes factor for the binned data, because the binned data are truly Poisson-distributed regardless of the variable that is employed for binning. 

It is interesting to notice that this binned Bayesian approach to exclusions is in fact very similar to its frequentist counterpart based on the maximum likelihood ratio that we defined in the previous section. Consider estimating the value of the Bayes factor that is expected to be observed if the data are distributed as predicted by the SM, i.e. for $c=0$. A reasonable estimate is to evaluate the Bayes factor on the Asimov data set defined by eq.~(\ref{eq:asds}). This gives
\beq
\log{\cal{B}}_{01}^{\textrm{B}}(\data_{\textrm{B}}^{\cal{A}})=\sum\limits_{{\textrm{b}}\in{\textrm{B}}}
\left[
N_{\textrm{b}}(c)-N_{\textrm{b}}(0)-
N_{\textrm{b}}(0)
\log\frac{N_{\textrm{b}}(c)}{N_{\textrm{b}}(0)}
\right]
\,.
\eeq
This expression is equal to half the maximum likelihood ratio test statistics evaluated on the Asimov set, as computed in eq.~(\ref{eq:tasi}). The $t_c^{\textrm{\sc{ml}}}(\data_{\textrm{B}}^{\cal{A}})=3.84$ contour lines in Figure~\ref{fig:asimovContour}, which correspond to $95\%$~CL frequentist exclusions, are thus also contours where the expected Bayes factor equals $\exp(3.84/2)=6.8$. The two approaches thus give the same results up to the conventional choice of the threshold for exclusion. On the other hand, notice that there is no exact correspondence between the value of $t_c^{\textrm{\sc{ml}}}$ that will be observed on the truly collected data set. Furthermore, nuisance parameters are treated differently in the Bayesian and in the frequentist frameworks leading to potentially different findings in the two cases.

\section{Conclusions and outlook}\label{sec:Conclusions}

We studied the advantages of employing reweighted Monte Carlo events for learning the dependence of the data distribution on the parameters of interest. We focused in particular on the Wilson coefficients of EFT interaction operators, in the perspective of their extensive investigation with the LHC and the forthcoming HL-LHC collider data. In Section~\ref{subsec:lfw} we defined our methodology. This extends our previous proposal in Ref.~\cite{Chen:2020mev} to exploit reweighted Monte Carlo data sets for training. In Section~\ref{sec:PS} we studied the performances of the new methodology on the same benchmark problems of Ref.~\cite{Chen:2020mev}, enabling direct comparisons. 

Our first finding is that reweighted training data enable an accurate learning of the distribution ratio even when the effect of the Wilson coefficients on the distribution is a small correction to the SM prediction. This can be crucial in order to learn effects---within or outside the EFT context---that cannot be parametrized and require point-by-point learning with realistic values of the parameters, for which the effects are indeed small. Furthermore, even if the effects can be parametrized like in the quadratic classifier setup, the capability of reweighted training to capture small effects lies at the heart of the other major advantage with respect to the methodology of Ref.~\cite{Chen:2020mev}: the weak sensitivity of the performances to the choice of the values of the Wilson coefficients used for the training of the model (see Section~\ref{sec:hyps}). The method of~\cite{Chen:2020mev} is on the contrary very sensitive to this choice. Even if the origin of this sensitivity is understood, and criteria are defined in Ref.~\cite{Chen:2020mev} for an optimal choice, the need of optimizing these hyper-parameters makes the proposal of Ref.~\cite{Chen:2020mev} harder to implement systematically or to automate. For the methodology of the present paper instead, a loose criterion of selecting the training points close to the sensitivity reach of the experiment is enough for  accurate learning.

In the table on the right panel of Figure~\ref{fig:idealContour} we report (in the ``$\hat{r}$'' entries) $95\%$~CL expected single-operator limits for the Wilson coefficient $G_W$ and $G_\varphi$ on ideal simulations. They are essentially identical in all directions of the plane to the limits obtained using the knowledge of the exact $r$ ratio (the ``$r$'' entries of the table). On the contrary, in Ref.~\cite{Chen:2020mev}, a slight gap in performances with the optimal reach could be observed in the $G_W$ direction.~\footnote{The relevant results are in Table~1 (specifically, the ``QC'' entries) of Ref.~\cite{Chen:2020mev}. Notice that the optimal reach was estimated in Ref.~\cite{Chen:2020mev} using the near-Gaussian approximation rather than using bins (see Section~\ref{sec:nppcalc}). This results in a slight discrepancy between the ``ME'' limits of Ref.~\cite{Chen:2020mev} and the ``$r$'' limits of Figure~\ref{fig:idealContour}.} While the sensitivity improvement is marginal, it should be noted that a total of 12M training points were used in Ref.~\cite{Chen:2020mev}, while only 3M are employed here. Furthermore, we saw in Section~\ref{sec:hyps} that almost identical performances would have been obtained using only 1M training points. This shows that reweighted training enables a more accurate learning with fewer points.

All the advantages of reweighting listed above are specific to the problem of learning the distribution ratio. On the other hand, our methodology also benefits from the generic advantage of reweighting, which is to enable predictions in the entire space of the parameters of interest using one single Monte Carlo sample generated at one point. 

In the paper we also advanced, in Section~\ref{subsec:val}, the design of performance metrics to assess the quality of the distribution ratio reconstruction. These metrics are extensions of ideas from Ref.~\cite{Chen:2020mev}, which however could be efficiently evaluated and employed systematically only thanks to the fast and accurate Monte Carlo predictions obtained with reweighting. The main performance indicator is the median $p$-value for the exclusion of EFT interactions. Its main advantage is to assess the quality of the distribution ratio reconstruction in relation with the actual experimental setup where it will be eventually employed for statistical inference. The sa\-tu\-ra\-tion of this quantity towards the absolute minimum---which can be only reached with perfect reconstruction---signals that the quality of the reconstruction is sufficient for optimal statistical inference using the data of the experiment under consideration. This defines a stopping criterion for the quality of the reconstruction, which should otherwise be indefinitely improved with larger neural networks and training data sets until when computationally feasible.

Finally, in Section~\ref{sec:PLS} we described how the learned distribution ratio could be used in practice for statistical inference on real data, or for sensitivity projections. We focused in particular on the task of setting exclusion limits point-by-point in the Wilson coefficient parameter space by comparing the agreement with data of the SM hypothesis---where the EFT Wilson coefficients vanish---with the hypothesis that they assume a given non-vanishing value. The exclusion will be set where the SM agreement strongly exceeds the EFT agreement. We saw that the problem can be addressed in a robust and computationally manageable manner, without major sensitivity loss, by using the reconstructed ratio to bin the data. A Bayesian approach to the problem is found to give the same result---in terms of sensitivity projections, and in the absence of nuisance parameters---of a classical approach that exploits Asymptotic formulas and the Asimov data set.

It should be noted that our ratio could be also used to reconstruct the unbinned likelihood and employed for a global EFT fit, like in Ref.~\cite{GomezAmbrosio:2022mpm}. On the other hand, performing a global fit is not necessarily the only or the most useful statistical inference task. The result of the fit depends on the prior expectations on the Wilson coefficients, which in turn depend on the microscopic physical model that gives origin to the effective interactions. Each microscopic physics scenario gives origin to one subset of all possible dimension-six EFT operators, or more generally it produces tight correlations between the Wilson coefficients of different operators. The correct prior that corresponds to each microscopic physics scenario should be used for a meaningful fit. A flat prior on all dimension-six operators, which is typically employed in EFT global fits, is not a valid prior because it does not correspond to the expectations of any known scenario: it never happens in concrete models that all operators are generated with comparable coefficients and no correlation among them. Point-by-point information in the Wilson coefficient parameter space, like the one we obtain with the strategy of Section~\ref{sec:PLS}, is instead meaningful regardless of the microscopic origin of the EFT interaction operators. The information can be transferred easily to the parameter space of any specific microscopic physics model.

Our methodology could be used for two distinct classes of applications. One is performing systematic or even automated EFT sensitivity projections. The other, is performing real analyses of the LHC data. The perspectives for progress in these two directions are discussed in turn.

Automated nearly-optimal sensitivity projections would be extremely beneficial because the number of EFT interaction operators, as well as the number of promising LHC processes to detect their presence, are both large. Furthermore, each process is characterized by a typically large number of potential observables that could help probe some of the possible EFT operators. One out of many examples of non-trivial observables are the decay angles of the bosons in the ZW process we studied in the present paper, and in Ref.~\cite{Chen:2020mev}. Their measurement offers access to the leading contribution from the ${\mathcal{O}}_W$ operator, which would on the contrary cancel exactly if only the kinematical distributions of the two bosons were measured. We would have discovered this fact empirically, if we had not known it theoretically, by observing strong sensitivity improvement due to the inclusion of the decay angles in the automated nearly-optimal analysis, and/or by observing improvement in comparison with the sensitivity of a regular analysis binned, for instance, on the transverse momentum of the bosons. A similarly intricate interplay between observables and operators is definitely at work in many other cases that have not been subject to a careful theoretical investigation. The discovery of these phenomena enabled by automated sensitivity projections might eventually lead to the design of suitable observables and binning strategies by which nearly optimal sensitivity could be attained without using the learned distribution ratio in the actual analysis of the data. Notice however that this has not yet been achieved for the $G_W$ operator in the ZW process: the binned analysis performed in Ref.~\cite{Chen:2020mev} including binning over the decay angles in order to access the relevant information about the decay of the bosons has a reach on $G_W$ that is a factor of two larger than the nearly-optimal reach obtained with our methodology.

We consider that all elements are in place for the deployment of our methodology for automated sensitivity projections. The scale of the problems to be faced is not different from the one we addressed in the present paper with relatively limited computational resources. It is likely that one would like to extend the analysis to a larger number of interaction operators than the two operators we considered in the present paper. However, as discussed in Section~\ref{qc-2}, the fact that the dependence of the distribution ratio on the Wilson coefficient is a quadratic polynomial implies that one can learn the dependence on any number of Wilson coefficients without the need of training with more than two non-vanishing coefficients. A parallelizable learning strategy was defined in Section~\ref{qc-2}, based on which we expect to be able to deal with tens of Wilson coefficients with reasonably available computational resources. It is possible that further methodological advances will be achieved in the future. A direct performance comparison with other approaches~\cite{Chatterjee:2022oco,GomezAmbrosio:2022mpm} would facilitate the emergence of new ideas for better and/or more efficient learning. However, the current state-of-the-art already enables the systematic exploration of several LHC processes and EFT operators.

The actual perspectives for systematically employing the learned distribution ratio in the analysis of the LHC data are instead still to be assessed. A major concern is the availability of synthetic data for training, validation and testing. For the present paper, we used a a factor of few tens of thousands more data than the expected number of data points at the HL-LHC. The majority of them were employed for testing purposes, namely for the evaluation of performance metrics to assess the quality of the reconstruction ratio. Less accurate quality studies than the one we performed here would require fewer testing points, and we saw that also the size of the training data set could have been reduced without major sensitivity degradation. Still, it is expected that order thousands of times more statistics than the actual data would be required for the implementation of our strategy. Monte Carlo simulations that simulate the response of the detector accurately are computationally costly, potentially preventing the generation of such large data sets. Notice however that detector effects are often modeled accurately by simplified simulation tools like Delphes~\cite{deFavereau:2013fsa}, which are fast to run. Training, validation and testing could be performed on Delphes samples, with the caveat that the learned ratio will be an approximation of the exact ratio that corresponds to the Delphes predictions, which is close but different from the true distribution ratio. Employing the Delphes ratio for statistical inference on the actual data will not entail a strong departure from optimality, if the Delphes simulation is accurate. Clearly it will be essential to employ a statistical inference methodology that is correct and internally consistent regardless of the agreement between the reconstructed and the true ratio, like the one we outlined in  Section~\ref{sec:PLS}. It is also essential that full detector simulations and not Delphes simulations are used to perform the analysis of the actual data and for sensitivity projections. The amount of simulated data required for this task is however expectedly not dissimilar from the one needed for a regular analysis. A realistic treatment of nuisance parameters will be also needed and this could challenge, for instance, the practical implementation of our proposal in Section~\ref{sec:PLS} for real data analysis. The viability of this strategy should be assessed for individual processes before drawing conclusions on the perspectives for the applicability of our strategy to real data. It should also be noted that the SM background data sets in real analyses of the LHC data often contain components that are estimated by data in a control region and not by running a Monte Carlo code. These data can be used like Monte Carlo data, but they must be available with sufficient statistics and correctly model the distribution of all the relevant observables.

It is worth mentioning that strategies like ours aimed at learning the dependence of the data distribution on parameters can find applications also outside the context of EFT searches. One such application is to learn the dependence on nuisance parameters locally in the space of observables $x$. Our parametrized classifier approach is perfectly suited for this task because the nuisance parameters are precisely a parametrization of the effects of imperfections in the theoretical predictions. Their effects are relatively small, hence they are typically parametrized with a linear polynomial or other functional forms like an exponential scaling factor to enforce cross section positivity strictly. All these options can be readily implemented in a parametrized classifier as done in Ref.~\cite{dAgnolo:2021aun} to learn the unbinned SM likelihood in order to incorporate the effects of nuisance in an unbinned strategy for agnostic new physics searches. When available, reweighted Monte Carlo data sets can facilitate the learning task. A clear case is the one of uncertainties on the parton distribution functions, which are incorporated in Monte Carlo event generators by event reweighting. It is expected that our methodology could learn the effect of the many nuisance parameters associated with parton distribution functions extremely efficiently.

\subsection*{Acknowledgements}
A.W.~is supported by the grant PID2020-115845GB- I00/AEI/10.13039/501100011033. G.P.~was supported in part by the MIUR under contract 2017FMJFMW (PRIN2017).

\appendix

\section{Fully-leptonic ZW at high energy}\label{app}

We consider the production of a Z and of a W boson at the 14~TeV LHC, with a lower cut of $300$~GeV on their transverse momentum. The process is dominantly mediated by the reactions
\beq\label{eq:partp}
u_L\,{\overline{d}}_L\to Z\,W^+\;\;
{\textrm{and}}\;\;d_L\,{\overline{u}}_L\to Z\,W^-\,,
\eeq
depending on the charge of the produced W boson. Leptonic decays are considered for both bosons, namely
\beq\label{eq:bosdec}
Z\to \ell^+_\chi\ell^-_\chi\,,\;\;\;\;\;W^{\pm}\to
\ell^\pm_L \overset{\scriptscriptstyle(-)}{\nu}_{\hspace{-3pt}{L}}\,,
\eeq
where $\ell$ is either an electron or a muon and $\chi=L,R$ denotes fermion chirality.

The observable final state contains three leptons. The lepton pair that emerges from the decay of the Z boson can be identified as having opposite charge, same flavor (electron or muon) and, in case of ambiguity, the closest invariant mass to the one of the Z particle. The lepton chirality cannot be measured, hence the final states that are relevant for our analysis are labeled by a total of three discrete indices: the total charge $Q=\pm1$ of the three leptons, which corresponds to the W boson charge; the flavor $\ell_Z=e,\mu$ of the leptons emerging from the decay of the Z; the flavor $\ell_W=e,\mu$ of the W boson decay. These three labels can be given as input to the neural network.

The neutrino escapes detection, however its momentum can be reconstructed starting from the total missing transverse energy of the event by imposing that it originates, together with an observable lepton, from the decay of a W boson. The reconstruction is performed by imposing that the invariant mass of the pair formed by the neutrino and the visible lepton is equal to the mass of the W particle. Since this defines a quadratic equation with two solutions, the neutrino momentum reconstruction is possible only up to a twofold ambiguity. We resolve this ambiguity by a random choice of the two solutions with equal probability on an event-by-event basis.\footnote{The on-shell equation has no solution when the lepton transverse mass exceeds $m_W$ because of experimental errors or because the W resonance was far from its mass-shell. In this case, the neutrino momentum is reconstructed by taking its rapidity to be equal to the one of the lepton. See~\cite{Panico:2017frx} for details.} After the reconstruction, the following 7 variables are known (see~\cite{Chen:2020mev} and references therein) to be useful to characterize the kinematics of the event
\beq
\{ s,\,p_{T,{\textrm{ZW}}},\,\Theta,\, \theta_W,\, \varphi_W,\, \theta_Z,\, \varphi_Z\}\,.
\eeq
The first three variables describe the kinematics of the ZW pair, namely its total invariant mass squared, its total transverse momentum, and the angle of the Z boson in the ZW rest frame, relative to the direction of its boost in the lab frame. The remaining four variables are the polar ($\theta$) and azimuthal ($\varphi$) decay angles of the bosons in their rest frames. The precise definition of these variables is detailed in~\cite{Chen:2020mev,Panico:2017frx}. 

We aim, as in~\cite{Chen:2020mev}, at studying the two specific dimension-six interaction operators~\footnote{We use the definition $H^ \dagger {\scriptstyle \overleftrightarrow{\rule{0pt}{.75em}}}\hspace{-.95em}{D}_\mu H = H^\dagger D_\mu H - (D_\mu H)^\dagger H$.}
\beq
{\cal O}_{\varphi} = G_{\varphi} \left(\overline Q_L \sigma^a \gamma^\mu Q_L\right)
(i H^ \dagger {\scriptstyle \overleftrightarrow{\rule{0pt}{.75em}}}\hspace{-.95em}{D}_\mu H)\,,\qquad\quad
{\cal O}_{W} = G_{W} \varepsilon_{abc} {W^{a\,\nu}_{\mu}} {W^{b\,\rho}_{\nu}} {W^{c\,\mu}_\rho}\,,
\label{operators}
\eeq
by modeling their effects on the distributions with two different Monte Carlo event generators: the ideal and the NLO generators. The ideal generator is based on a simplified description of the process, relying on approximations that are not sufficiently accurate for the description of the actual experimental data. However, it produces distributions that are similar to those obtained with realistic simulations, and defines a learning problem with the same scale of complexity. At the same time, its simplifying assumptions enable the exact analytical calculation of the distribution ratio $r(x;c)$, to be compared with the reconstructed ratio ${\hat{r}}(x;c)$. This comparison provides an important validation of our methodology. The NLO generator offers instead an accurate description of the process, including radiative corrections from QCD at the one-loop order. This additional theoretical complexity does not change the complexity of the learning problem: the NLO distributions are similar to the ideal ones, the number of variables is (almost, see below) the same and no new relevant sub-process kicks in at NLO in the phase-space region that is relevant for the analysis. On the other hand, introducing NLO effects confronts our methodology with the novel challenge of negative weights. In fact, NLO corrections can be included in Monte Carlo data sets only by allowing for some of the events to have negative weight $w_{\textrm{e}}(c)<0$. Such events give a negative contribution to the loss function, encouraging overfitting. It is thus important to validate our methodology also on NLO data. No performance degradation will be observed, showing that our approach is perfectly suited to deal with negative-weight Monte Carlo data, at least if the fraction of negative weight is as small as for the state-of-the-art NLO generator that we employ in our study.

The next two sections provide a technical description of the ideal and of the NLO generators.

\subsection{Ideal generator}

The ideal generator relies, in the first place, on the narrow-width approximation for the decay of the two bosons. This enables us to factorize the scattering amplitude for the partonic processes~(\ref{eq:partp}) in terms of a ``hard'' amplitude describing the production of on-shell bosons, times the amplitudes that describe the decay of the bosons. The contributions of intermediate on-shell bosons with different helicities need to be summed up at the amplitude level, leading to the following expression for the differential cross section~\cite{Chen:2020mev,Panico:2017frx}
\beq\label{acs}
d \sigma(\xi;c) = \sum\limits_{h,h^\prime} d \rho^{\textrm hard}_{h_Z^{\phantom\prime} h_W^{\phantom\prime} h'_W h'_Z} d \rho^{Z}_{h_Z^{\phantom\prime} h'_Z}
d \rho^{W}_{h_W^{\phantom\prime} h'_W}\,.
\eeq

In the equation, $d \rho^{\textrm hard}$ is the density matrix for polarized WZ production, namely
\beq\label{eq:hardd}
d \rho^{\textrm hard}_{h_Z^{\phantom\prime} h_W^{\phantom\prime} h'_Z h'_W} =\frac1{24\,s}{\cal M}_{h_Z^{\phantom\prime} h_W^{\phantom\prime}} ({\cal M}_{h_Z^{\prime} h_W^{\prime}})^*\, d \Phi_{\textrm{ZW}} \,,
\eeq
where $d \Phi_{\textrm{ZW}}$ is the phase-space factor. Since we are studying high-energy production, we can consider the limit of massless vector bosons. The non-vanishing production amplitudes in this limit read
\begin{eqnarray}\label{eq:amplitudes}
&\displaystyle{\cal M}_{00} = - \frac{g^2\sin \bar\Theta}{2 \sqrt{2}} - \sqrt{2} G_{\varphi }  \bar{s} \sin \bar\Theta\,,\qquad\quad
{\cal M}_{++}={\cal M}_{--}= \frac{3 g c_{\textrm{w}} G_W  \bar{s} \sin \bar{\Theta}}{\sqrt{2}}\,,&\\
&\displaystyle{\cal M}_{-+} = - \frac{g^2(s_{\textrm{w}}^2- 3 \, c_{\textrm{w}}^2 \cos \bar\Theta)}{3 \sqrt{2} c_{\textrm{w}}} \cot \frac{\bar\Theta}{2}\,,\qquad\quad
{\cal M}_{+-} = \frac{g^2(s_{\textrm{w}}^2 - 3 c_{\textrm{w}}^2 \cos \bar\Theta)}{3 \sqrt{2} c_{\textrm{w}}} \tan \frac{\bar\Theta}{2}\,,&\nonumber
\end{eqnarray}
where $g$ is the SU$(2)_L$ coupling, $c_{\textrm{w}}$ and $s_{\textrm{w}}$ are the cosine and the sine of the Weak angle.\footnote{An overall factor equal to the cosine of the Cabibbo angle has not been reported for shortness. The ones above are the amplitudes for $W^+$ production. Those of the $d{\overline{u}}\to ZW^-$ process can be obtained with the formal substitutions $\bar\Theta \rightarrow - \bar\Theta$ and $s_{\textrm{w}}^2 \rightarrow - s_{\textrm{w}}^2$.} We denote as $\bar{s}$ and $\bar\Theta$ the center of mass energy squared of the ZW system and the polar angle of the Z boson in order to distinguish them from the observable variables ${s}$ and $\Theta$. The relation between these quantities will be discussed later.

The density matrices $d \rho^{W,Z}$ describe the decay of the bosons. They take the generic form
\beq
d \rho^{V}_{h_V^{\phantom\prime} h'_V} = \frac{1}{2m_V\Gamma_V} {\cal A}_{h_V^{\phantom{\prime}}}^{\phantom{*}} {\cal A}^*_{h'_V}\,,\;\textrm{with}\;\;{\cal A}_{h} = - \sqrt{2} g_{V} m_{V} e^{i h \bar\varphi} d_{h}(\bar\theta)\,,
\eeq
where $d_h$ are Wigner-$d$ functions, $\bar\theta$ and $\bar\varphi$ are the angles of the fermion or anti-fermion of $+1/2$ helicity produced in the decay in the rest frame of each boson. Once again, they are denoted with a bar to distinguish them from the observable quantities. See Ref.~\cite{Chen:2020mev} for details and for the explicit expression of the $g_{V}$ couplings. Notice that the coupling for the Z boson decay depends on the chirality of the final-state leptons, denoted as $\chi$ in eq.~(\ref{eq:bosdec}).

We saw in Section~\ref{sec:intro} that Monte Carlo generators work by sampling in a space of latent variables $\xi$, which are later projected into the space of the observable variables $x$. It is instructive to illustrate these generic considerations in the concrete case at hand, starting from listing the latent-space variables and next discussing their relation with the observables. The partonic cross section~(\ref{acs}) depends on the following latent variables
\beq\label{eq:latent}
\xi=\{
Q,\,
\chi,\,
\bar{s},\,
\bar\Theta,\,
\bar\theta_W,\, 
\bar\varphi_W,\, 
\bar\theta_Z,\, 
\bar\varphi_Z\,,y
\}\,.
\eeq
Notice that the list does not include the total ZW transverse momentum $p_{T,{\textrm{ZW}}}$, which vanishes exactly at tree level, and the flavor of the leptons produced in the decays, $\ell_{W,Z}$. The latter variables are excluded because electrons and muons behave identically in our approximation. The latent variable vector also includes the rapidity, $y$, of the partonic system center of mass frame. The partonic cross section~(\ref{acs}) is boost-invariant and hence independent of $y$. However, the rapidity is needed for the complete characterization of the kinematics of the event.

The electric charge of the W boson, $Q$, is the only variable in eq.~(\ref{eq:latent}) that can be measured experimentally, as the sum of the lepton charges. The chirality of the leptons from the Z boson decay, $\chi$, is not observable. All the kinematical variables including $\bar{s}$ and $\bar\Theta$ depend on the momentum of the neutrino, which is not measured directly. Actually, even a direct measurement of the neutrino momentum would not enable the complete determination of all the variables. In particular, the Z boson angle $\bar\Theta$ must be measured relative to the direction of the incoming quark, which cannot be determined from the kinematics of the final-state particles. We circumvent this issue by assuming---and enforcing in the simulation---that the incoming quark is always more energetic than the anti-quark. This is a reasonable approximation given the shape of the corresponding parton distribution functions. The approximation enables us to determine the direction of motion of the quark as the direction of the boost of the ZW system.

The difficulty in relating the latent-space variables to the observables is immaterial for event generation. We can merely sample the $\xi$ variables according to their distribution~(\ref{acs})~\footnote{We duly take into account parton distribution functions, using the nCTEQ15~\cite{Kusina:2015vfa} set and the  {\tt{ManeParse}}~\cite{Clark:2016jgm} Mathematica implementation. See~\cite{Chen:2020mev} for details.}, compute the momentum of the three leptons and obtain the observables
\beq\label{feat}
x=\{ Q\,, s,\,\Theta,\, \theta_W,\, \varphi_W,\, \theta_Z,\, \varphi_Z\}\,,
\eeq
by running on the Monte Carlo data the same reconstruction algorithm that we would employ on the real data. The reconstruction involves the determination of the momentum of the neutrino which we perform, as previously explained, by imposing the on-shell condition for the W boson. A random choice is performed between the two solutions. The corresponding binary random variable should be regarded as an additional component of the latent-space vector~(\ref{eq:latent}).

Unlike in Ref.~\cite{Chen:2020mev}, we are interested in Monte Carlo data sets that incorporate the dependence on the Wilson coefficients $c=(G_{\varphi},\, G_{W})$ by the technique of event reweighting. We thus generate unweighted events at the SM point $c=0$ and assign to each of them a weight
\beq\label{rw:lat}
w_{\textrm{e}}(c)=w_{\textrm{e}}(0)\frac{d\sigma(\xi_{\textrm{e}},\;c)}{d\sigma(\xi_{\textrm{e}},\;0)}\,,
\eeq
where $w_{\textrm{e}}(0)$ is equal to the total cross section---as obtained from the Monte Carlo integration---divided by the number of events in the data set. The determination of the weights is trivial, owing to the knowledge of the value assumed by the latent variables for each event, $\xi_{\textrm{e}}$, and to the analytic knowledge of the cross section~(\ref{acs}) in the latent space. On the other hand, the observable variables $x$ do not contain enough information to compute the weights. In particular, this means that the weights cannot be computed on real data points. Notice  $d\sigma(\xi,\,c)$, and in turn $w_{\textrm{e}}(c)$---is a quadratic polynomial of $c$. The coefficients of this polynomial are stored in the event file, providing the complete representation of the $c$-dependent weight function $w_{\textrm{e}}(c)$ for each event.

The existence of a complicated relation between the latent and the observable variables is immaterial for event generation and for event reweighting, but it prevents in general the analytical calculation of the differential cross section, $d\sigma(x;c)$, in the space of observables. In turn, this prevents the determination of the cross section ratio $r(x;c)$. However, the relation between latent and observable variables is in fact simple within the assumptions that underpin the ideal description of the ZW process. In the narrow width approximation, the W particle is exactly on-shell, therefore the true momentum of the neutrino obeys the on-shell condition exactly. One of the two solutions to the on-shell equation---call them $p_\nu^{(a)}$ and $p_\nu^{(b)}$---is thus exactly equal to the true momentum. Furthermore, it was noticed in Ref.~\cite{Panico:2017frx} that a very simple relation exists between the two solutions in the massless limit $m_W\to0$. The reconstructed neutrino momenta approach each other in the limit, $p_\nu^{(a)} = p_\nu^{(b)}$, but they become collinear to the momentum of the lepton produced in the W decay. The W decay angles determined on the two solutions depend on how the collinear limit is approached, and they are not identical. One finds that
\beq\label{eq:wamb}
\varphi_W^{(a)}=\pi-\varphi_W^{(b)}\,,
\eeq
while the two determinations of the polar angle are instead equal, $\theta_W^{(a)}=\theta_W^{(b)}$. All the other kinematical variables are identical on the two solutions because they both reconstruct the correct W boson momentum $p_W=p_\nu+p_\ell$. The reconstruction of the observable variables~(\ref{feat}), starting from the latent variables~(\ref{eq:latent}), thus simply amounts to identify $s,\,\Theta,\, \theta_W,\, \theta_Z$ and $\varphi_Z$ with their barred version and to assign
\beq\label{eq:wamb1}
\varphi_W =\bar\varphi_W\;\;{\textrm{or}}\;\;
\varphi_W =\pi - \bar\varphi_W\,,
\eeq
at random with equal probability.

These simplifications allow us to compute the differential cross section $d\sigma(x;c)$ easily, obtaining an exact analytical expression for the distribution ratio
\beq
r(x;c)=\frac{d\sigma(x;c)}{d\sigma(x;0)}\,.
\eeq
As discussed in~\cite{Chen:2020mev}, one simply needs to average over the twofold ambiguity~(\ref{eq:wamb1}) in the reconstruction of $\varphi_W$, and to sum over the two possible chiralities $\chi=L,R$ of the Z boson decay. The knowledge of the exact distribution ratio will be used to validate the performances of our method on ideal data.

Given that $d\sigma(x;c)$ can be computed, our strategy for event generation appears unnecessarily complicated: we could have sampled directly in the $x$ space and ignored latent variables altogether. Event reweighting could have been also easily implemented, leading to weights
\beq\label{rw:nolat}
\tilde{w}_{\textrm{e}}(c)=w_{\textrm{e}}(0)\frac{d\sigma(x_{\textrm{e}};c)}{d\sigma(x_{\textrm{e}};0)}=
w_{\textrm{e}}(0)\,r(x_{\textrm{e}};c)\,.
\eeq
Unlike the ${w}_{\textrm{e}}$ weights that we will employ in the rest of this paper, defined by eq.~(\ref{rw:lat}), the $\tilde{w}_{\textrm{e}}$ weights are uniquely determined by the observable variables $x$. Furthermore, they are proportional to the $r$-ratio computed on the event. However, this alternative approach to event generation that does not employ latent variables is not viable, because our purpose is to define an idealized but still realistic setup to validate our ability to learn $r$. The cross section in the $x$ space cannot be computed for realistic simulators. The usage of latent variables is mandatory and we do not have access to the $\tilde{w}_{\textrm{e}}$ weights in eq.~(\ref{rw:nolat}), but only to the $w_{\textrm{e}}$ weights~(\ref{rw:lat}) in the latent space. Using the knowledge of the $\tilde{w}_{\textrm{e}}$ weights would define an unrealistically simple problem that is not suited for the validation of our methodology.

\begin{figure}[t]
    \centering
\includegraphics[width=0.5\columnwidth]{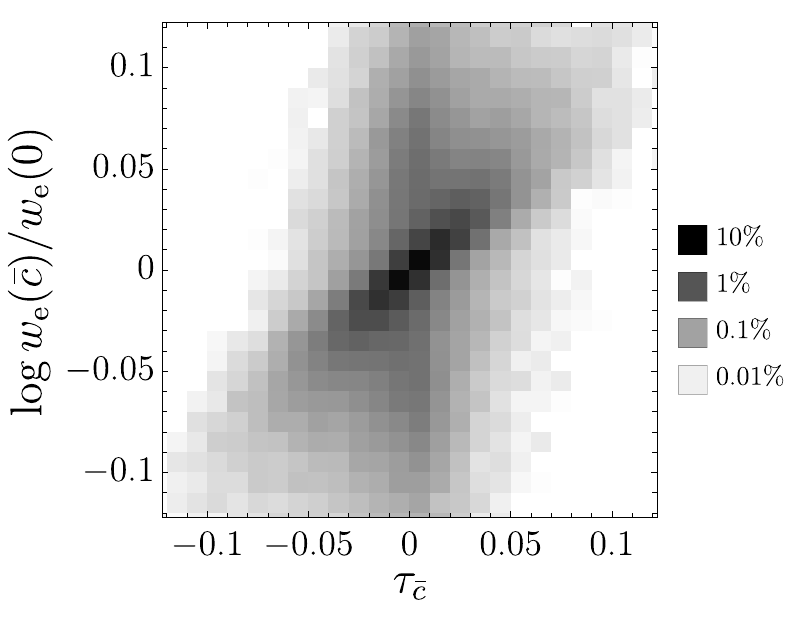}
    \caption{Density histogram of the correlation, on ideal Monte Carlo data, between the true distribution log-ratio $\tau(x_{\textrm{e}})=\log r(x_{\textrm{e}}
    ,\bar{c})$ and the logarithm of the weight ratio ${w}_{\textrm{e}}(\bar{c})/{w}_{\textrm{e}}(0)$. The $\bar{c}$ point has $G_{\varphi}=0$ and $G_{W}=10^{-2}\,{\textrm{TeV}}^{-2}$.} \label{fig:scatterreweights}
\end{figure}

It is instructive to discuss in more detail why employing the $\tilde{w}_{\textrm{e}}$ weights in place of the ${w}_{\textrm{e}}$ weights would define an overly simple problem. Consider learning the distribution ratio at a specific point $c=\bar{c}$ in the parameter space, as described in Section~\ref{subsec:sc}. We have shown that by employing a flexible set of trainable functions to model the classification function $f(x)\in(0,1)$ we obtain, after training with the loss function in eq.~(\ref{eq:lossw}), a good approximation ${\hat{r}}(x;\bar{c})$ of the distribution ratio. Owing to eq.~(\ref{eq:opc}), after training 
\beq\label{eq:tarf}
f(x)=\frac1{1+\hat{r}(x;\bar{c})}\,.
\eeq
This result holds provided the training data offer a consistent representation of the $x$ distribution, which is the case both if the reweighting is performed in the space of latent variables with the ${w}_{\textrm{e}}$ weights, or if instead the $\tilde{w}_{\textrm{e}}$ weights are used. However, in the latter case the training problem becomes trivial because the loss function in eq.~(\ref{eq:lossw}) can be rewritten as 
\beq\label{eq:lossreg}
\ell[f(\cdot)]=
\sum\limits_{{\textrm{e}}\in {\textrm{S}}}
\big[w_{\textrm{e}}({\bar{c}})+w_{\textrm{e}}(0)\big]\left[
f(x_{\textrm{e}})-\frac1{1+w_{\textrm{e}}({\bar{c}})/w_{\textrm{e}}({{0}})}
\right]^2
+\sum\limits_{{\textrm{e}}\in {\textrm{S}}}\frac{w_{\textrm{e}}(c)\,w_{\textrm{e}}(0)}{w_{\textrm{e}}(c)+w_{\textrm{e}}(0)}\,.
\eeq
The second term is a constant, irrelevant for the minimization. Consider the first term, and suppose employing the $\tilde{w}_{\textrm{e}}$ weights in place of the ${w}_{\textrm{e}}$ weights. Since $\tilde{w}_{\textrm{e}}(c)/\tilde{w}_{\textrm{e}}(0)=r(x_{\textrm{e}},\,\bar{c})$, by eq.~(\ref{rw:nolat}), we see that the first term of the loss function is proportional to the square of the difference between the trainable function $f$ and the target function $1/(1+r)$, evaluated on the training points. This is the typical form of the loss function one would employ for regression. In fact, eq.~(\ref{eq:lossreg}) shows that our learning problem reduces to a trivial problem of regression without noise, if the $\tilde{w}_{\textrm{e}}$ weights were used for training. This would not offer a fair representation of the realistic problem.

In order to appreciate how far our actual problem is from a regression, we show in Figure~\ref{fig:scatterreweights} a density histogram that displays the correlation between the true distribution ratio $r(x_{\textrm{e}};\bar{c})$ and the ratio of the weights, $w_{\textrm{e}}(\bar{c})/w_{\textrm{e}}(0)$, on the generated events. More precisely, we plot $\tau=\log{r}$ versus the logarithm of the weights ratio. The $\bar{c}$ point used for illustration has $G_{\varphi}=0$ and $G_{W}=10^{-2}\,{\textrm{TeV}}^{-2}$. We observe a loose correlation between the two quantities. Our problem is thus very far from a regression. It could still be described as a regression, but one with large and non-Gaussian noise. We can compare Figure~\ref{fig:scatterreweights} with the central and right panels of Figure~\ref{fig:scatter} in Section~\ref{sec:idd}, which displays the much tighter correlation between the true $\tau$ and the reconstructed $\hat\tau=\log\hat{r}$, obtained of course from training with the ${w}_{\textrm{e}}$ weights and not with the $\tilde{w}_{\textrm{e}}$ weights. 

\subsection{NLO generator}\label{sec:nlogen}

Our NLO event generator is based on the {\sc{MadGraph\_aMC@NLO}}~\cite{Alwall:2014hca} software suite, version 2.9.3, using Pythia~8.245~\cite{Sjostrand:2006za, Sjostrand:2007gs} for showering. We generate the complete $2\to 4$ processes
\beq
p p \to \ell^+ \ell^+ \ell^- \nu_\ell \;\; \textrm{and}\;\; p p \to \ell^- \ell^- \ell^+ \bar\nu_\ell\,,
\eeq
with $\ell = e, \mu$. Loose generation-level cuts are imposed to avoid singularities. The generated events are passed through Delphes~\cite{deFavereau:2013fsa} for clustering. The Delphes ``GenJet'' objects, which do no include detector effects, are used for the analysis. Clustering is performed with the anti-$k_T$ algorithm with cone radius $\Delta R = 0.1$. If leptons and photons carry more than 90\% of the GenJet cluster, the cluster is labeled as a lepton with the same flavor of the most energetic lepton. Otherwise, it is labeled as a QCD jet. 

We impose the following acceptance cuts on the reconstructed leptons 
\beq
    p_T^\ell > 25\text{ GeV} \,, \quad \eta^\ell < 2.5 \,,
    \eeq
and a minimum separation $\Delta R > 0.4$ between the lepton and the jets. Only events with exactly three leptons that satisfy these requirements---two of which with opposite charges but the same flavor---are kept for the analysis. We further ask for a minimum missing energy ${\slashed{E}}_T >30\text{ GeV}$. 

The leptons coming from the $Z$ boson have same flavor and opposite charge. If more than one lepton pair exists with these properties---because all three leptons have the same flavor---the pair with invariant mass closest to the mass of the $Z$ boson is chosen. The other lepton is assumed to be coming from a $W$ boson with the same charge as the lepton. After the Z and W decay products are identified, we compute the invariant mass $m_{Z^*}$ of the virtual Z, the transverse mass $m_{T,W^*}$ of the W boson and the transverse momenta of the Z and of the W, and we impose the selection cuts
\beq
70\text{ GeV} < m_{Z^*} < 110\text{ GeV} \,,\quad m_{T,W^*} < 90\text{ GeV} \,,
\quad \mathrm{min} [p_{T,W}, p_{T,Z}] > 300 \text{ GeV}\,.
\eeq
We finally reconstruct the neutrino momentum as previously explained and compute the vector of observables
\beq\label{featNLO}
x=\{Q,\,\ell_Z,\,\ell_W\,, s,\,\Theta,\, \theta_W,\, \varphi_W,\, \theta_Z,\, \varphi_Z,\, p_{T,{\textrm{ZW}}}\}\,.
\eeq

The events are generated according to the SM hypothesis $c=0$ and are then reweighted. We use the automated reweighting at NLO accuracy that is available in {\sc{ MadGraph\_aMC@NLO}}, with the implementation of the EFT operators~\eqref{operators} that is provided in the SMEFTatNLO model~\cite{Degrande:2020evl}. The coefficients of the quadratic polynomial that defines $w_{\textrm{e}}(c)$ are computed and stored in the event file.

In comparison with the ideal data in eq.~(\ref{feat}), more observables are considered at NLO~(\ref{featNLO}). The flavor $\ell_{Z,W}$ of the leptons are included because the NLO generator accounts for the effect of QED showering, through Pythia. The electron and the muon shower differently, hence their distributions are not identical unlike in ideal data. These additional discrete labels, like the charge $Q$, will be given as input to the neural network. NLO data also include the total transverse momentum of the ZW system, $p_{T,{\textrm{ZW}}}$, which emerges from the real QCD radiation.


\providecommand{\href}[2]{#2}\begingroup\raggedright\endgroup

\end{document}